\pdfoutput=1
\documentclass[%
 reprint,
 amsmath,amssymb,
 aps,
pra,
floatfix,
]{revtex4-1}

\usepackage{graphicx}
\usepackage{dcolumn}
\usepackage{bm}
\usepackage{subfigure}
\usepackage{hyperref}
\usepackage{wrapfig}
\usepackage{placeins}
\usepackage{natbib}
\usepackage{morefloats}
\usepackage{bbold}
\usepackage{xcolor}

\begin{document}


\title{Diverse tunable dynamics of two quantum random walkers 
}
\author{Shrabanti Dhar}
\affiliation{Department of Physics, Gokhale Memorial Girls College, 1/1 Harish Mukherjee Road, Kolkata-700020, India}
\affiliation{Department of Physics, Aliah University, II-A/27, Newtown, Kolkata-700160, India.}
\author{Abdul Khaleque}
\affiliation{Department of Physics, Bidhan Chandra College, Hooghly-712248, India.}
\author{Tushar Kanti Bose}
\email{tkb.tkbose@gmail.com}
\affiliation{School of Physical Sciences, Indian Association for the Cultivation of Science, 2A \& 2B Raja S.C. Mullick Road, Jadavpur, Kolkata-700032, India.}

\date{\today}
\begin{abstract}
Quantum walk research has mainly focused on evolutions due to repeated applications of time-independent unitary coin operators. However, the idea of controlling the single particle evolution using time-dependent unitary coins has still been a subject of multiple studies as it not only hosts interesting possibilities for quantum information processing but also opens a much richer array of phenomena including static and dynamic localizations. So far, such studies have been performed only for single quantum walkers. In case of multi-walker systems, time-dependent coins may generate measurable phenomena not described by the single-particle model, due to entanglement and interaction among the walkers. In this context, we present here a thorough numerical study of an one dimensional system of two quantum walkers exhibiting rich collective dynamics controlled by simple time­-dependent unitary coins proposed in [Phys. Rev. A \textbf{80}, 042332(2009)] and [Phys. Rev. A \textbf{73},062304(2006)]. We study how the interplay of coin time-dependence, simple interaction schemes, entanglement and the relative phase between the coin states of the particles influences the evolution of the quantum walk. The results show that the system offers a rich variety of collective dynamical behavior while being controlled by time dependent coins. In particular, we find and characterize fascinating two-body localization phenomena with tunable quasiperiodic dynamics of correlations and entanglements which are quantities of quantum origin. 





\end{abstract}

\pacs{Valid PACS appear here}
\maketitle

\section{Introduction}

Quantum random walk or quantum walk (QW) in its original form is simply the dynamics of a quantum particle that has a spin-1/2-like internal degree of freedom in addition to its position and momentum \cite{first}. Being a natural quantum version of the classical random walk that appears in statistics, computer science, finance, physics, chemistry and biology, it has been a topic of fundamental interest \cite{review}. Moreover, QW research now enjoy broader interest due to its widespread applications in the areas of quantum algorithms \cite{kempe}, quantum computing \cite{q_computing}, quantum biology \cite{q_biology}, and quantum simulation \cite{q_simulation}.\\ 

The dynamics of a quantum walker is usually controlled by two unitary operators : a rotation operator \(\hat{C}\) (called ``quantum coin") and a shift operator \(\hat{S}\). The coin operator acts on the walker's internal degrees of freedom, leaving it generally in a superposition of spin up and spin down. The shift operator then shifts the position according to the walker's internal degree of freedom. Hence, the internal and external degrees of freedom becomes entangled. Successive applications of the two operators (\(\hat{C}\) \& \(\hat{S}\)) generate discrete time evolution of the walker. This is what we call the one-dimensional discrete time QW. One major advantage of QW over the classical random walk is that a quantum walker spreads over the line linearly in time (standard deviation \(\sigma\sim t\)), while the classical random walk spreads in a slower fashion (\(\sigma\sim t^{1/2}\)).\\

 A QW with multiple particles contains quantum resources like multi-particle quantum correlations and multi-partite entanglement which have no classical analogue.  Moreover, in case of identical particles, quantum statistics gives an additional feature to QWs that can also be exploited. In 2006, Omar et al. first extended the idea of single particle QW to the case of two particles \cite{Omar}. They showed that a QW with two particles can indeed behave very differently from two independent single-particle QWs even in the absence of any inter-particle interactions \cite{Omar}. In particular, the probability to find at least one particle in a certain position after some steps of the walk, as well as the average distance between the two particles, was shown to be larger or smaller than the case of two unentangled particles, depending on the initial conditions \cite{Omar}. Thereafter, the topic of two-particle QW has attracted significant attention. Berry and Wang considered simple interaction schemes between two particles and showed that the interactions lead to a diverse range of probability distribution that depend on correlations and relative phases between the initial coin states of the two particles \cite{Berry}. They also showed that two interacting walkers can be used to distinguish all nonisomorphic strongly regular graphs \cite{Berry}. Stefanak et al. showed that the directional correlations between two interacting particles can exceed the limits for non-interacting particles \cite{Stefanak2}. Shu et al. studied the effect of coin parameters on two-particle QWs for different initial states and showed that the coin parameters can be used to tune the entanglement between the particles \cite{Shu}. Pathak and Agarwal reported the QWs of two photons in separable and entangled states \cite{Pathak}. In recent years, different experimental implementations of two-particle QWs have been reported \cite{expt1,expt2,expt3}.\\

QW research, including the above mentioned works, has mainly focused on evolutions due to repeated applications of time-independent unitary coin operators whereas the time-dependent coins have attracted much less attention. However, works on single particle QWs with time-dependent unitary coins have found a rich array of phenomena \cite{19,coin2,coin1,interferometer}. In those works, the time dependences were introduced either by choosing coin parameters having explicit time-dependence or by selecting one coin at every step from a deterministic aperiodic sequence of two coins. Ba{\~n}uls et al. first prescribed a coin operator with explicit time dependence in ref.\cite{coin2}. They showed that the operator generates dynamical localization and quasiperiodic dynamics. Such fascinating behavior was also realized separately in a QW with a time independent coin and position-dependent phases at every step \cite{15,16}. Ba{\~n}uls et al. also showed that the time-dependent coin can be used as a control mechanism to compensate for the phases arising from some external influence \cite{coin2}. A different type of explicit time-dependence was introduced by Romanelli who actually generalized the discrete time QW on the line using a time-dependent unitary coin operator \cite{coin1}. He showed that the time dependent coin allows the particle to exhibit a variety of predetermined asymptotic wave-function spreadings : ballistic, sub-ballistic, diffusive, sub-diffusive and localized. These coherent intermediate situations might be useful for controlling quantum information and for the development of quantum algorithms \cite{coin1}.  In recent experiments, Broome et al. simulated another different type of time-dependent coin control by setting different coin parameters for different steps, which were effected in different locations along the longitudinal axis within their photonic beam-displacer interferometer \cite{interferometer}. The linearly-ramped time-dependent coin operation generated two periodic revivals of the walker distribution. On the other hand, a QW where the coin at every step is obtained from a deterministic aperiodic sequence of two coins was  first introduced by Ribeiro et al. in ref. \cite{19}. This type of time-dependence generates different types of wave function spreadings  e.g., sub-ballistic, diffusive etc. depending on the nature of the aperiodic sequences \cite{19,191}.\\

 The above described studies have shown that time-dependent coins opens a rich array of phenomena. However, such studies have been performed only on single particle systems. Two-particle systems are yet to be explored. This has been our primary motivation for the present work.\\
 
We also intend to generalize here the dynamical behavior of two quantum walkers. Generalizing two-particle QWs is in itself a topic of interest as multi-walker systems are expected to generate measurable phenomenon not described by the single-particle model, due to entanglement and interaction among the walkers. For example, non-trivial effects were found in case of two particle QW on a disordered lattice \cite{loc}. It is quite difficult to predict such effects from our knowledge of the related single particle models. Since the time dependent coins allow both the particles to exhibit a variety of dynamic behavior, we can generalize two-particle QW evolution using time-dependent coins. The advantage of using time-dependent coin in studying various wave function spreadings is that the modification has to be done only on the coin whereas with time-independent coins, the modifications are required to be performed on a much larger part of the system. For example, to generate dynamic localization without time dependent coins, different position dependent phases are required to be introduced at every step \cite{15,16}. \\

The time-dependent coins also allow us to numerically study the collective dynamics of two quantum walker of different nature. Such studies can be instrumental for developing a deeper understanding of two-particle QW. For example, using time-dependent coins we can study the dynamics of two particles where one of them has a predetermined ballistic wave function spreading whereas the other one has a predetermined localized wave function spreading.  Interesting dynamical correlations can be found in those cases. Such study can not directly be performed using time-independent coins in the presence(absence) of disorder as both particles will then perform localized(ballistic) evolution. \\

We present here a thorough numerical simulation study of a one dimensional system of two quantum walkers exhibiting rich collective dynamics controlled by simple time­-dependent unitary coins proposed by Romanelli \cite{coin1} and Ba{\~n}uls et al. \cite{coin2}. We investigate how the interplay of time-dependence, simple interaction schemes, entanglement and the relative phase between the coin states of the particles will influence the evolution of the QW. We demonstrate and characterize the wide-spectrum of tunable dynamical behavior offered by the two particle QW evolving under the influences of quantum coins having explicit time dependence.\\ 

The paper is organized as follows. In the section \ref{two}, we describe the formalisms of single and two particle QWs. The \(\mathbb{1}\) and \(\pi\)-phase interaction schemes considered here are described in section \ref{three}. Section \ref{four} has a description of the time-dependent coins used in the present work. The formalism of two particle QW with time-dependent coins has been described in section \ref{five}. Section \ref{six} describes the observables. All the numerical results of our study are presented in section \ref{seven}. In section \ref{eight}, we draw the conclusions and present future pathways.

\section{ Standard single particle and two particle QW \label{two}}

\subsection{Single particle QW}

The relevant degrees of freedom for a single particle discrete-time QW on a line are the particle’s position \(x\) (with \(x \in z\)) on the line, as well as its coin state. The total Hilbert space is given by \(H_{Total} \in H_{P}\otimes  H_{C}\) , where \(H_{P}\) is spanned by the orthonormal position vectors \(\{|x\rangle\}\) and \(H_{C}\) is the two-dimensional coin space spanned by two orthonormal vectors which we denote as \(|\uparrow\rangle\) and \(|\downarrow\rangle\). Each step of the QW consists of two subsequent operations: the coin operation and the shift-operation. The coin operation, given by \(\hat{C}\), and acting only on \(H_{C}\) , allows for superpositions of different alternatives, leading to different moves. This operation is the quantum equivalent of randomly choosing which way the particle will move in case of classical random walk. Then, the shift operation \(\hat{S}\) moves the particle according to the current coin state, transferring this way the quantum superposition to the total state in \(H_{Total}\). The evolution of the system at each step of the walk can then be described by the total unitary operator.\\
\begin{equation}
\hat{U}\equiv \hat{S}(\hat{I}\otimes\hat{C})  
\end{equation}
where \(\hat{I}\) is the identity operator acting on \({H_{P}}\). A popular choice for \(\hat{C}\) is the Hadamard operator \(\hat{C}_{H}\):
\begin{center}
\begin{equation}
\hat{C}_{H}=\frac{1}{\sqrt{2}} \begin{pmatrix}
       1 & 1 \\[0.3em]
       1 & -1 
     \end{pmatrix}
\end{equation}
\end{center}
The shift operator is given by
\begin{equation}
\hat{S} = ( \sum\limits_{x} |x+1\rangle \langle x|)\otimes |\uparrow\rangle\langle\uparrow| + ( \sum\limits_{x} |x-1\rangle \langle x|)\otimes |\downarrow\rangle\langle\downarrow|  \end{equation}

\subsection{Two particle QW}

A two-particle QW takes place in the Hilbert space \(H = H_{1} \otimes H_{2}\) , where \(H_{i} = (H_{P}\otimes H_{C} )_i\) (\(i\)=1,2). Let \(|x,\alpha ; y,\beta \rangle= |x,\alpha\rangle_1 \otimes |y,\beta\rangle _2\) be a two-particle basis state, where \(x,y\) represent the positions of the two particles on the same axis and \(\alpha,\beta \in \{\uparrow, \downarrow\}\) represent their respective coin states. The time-evolution operator is defined as \(\hat{U} = \hat{S}(\hat{I} \otimes \hat{C})\), where \(\hat{S}\) is defined in the two-particle basis by
\begin{center}\begin{equation} \begin{matrix}
       \hat{S}=|x+1, \uparrow ; y+1, \uparrow\rangle \langle x, \uparrow; y, \uparrow|  \\[0.3em]
       +|x+1,\uparrow;y-1,\downarrow\rangle\langle x,\uparrow;y,\downarrow|  \\[0.3em]
       +|x-1,\downarrow;y+1,\uparrow\rangle\langle x,\downarrow;y,\uparrow|  \\[0.3em]
       +|x-1,\downarrow;y-1,\downarrow\rangle\langle x,\downarrow;y,\downarrow|  
      \end{matrix}\end{equation}
\end{center}
The coin operator can be represented as a \(4 \times 4\) matrix \(\hat{C}_{1,2} = C_{1} \otimes \hat{C}_{2}\) where \(\hat{C}_{1}\) and \(\hat{C}_{2}\) act on two different particles. For example, if \(\hat{C}_{1}\) and \(\hat{C}_{2}\) are both equal to the standard \(2 \times 2\) Hadamard matrix \(\hat{C}_{H}\) then \(\hat{C}_{1,2}\) acts on the coin Hilbert space as\\
\begin{center}
\begin{equation}
\hat{C}_{1,2}\begin{pmatrix}
       a_{\uparrow\uparrow}            \\[0.3em]
       a_{\downarrow\uparrow}           \\[0.3em]
       a_{\uparrow\downarrow}           \\[0.3em]
       a_{\downarrow\downarrow}            \\[0.3em]
     \end{pmatrix} = \frac{1}{2}\begin{pmatrix}
       1 & 1 & 1 & 1           \\[0.3em]
       1 & -1 & 1 & -1           \\[0.3em]
       1 & 1 & -1 & -1           \\[0.3em]
       1 & -1 & -1 & 1           \\[0.3em]
     \end{pmatrix}\begin{pmatrix}
       a_{\uparrow\uparrow}            \\[0.3em]
       a_{\downarrow\uparrow}           \\[0.3em]
       a_{\uparrow\downarrow}           \\[0.3em]
       a_{\downarrow\downarrow}            \\[0.3em]
     \end{pmatrix}
\end{equation}
\end{center}
where \(a_{\alpha\beta}= \langle x,\alpha ; y,\beta|\psi\rangle\). Here \(|\psi\rangle\) represents the current state of the system. 
 The two-particle probability distribution, \(P (x,y,t)\), is the probability of finding particle \(1\) at position \(x\) and particle \(2\) at position \(y\) after \(t\) steps of the two-particle QW, i.e., \(P(x,y,t)=\sum\limits_{\alpha,\beta=\uparrow,\downarrow}|\langle x,\alpha ; y,\beta|(U )^t |\psi_0 \rangle|^2 \), where \(|\psi_{0}\rangle\) is the initial state of the system. The evolution of the system crucially depends on the choice of the initial states \cite{Omar}.\\

Here, we study the QW evolutions starting from three different initial states. One is a separable product state \(|Sep\rangle\) formed from two particles in unbiased states, i.e., \(  |Sep\rangle = \frac{1}{2} (|0, \uparrow\rangle_{1} + i|0, \downarrow\rangle_{1} ) \otimes (|0, \uparrow\rangle_{2} + i|0, \downarrow\rangle_{2} ) = \frac{1}{2} (|0, \uparrow ; 0, \uparrow\rangle + i|0, \uparrow ; 0, \downarrow\rangle+ i|0, \downarrow ; 0, \uparrow\rangle - |0, \downarrow ; 0, \downarrow\rangle ).  \) The other two are the two Bell states \(|\psi^{+}\rangle\),\(|\psi^{-}\rangle\) in which the coin states of the two particles are maximally entangled : \(|\psi^{+}\rangle= \frac{1}{\sqrt{2}} (|0, \uparrow ; 0, \downarrow\rangle + |0, \downarrow ; 0, \uparrow\rangle ),\) \(|\psi^{-}\rangle= \frac{1}{\sqrt{2}} (|0, \uparrow ; 0, \downarrow\rangle - |0, \downarrow ; 0, \uparrow\rangle )\). These two entangled states differ by a relative phase which creates the differences in the resultant behaviors.\\


\section{\(\mathbb{1}\) and \(\Pi\)-phase interaction schemes \label{three}}
\begin{figure*}[th]
\begin{center}
\subfigure[\label{fig:110a}]{\includegraphics[scale=0.70]{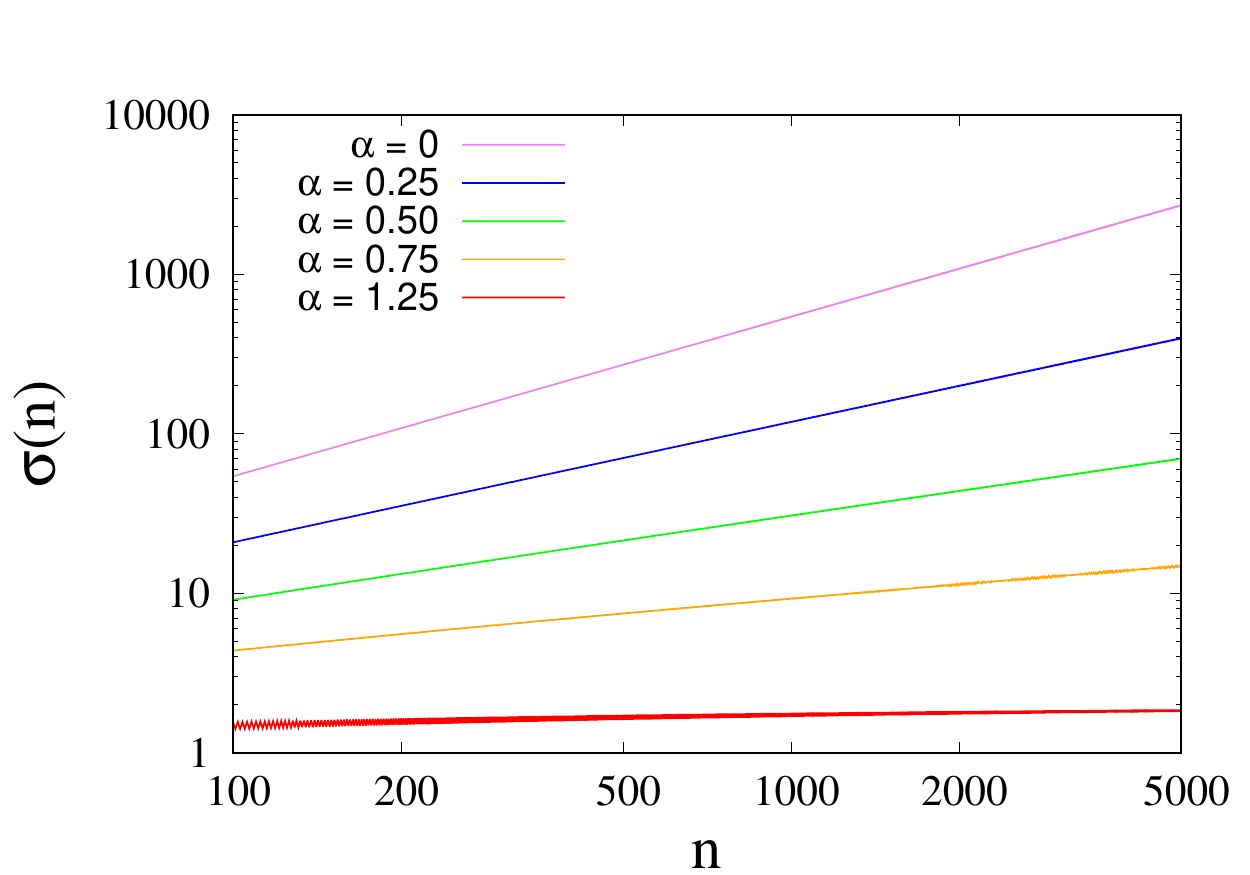}}\hspace*{-.15cm}
\subfigure[\label{fig:110b}]{\includegraphics[scale=0.70]{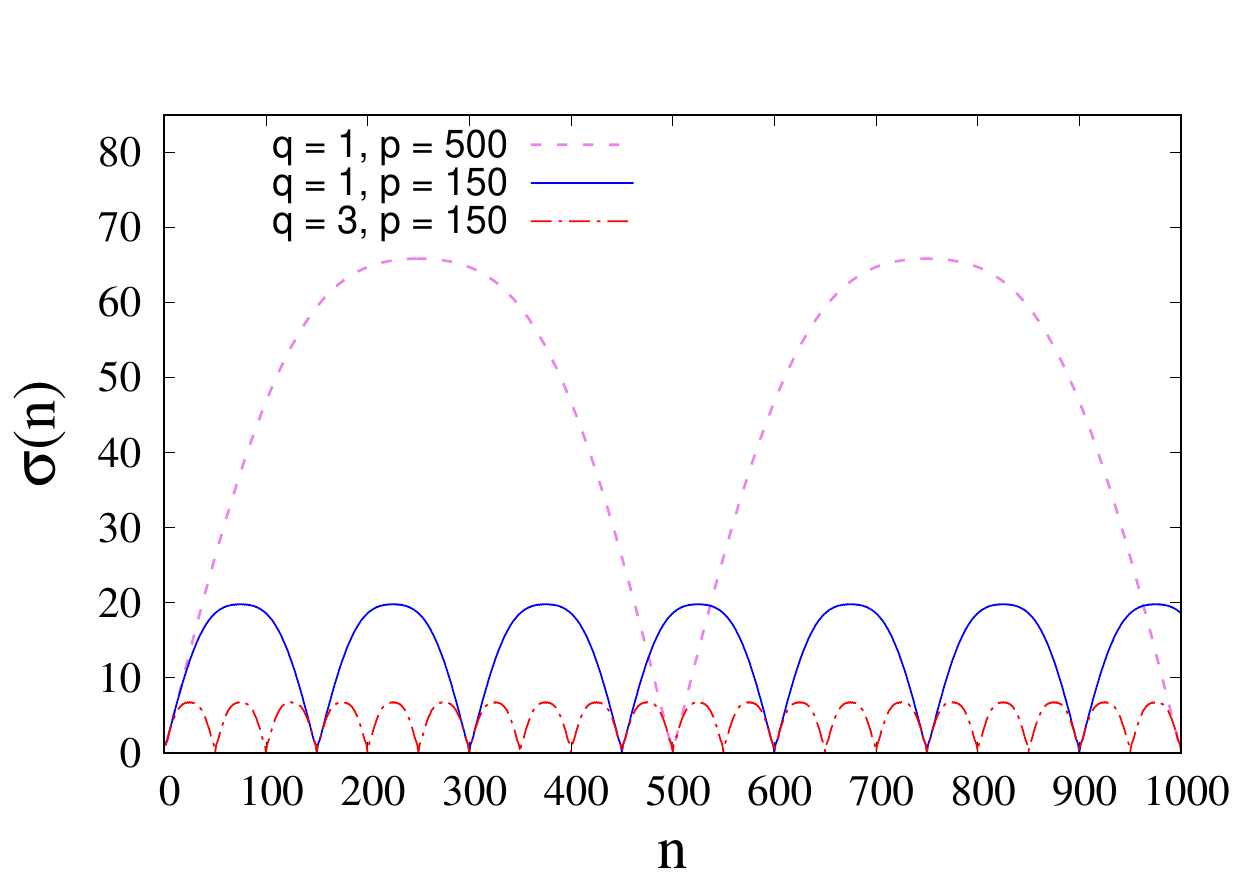}}
\end{center}
\caption{(a) The standard deviation ${\protect\sigma }(n)$ for coin \(C_{\alpha}(t)\) as a function of the
dimensionless time $n$ in log-log scales. The values of $\protect\alpha$ are as follows : $0$ (top), 0.25, 0.50(middle), 0.75 to $1.25$ (bottom). (b) The standard deviation ${\protect\sigma }(n)$ for coin \(C_{\Phi}(t)\) as a function of the dimensionless time $n$ for different choice of the two parameters \(p\) and \(q\) as described inside the figure.}
\label{f1}
\end{figure*}
For non-interacting walks, the quantum coin \(C\) is taken to be identical for all two-particle position states. 
The situation becomes more interesting when two particles interact with each other. Even simple interaction schemes can generate quite different behavior compared to the non-interacting case. Berry et al. introduced two simple interaction schemes, which are known as the \(\mathbb{1}\) interaction and the \(\pi\)-phase interactions \cite{Berry}. 
For two-particle quantum walks, the \(\mathbb{1}\) interaction is implemented by substituting the standard coin operator with the negative identity operator when both the particles are in the same position state. For example, in the two-particle QW on the line with the Hadamard coin \(\hat{C}_{H}\), the coin operator for the states \(\{|x,\alpha; y,\beta \rangle\}\) becomes \(\hat{C}=\hat{C}_{H} \otimes \hat{C}_{H} \) when \(x\neq y\) and \(\hat{C} = -\mathbb{1} \otimes - \mathbb{1} =\mathbb{1}\) when \(x=y\), This interaction was introduced by analogy with the QW based search procedure described in \cite{Shenvi}, in which a quantum oracle was implemented as a substitution of the Grover coin operator at the ``marked" vertices. In some sense, the \(\mathbb{1}\)-interacting two-particle walk is equivalent to the search procedure with all doubly occupied vertex states being ``marked."\\
In case of the \(\pi\)-phase interacting two-particle QW on the line with the Hadamard coin \(\hat{C}_{H}\),  coin operator becomes \(\hat{C}=\hat{C}_{H} \otimes \hat{C}_{H} \) when \(x\neq y\) and \(\hat{C} = e^{i\pi} \hat{C}_{H} \otimes \hat{C}_{H}\) when \(x=y\) \cite{Berry}.\\
Neither of these simple interactions are intended to represent particular physical situations. However, they have been considered in most studies on two particle interacting QW as they allow us to examine, in a simple way, the characteristics of the two-particle quantum walks that are affected by explicit spatial interactions between particles \cite{topological,another,Sun,Rodriguez}.
\section{ Time dependent quantum coins\label{coins} \label{four}}
In general, an arbitrary time-independent coin operator can be written as \cite{book}
\begin{equation}
\hat{C}=\left(\begin{array}{cc}
\mbox{cos} \theta & e^{-i\phi_{1}}\mbox{sin} \theta\\
e^{i\phi_{2}}\mbox{sin} \theta & -e^{i(\phi_{1}+\phi_{2})}\mbox{cos} \theta\end{array}\right).\label{Cgral}\end{equation}
For \(\phi_{1}=\phi_{2}=0\), the above form reduces to 
\begin{equation}
\hat{C}=\left(\begin{array}{cc}
\mbox{cos} \theta & \mbox{sin} \theta\\
\mbox{sin} \theta & -\mbox{cos} \theta\end{array}\right).\label{Cgral}\end{equation}
Romanelli and Ba{\~n}uls et al. considered separately the idea of a modified QW, where the coin elements change with time during the evolution.\\

Romanelli prescribed and studied in ref. \cite{coin1}, a deterministic angular time dependence $\theta =\theta (t)$ for the coin operator and studied the case where
\begin{equation}\label{eq:8}
\hat{C}={\hat{C}_{\alpha}(t)}=\left(\begin{array}{cc}
\mbox{cos} \theta(t) & \mbox{sin} \theta(t)\\
\mbox{sin} \theta(t) & -\mbox{cos} \theta(t)\end{array}\right)\end{equation} with
\begin{equation}\label{eq:9}
\cos \theta (t)=\frac{1}{\sqrt{2}}\left( \frac{\tau }{t+\tau }\right)
^{\alpha }  
\end{equation}%
He considered $\alpha \geq 0$ and also defined the discrete dimensionless time as $t=(n-1)\tau $ where \(n\) is the number of time steps and \(\tau\) is the unit of time \cite{coin1}. He found five different types of asymptotic behaviors depending on the values of the parameter \(\alpha\):\\
\hspace*{2cm}a) ballistic for $\alpha =0$,\\
\hspace*{2cm}b) sub-ballistic for $0<\alpha <0.5$,\\
\hspace*{2cm}c) diffusive for $\alpha =0.5$\\
\hspace*{2cm}d) sub-diffusive for $0.5<\alpha\leq 1$,\\
\hspace*{2cm}e) localized for $\alpha >1$. \\
We have shown the variations of the standard deviation \(\sigma\) against time for single particle QW under the influence of \(\hat{C}_{\alpha}(t)\) for five different values of \(\alpha\) in Fig. \ref{fig:110a}. It can be seen that the slope of the curves gradually decreases with increasing values of \(\alpha\) in case of single particle QW.\\

On the other hand, Ba{\~n}uls et al. \cite{coin2} prescribed and studied the effect of a time-dependent coin of the following special form \begin{equation}\label{eq:10}
\hat{C}=\hat{C}_{\Phi}(t)=\frac{1}{\sqrt{2}}\left(\begin{array}{cc}
e^{-i\Phi(t)} & e^{-i\Phi(t)}\\
e^{i\Phi(t)} & -e^{i\Phi(t)}\end{array}\right).\end{equation}
 Notice that the above coin can be obtained as the sequence of two
operations, i.e., \begin{equation}
\hat{C}_{\Phi}(t)=\hat{C}_{0}(t)\hat{C}_{H}\end{equation}
 with\begin{equation}
\hat{C}_{0}(t)=\left(\begin{array}{cc}
e^{-i\Phi(t)} & 0\\
0 & e^{i\Phi(t)}\end{array}\right).\end{equation}
Here \(\hat{C}_{H}\) is the time independent Hadamard coin and $\Phi(t)$ is a general function. Ba{\~n}uls et al. studied a particularly interesting case  where \(\Phi(t) = \Phi_{0} t \). For rational values of \(\Phi_{0}/2\pi\), dynamical localization around the origin was observed during a transient period. The standard deviation \(\sigma\) oscillates periodically with time during this period \cite{coin2}. 
After long enough times, ballistic diffusion starts \cite{coin2}. The duration of the transient regime is dependent on the coin parameters. For irrational values of \(\Phi_{0}/2\pi\), the long-time diffusion gets suppressed and the QW shows dynamical localization around the origin for arbitrarily long time \cite{coin2}.\\

Ba{\~n}uls et al. interpreted the dynamic localization as a propagating solution in the dispersive medium with null mean value of its group velocity \cite{coin2}. 
They considered that the value of \( \Phi_{0}(=2\pi \frac{q}{p})  \) depends on two parameters \(q\) and \(p\). The parameter \(p\) was found to control the period of primary oscillations whereas the period of secondary oscillations was controlled by both the parameters \(p\) and \(q\). Ba{\~n}uls et al. numerically showed oscillations of \(\sigma\) with quasiperiod `\(p\)' by considering a rational value of \(q/p\). The secondary oscillations were more pronounced the smaller is \(q\) and the larger is the quasiperiod \cite{coin2}.\\

We have shown the variations of \(\sigma\) against time for a single quantum walker evolving under the influence of  \(\hat{C_{\phi}}(t)\) for three different (\(q,p\)) combinations in Fig. \ref{fig:110b}. It can be seen that the standard deviation \(\sigma\) oscillates with time period `\(p\)'. For \(q=3\), it can be seen that there are secondary oscillations of period \(\frac{p}{q}\).\\

\section{ Different combinations of the time-dependent quantum coins \label{five}}
 We have studied the dynamics of two particle QW for the following different combinations of two time-dependent coins : 
\subsection{Combination-I} Both the particles are driven by the following time-dependent coin introduced in equations \ref{eq:8} and \ref{eq:9}.
\begin{equation}
\hat{C}_{\alpha}=\left(\begin{array}{cc}
cos \theta & sin \theta\\
sin \theta & -cos \theta\end{array}\right).\label{Cgral}\end{equation} where
\begin{equation}
\cos \theta (t)=\frac{1}{\sqrt{2}}\left( \frac{\tau }{t+\tau }\right)
^{\alpha },  \label{cos}
\end{equation}
Here the related \(4 \times 4\) coin matrix is written as \(\hat{C}_{\alpha,\alpha}= \hat{C}_{\alpha}\otimes \hat{C}_{\alpha}\)
\begin{center}
\begin{equation}
 =\begin{pmatrix}
       cos^{2}\theta(t) & \frac{sin 2\theta(t)}{2} & \frac{sin 2\theta(t)}{2} & sin^{2}\theta(t)          \\[0.3em]
       \frac{sin 2\theta(t)}{2} & -cos^{2}\theta(t) & sin^{2}\theta(t) & -\frac{sin 2\theta(t)}{2}           \\[0.3em]
       \frac{sin 2\theta(t)}{2} & sin^{2}\theta(t) & -cos^{2}\theta(t) &   -\frac{sin 2\theta(t)}{2}         \\[0.3em]
       sin^{2}\theta(t) & \frac{-sin 2\theta(t)}{2} & \frac{-sin 2\theta(t)}{2}  & cos^{2}\theta(t)           \\[0.3em]
     \end{pmatrix}\end{equation}\\
\end{center}
\begin{center}
\subsection{Combination-II} The particles are driven by two coins with different \(\alpha\) parameters i.e, the related \(4 \times 4\) coin matrix is written as \(\hat{C}_{\alpha_{1},\alpha_{2}}= \hat{C}_{\alpha_{1}} \otimes \hat{C}_{\alpha_{2}}\)\\ \begin{equation}= \begin{pmatrix}
       cos\theta_{1}cos\theta_{2} & cos\theta_{1}sin\theta_{2} & sin\theta_{1} cos\theta_{2} & sin\theta_{1} sin\theta_{2}          \\[0.3em]
       cos\theta_{1}sin\theta_{2} & -cos\theta_{1}cos\theta_{2} & sin\theta_{1} sin\theta_{2} & -sin\theta_{1}cos\theta_{2}           \\[0.3em]
       sin\theta_{1}cos\theta_{2} & sin\theta_{1} sin\theta_{2} & -cos\theta_{1} cos\theta_{2} &  -cos\theta_{1} sin\theta_{2}         \\[0.3em]
       sin\theta_{1}sin\theta_{2} & -sin\theta_{1} cos\theta_{2} & -cos\theta_{1} sin\theta_{2}  & cos\theta_{1} cos\theta_{2}           \\[0.3em]
     \end{pmatrix}\end{equation}\\
\end{center}

\begin{equation}
\mbox{ where }   \cos \theta_{1}=\frac{1}{\sqrt{2}}\left( \frac{\tau }{t+\tau }\right)
^{\alpha_{1} }  \label{cos}
\end{equation}

\begin{equation}
\mbox{ and }   \cos \theta_{2}=\frac{1}{\sqrt{2}}\left( \frac{\tau }{t+\tau }\right)
^{\alpha_{2} }  \label{cos}
\end{equation}
\subsection{Combination-III} Both the particles are driven by the following time-dependent coin introduced in equation \ref{eq:10}\\
\begin{equation}
\hat{C}_{\Phi}(t)=\left(\begin{array}{cc}
\sqrt{\frac{1}{2}}e^{-i\Phi(t)} & \sqrt{\frac{1}{2}}e^{-i\Phi(t)}\\
\sqrt{\frac{1}{2}}e^{i\Phi(t)} & -\sqrt{\frac{1}{2}}e^{i\Phi(t)}\end{array}\right)\label{Cgral}\end{equation}

Therefore the related \(4 \times 4\) coin matrix is written as

\begin{center}
\(\hat{C}_{\Phi,\Phi}= \hat{C}_{\Phi} \otimes \hat{C}_{\Phi}\)\\
\begin{equation} = \frac{1}{2}\begin{pmatrix}
       e^{-2i\Phi(t)} & e^{-2i\Phi(t)} & e^{-2i\Phi(t)} & e^{-2i\Phi(t)}           \\[0.3em]
       1 & -1 & 1 & -1           \\[0.3em]
       1 & 1 & -1 & -1           \\[0.3em]
       e^{2i\Phi(t)} & -e^{2i\Phi(t)} & -e^{2i\Phi(t)} & e^{2i\Phi(t)}          \\[0.3em]
     \end{pmatrix}\end{equation}
\end{center}
We consider \(\phi(t)=\phi_{0}t\) and \(\phi_{0}/2\pi=\frac{q}{p}\) following ref. \cite{coin2} and study the dynamics for different rational values of \(q/p\).
\subsection{Combination-IV} Hadamard coin \(\hat{C}_{H}\) is applied on one particle and the time-dependent coin \(\hat{C}_{\Phi}(t)\) is applied on the other particle. Therefore the related \(4 \times 4\) coin matrix is written as

\begin{center}
\(\hat{C}_{H,\Phi}= \hat{C}_{H} \otimes \hat{C}_{\Phi}\)\\
\begin{equation} =\frac{1}{2}\begin{pmatrix}
       e^{-i\Phi(t)} & e^{-i\Phi(t)} & e^{-i\Phi(t)} & e^{-i\Phi(t)}           \\[0.3em]
       e^{i\Phi(t)} & -e^{i\Phi(t)} & e^{i\Phi(t)} & -e^{i\Phi(t)}           \\[0.3em]
       e^{-i\Phi(t)} & e^{-i\Phi(t)} & -e^{-i\Phi(t)} & -e^{-i\Phi(t)}           \\[0.3em]
       e^{i\Phi(t)} & -e^{i\Phi(t)} & -e^{i\Phi(t)} & e^{i\Phi(t)}          \\[0.3em]
     \end{pmatrix}\end{equation}
\end{center}             

\section{ The dynamical observables \label{six}}
\vspace*{.5cm}

We have numerically studied the time evolutions of the following joint properties : \(C_{12}\), \(\Delta_{12}\) and \(E(|\psi\rangle )\), apart from the joint two particle probability distribution \(P(x,y)\). Here x,y represent the positions of the two particles on the same line. The first observable \(C_{12}\) is the positional correlation function which is given by \(C_{12}= \langle x y\rangle - \langle x \rangle \langle y \rangle\). It measures the positional correlation between the particles. It is also used to quantify the bunching or anti-bunching behavior of the two particles. Positive (negative) spatial correlation indicates bunching (anti-bunching) behavior of the particles.\\The second observable is the average distance between the two particles, defined as \(\Delta_{12} = \langle |x - y|\rangle \). The third observable \(E(|\psi\rangle )\) is the entanglement entropy. We have evaluated \(E(|\psi\rangle )\) following the procedure described in ref \cite{Berry}. For two-particle quantum walks, the composite space \(H\) can be divided into two single-particle subsystems, \(H_{1(2)} = (H_{P} \otimes H_{C} )_{1(2)}\), to measure the total entanglement between the two particles. The entanglement between two subsystems of a bipartite pure quantum state \(|\psi \rangle \) can be measured using the von Neumann entropy \(S\) of the reduced density matrix of either subsystem \cite{mintert}, \(E(|\psi\rangle ) = S(\rho_{1} ) = S(\rho_{2} ) = - Tr(\rho_{1} \log_{2} \rho_{1} )\). Since the trace is invariant under similarity transformation and the density matrix \(\rho_{1}\) has real, non-negative eigenvalues \(\lambda_{i}\) , the von Neumann entropy can easily be calculated as \(S(\rho_{1} ) = - \sum\limits_{i} \lambda_{i} \log_{2} \lambda_{i} \). \\
For a pure two-particle state, \( |\psi\rangle = \sum\limits_{xy} \sum\limits_{ij} a_{xiyj} |x,i ; y,j \rangle,\) the reduced density matrix \(\rho_{1}\) is obtained by tracing the density matrix \(\rho\) = \(|\psi\rangle \langle \psi|\) over subsystem 2,\\
 \( \rho_{1}=Tr_{2}(\rho)= \sum\limits_{xyzw}\sum\limits_{ijkl} a_{xiyj} a^{*}_{zkwl} |x,i\rangle \langle z,k| \langle w,l|y,j\rangle \)\\
where \(x,y,z,w\) represent points on the line, while \(i,j,k,l\) represent coin states and \(a_{xiyj}\) are coefficients of the two-particle basis states. Using the orthonormality condition \(\langle w,l|y,j\rangle =\delta_{yw}\delta_{jl}\), we obtain \(\rho_{1} = \sum\limits_{xz}\sum\limits_{ik} b_{xizk} |x,i\rangle \langle z,k|, \) where \(b_{xizk} = \sum\limits_{y}\sum\limits_{j} a_{xiyj} a^{*}_{zkyj}\).\\
We then numerically calculate the eigenvalues \(\lambda_{i}\) of \(\rho_{1}\). The entanglement \(E\) between the two particles can be obtained at each time step from the following relation \(S(\rho_{1} ) = - \sum\limits_{i} \lambda_{i} \log_{2} \lambda_{i} \). The maximum entanglement between two \(k\)-dimensional subsystems is \(E_{max} = \log_{2} k\). If both particles are initially placed at the origin, then \(k=2\) (as the coin-space is two dimensional and position space is one dimensional), so the Bell states \(|\psi^{\pm}\rangle,|\phi^{\pm}\rangle \) are maximally entangled \((E_{max} = \log_{2} 2 = 1)\). As the QW spreads at a rate of one lattice position per time step in each direction, the number of possible occupied states in a two-particle QW on the line increases linearly with the number of steps. The dimension of each of the single-particle subspaces is therefore \(k = 2(2n + 1)\), giving \(E_{max} = \log_{2} [2(2n + 1)] = 1 + \log_{2} (2n + 1)\) where \(n\) is the number of time steps. So the upper bound on entanglement grows logarithmically with the number of steps \(n\).\\
\section{ Diverse dynamics of two quantum walkers \label{seven}}
\subsection{The case of coin \(\hat{C}_{\alpha}(t)\)}
\FloatBarrier
\begin{figure*}[th]
\centering

\subfigure[\label{fig:1a} ]{\includegraphics[scale=0.25]{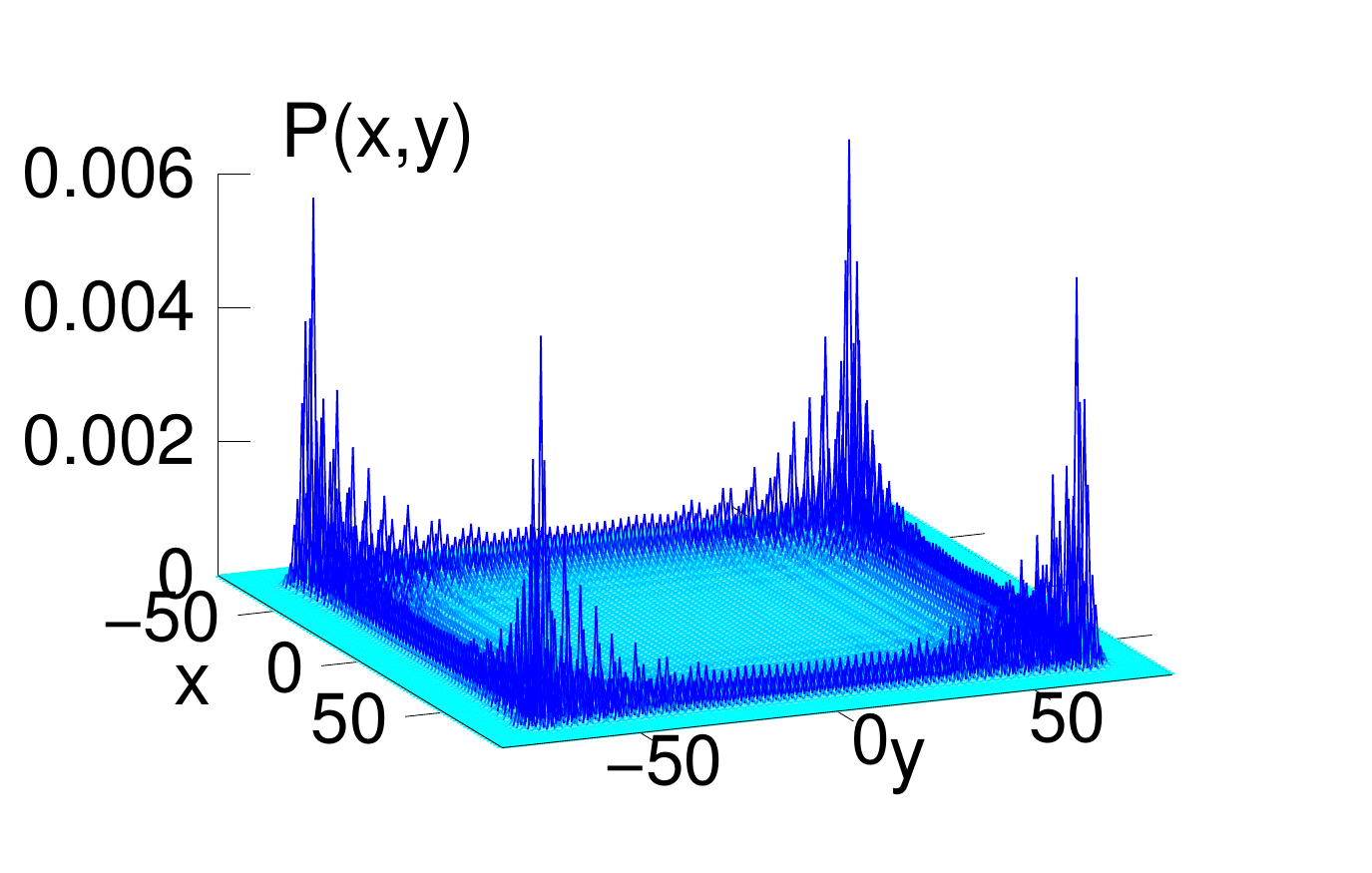}}\hspace*{-.35cm}
\subfigure[\label{fig:1b} ]{\includegraphics[scale=0.25]{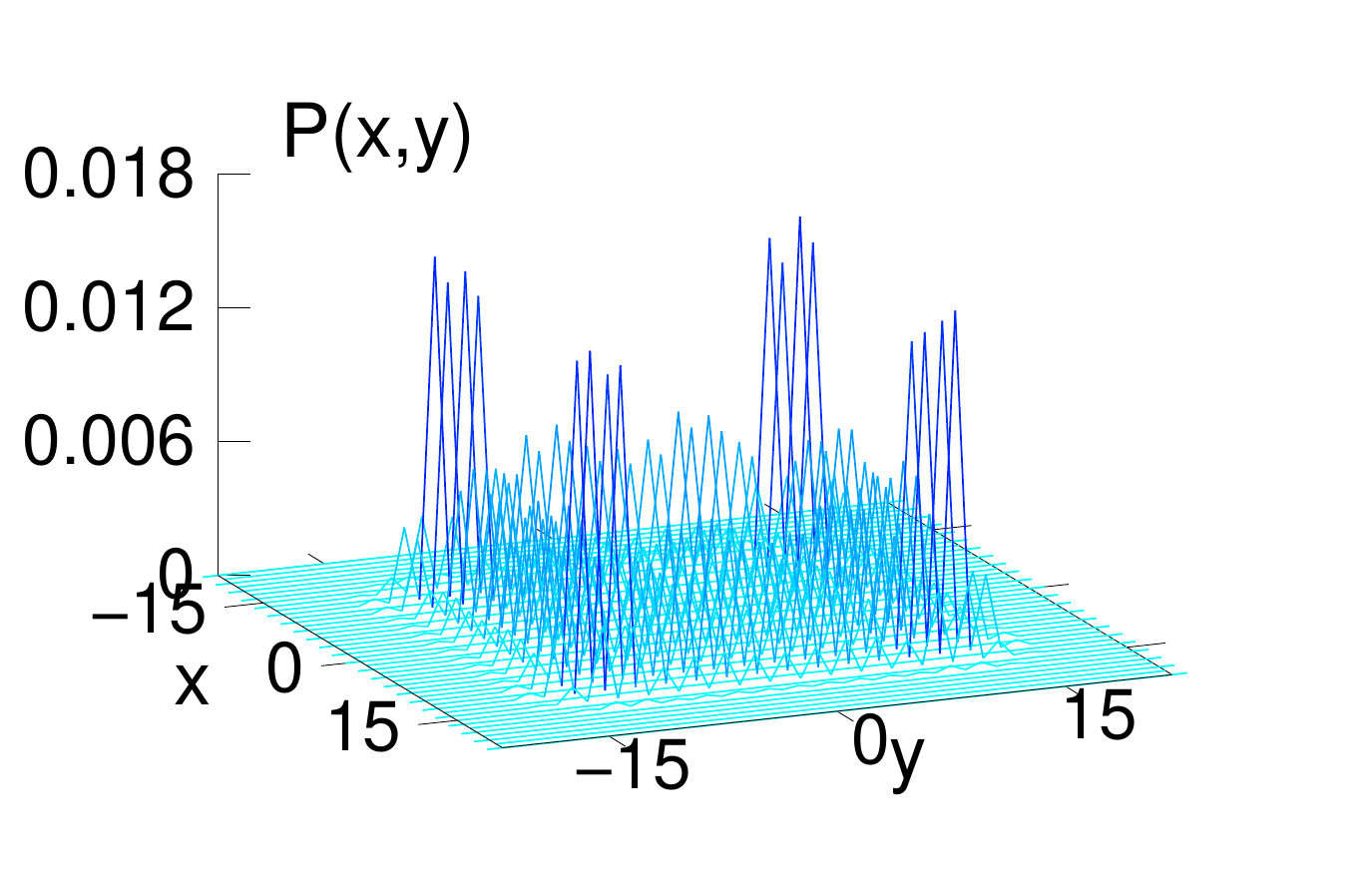}}\hspace*{-.35cm}
\subfigure[\label{fig:1c} ]{\includegraphics[scale=0.25]{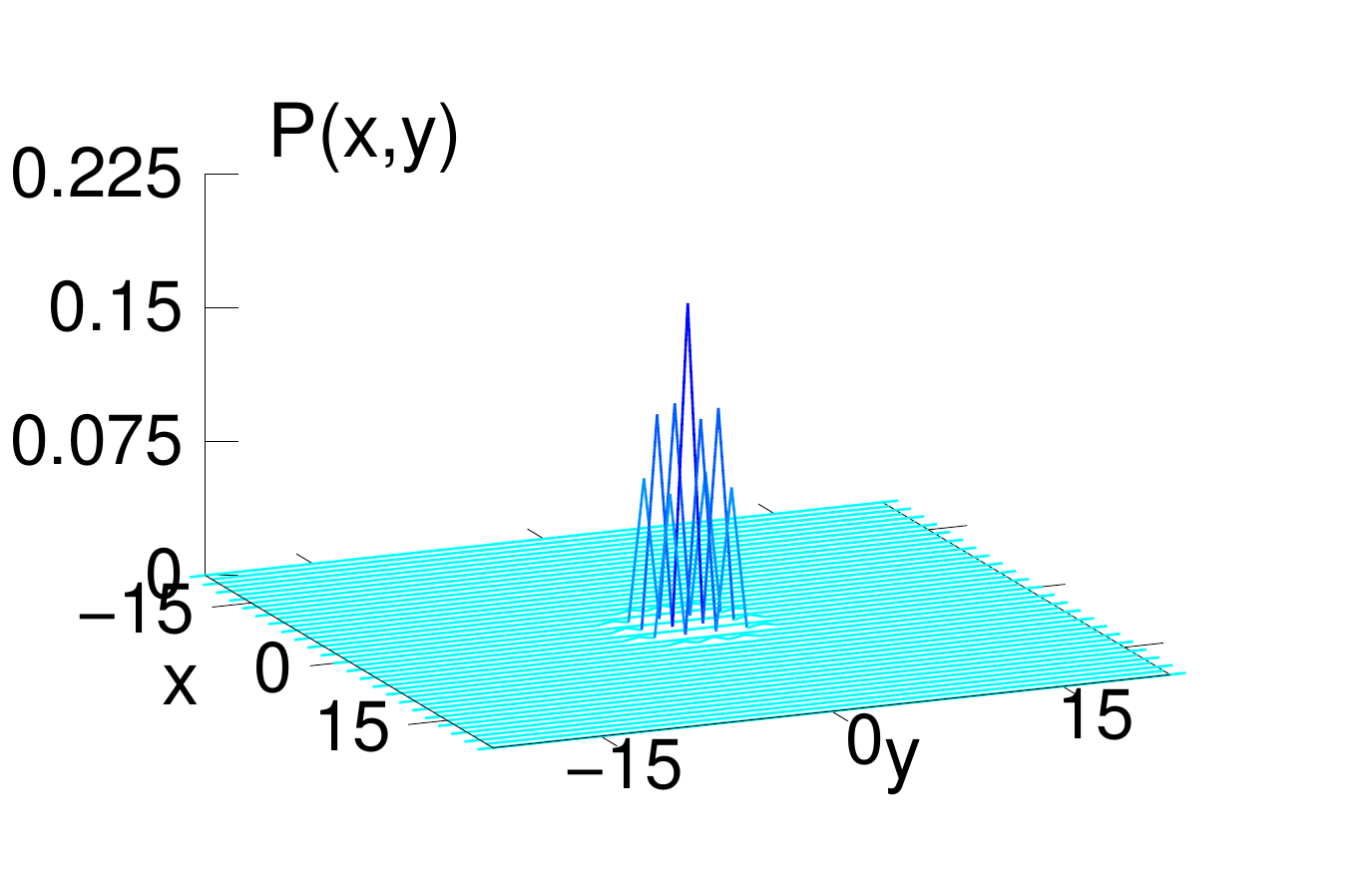}}\hspace*{-.35cm}
\subfigure[\label{fig:1d} ]{\includegraphics[scale=0.25]{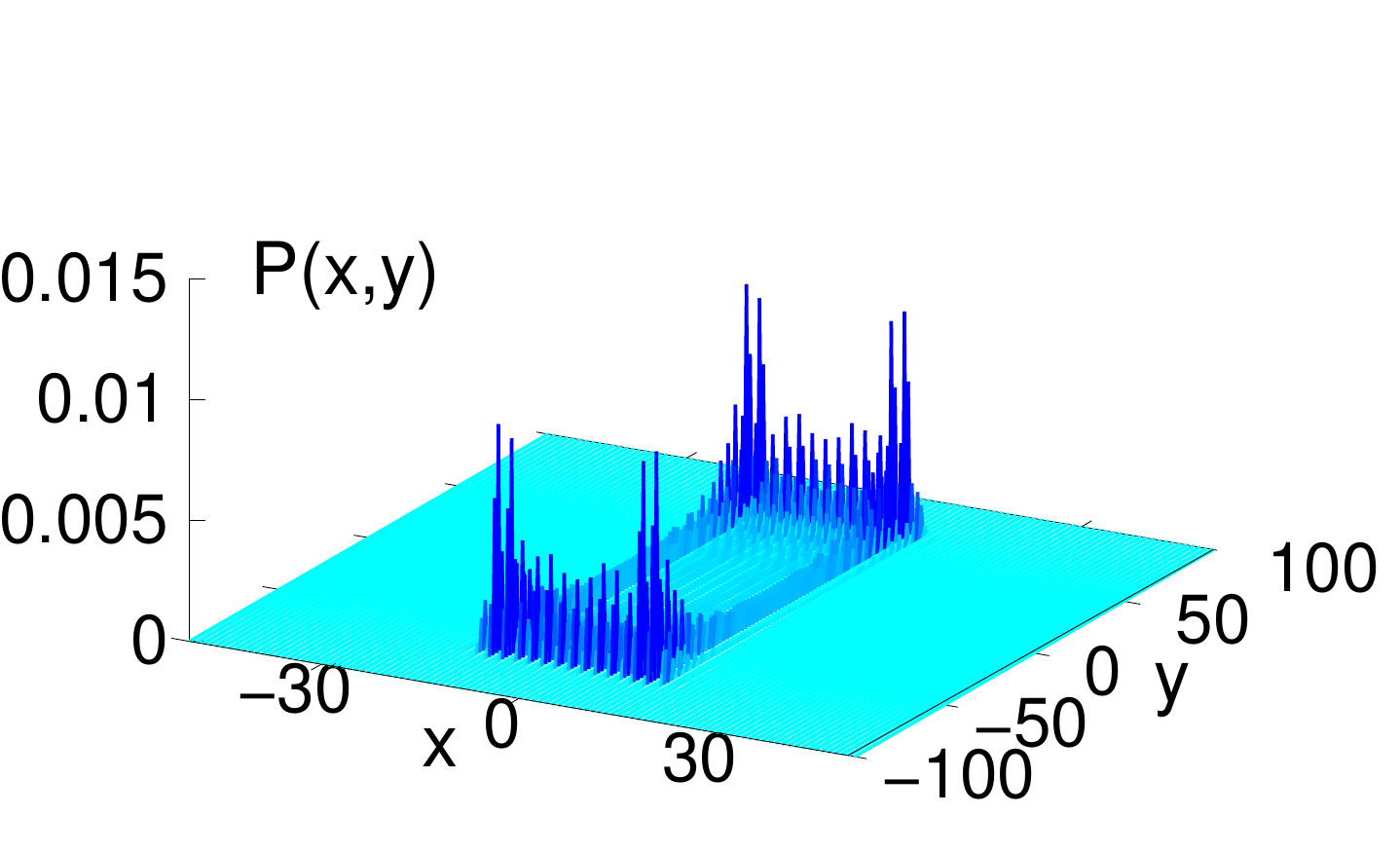}}\hspace*{-.15cm}
\subfigure[\label{fig:1e} ]{\includegraphics[scale=0.25]{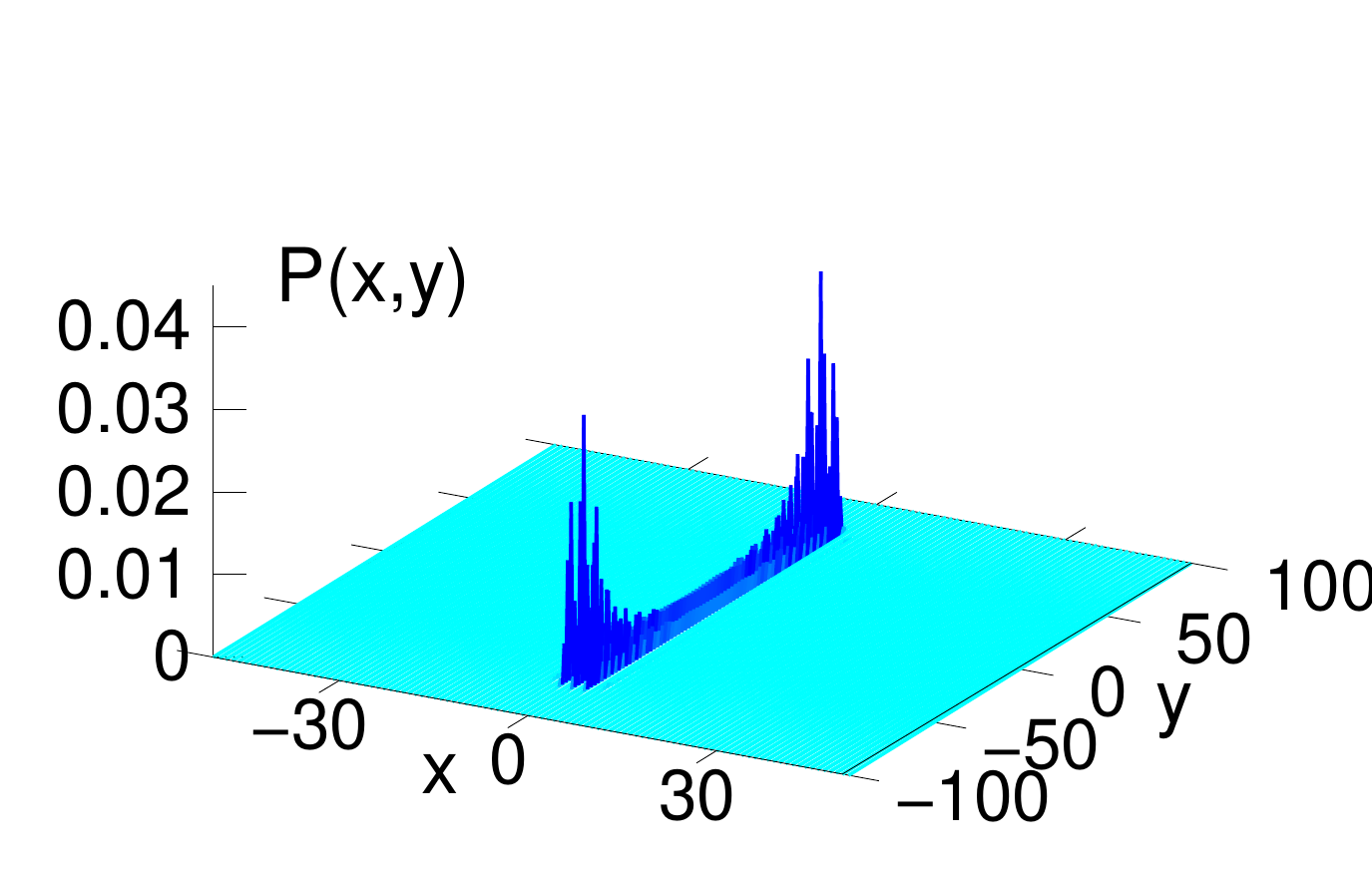}}
\subfigure[\label{fig:2a} ]{\includegraphics[scale=0.25]{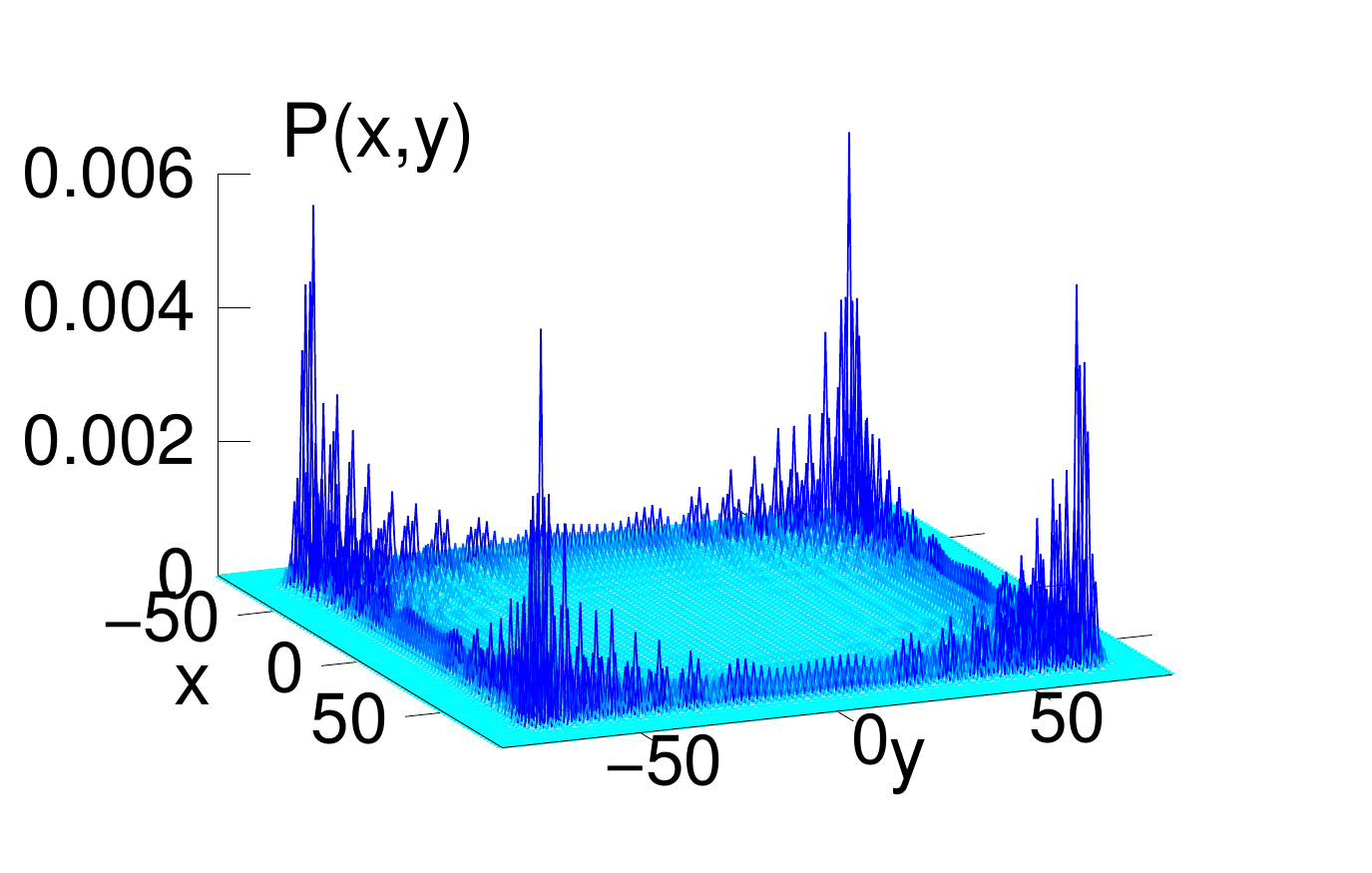}}\hspace*{-.35cm}
\subfigure[\label{fig:2b} ]{\includegraphics[scale=0.25]{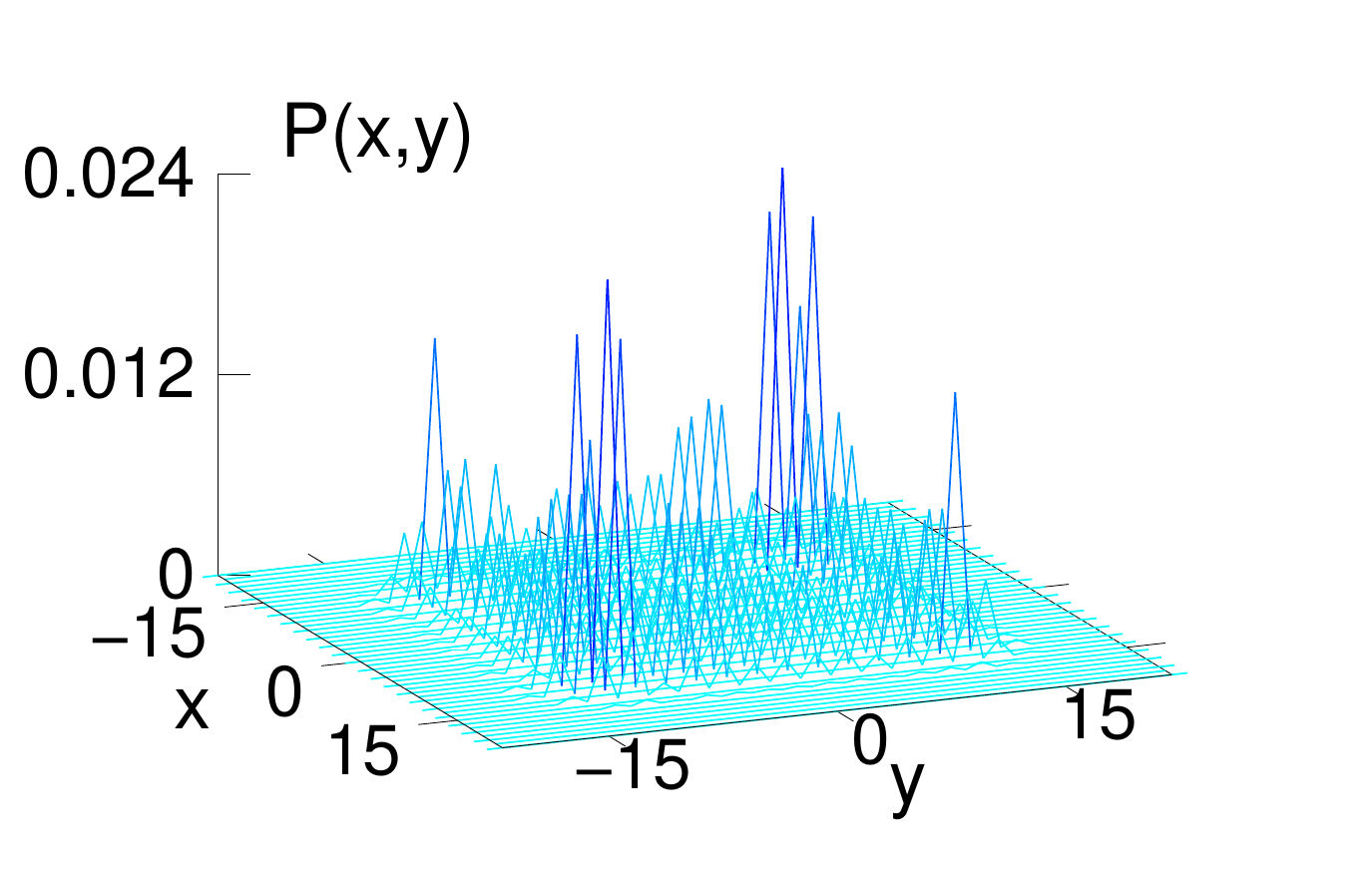}}\hspace*{-.35cm}
\subfigure[\label{fig:2c} ]{\includegraphics[scale=0.25]{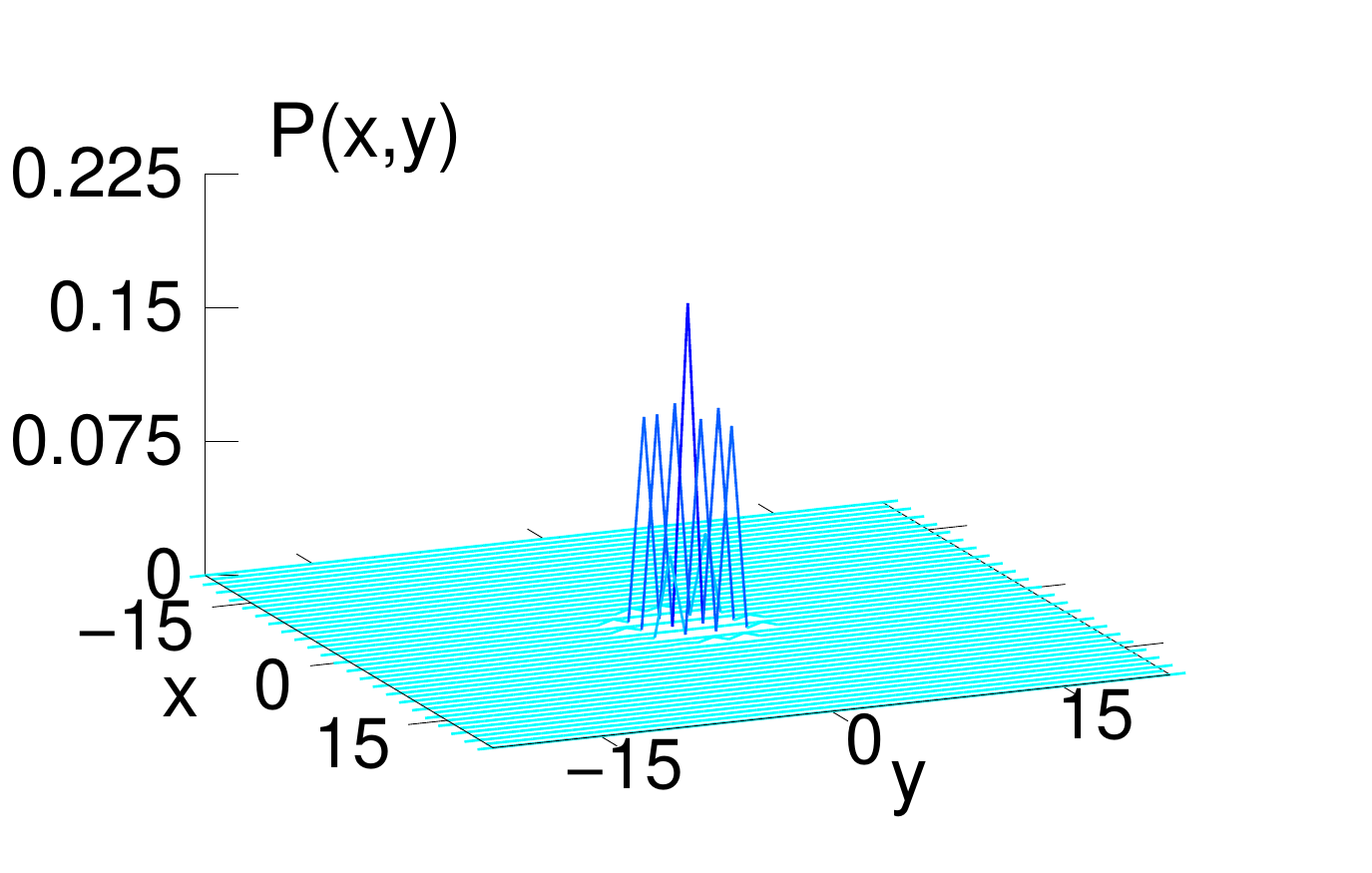}}\hspace*{-.35cm}
\subfigure[\label{fig:2d} ]{\includegraphics[scale=0.25]{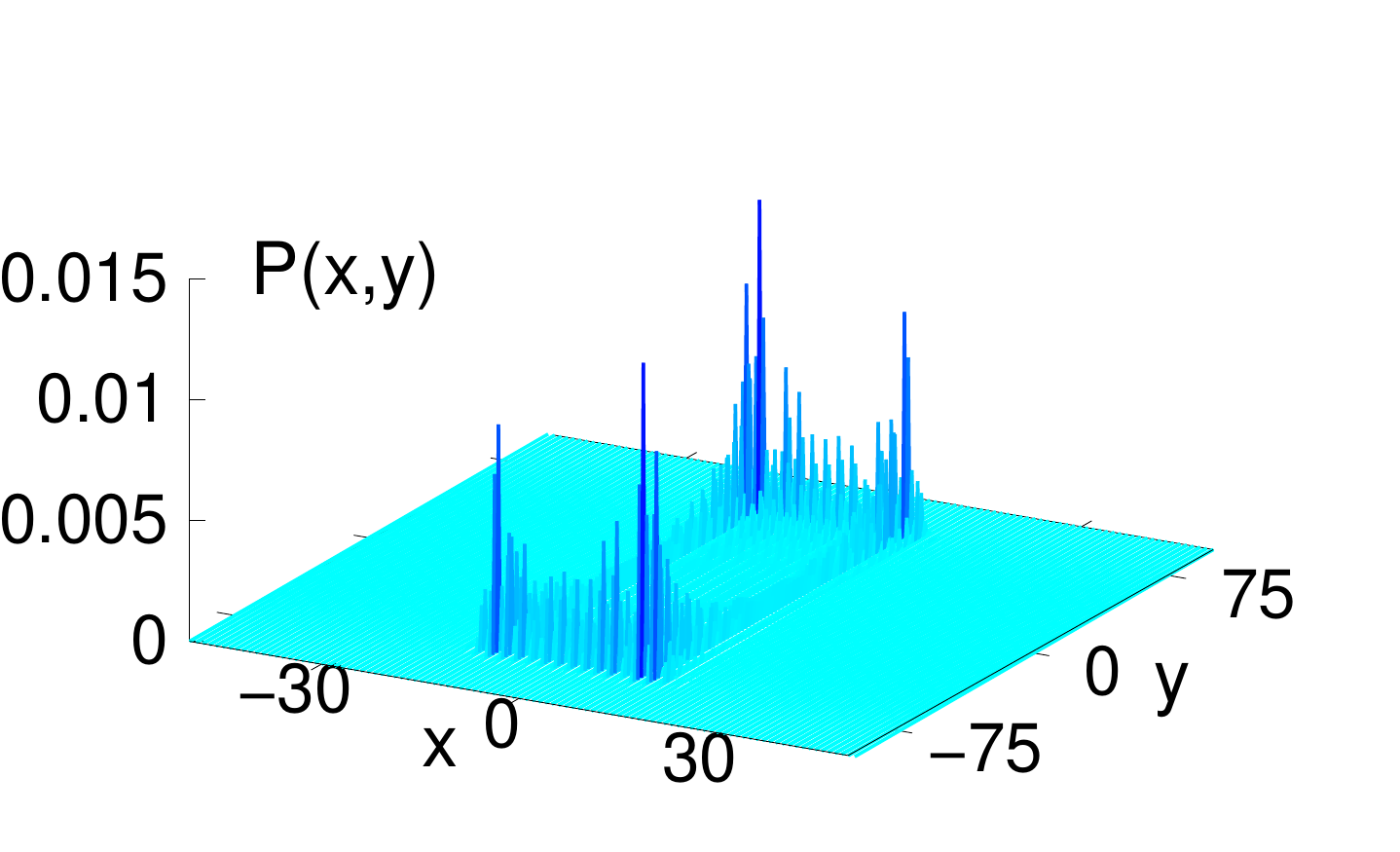}}\hspace*{-.15cm}
\subfigure[\label{fig:2e} ]{\includegraphics[scale=0.25]{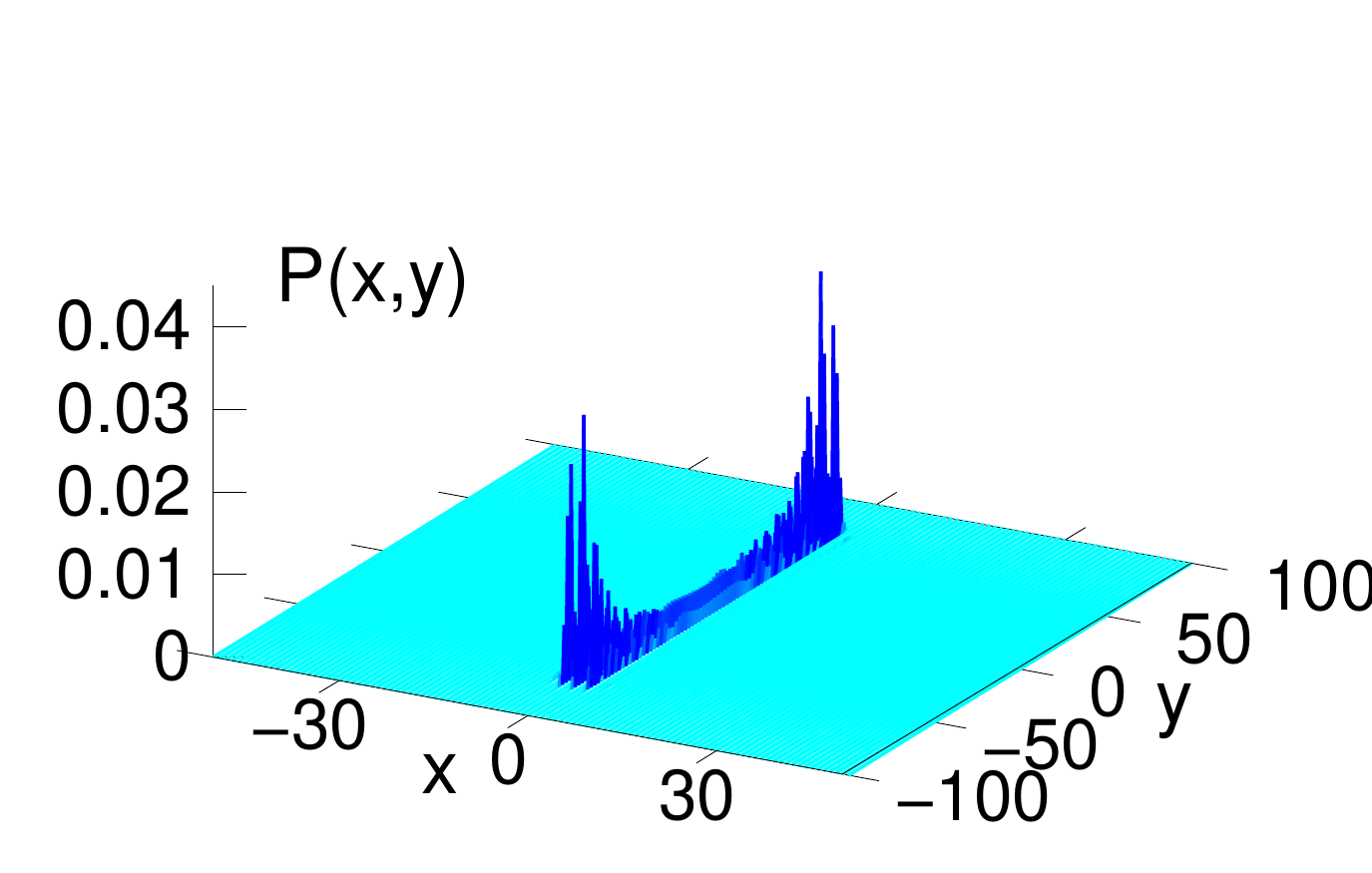}}
\subfigure[\label{fig:3a} ]{\includegraphics[scale=0.25]{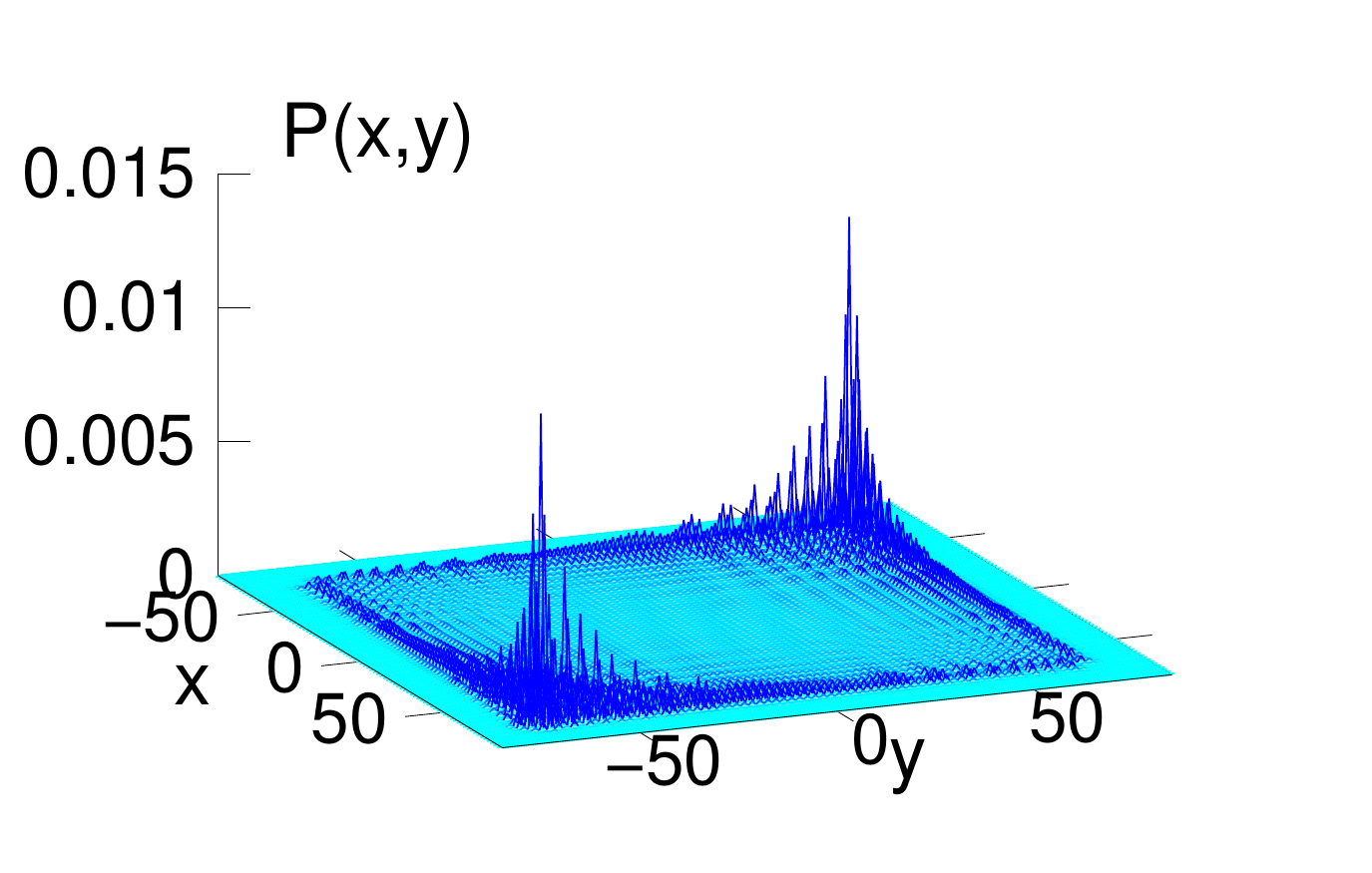}}\hspace*{-.35cm}
\subfigure[\label{fig:3b} ]{\includegraphics[scale=0.25]{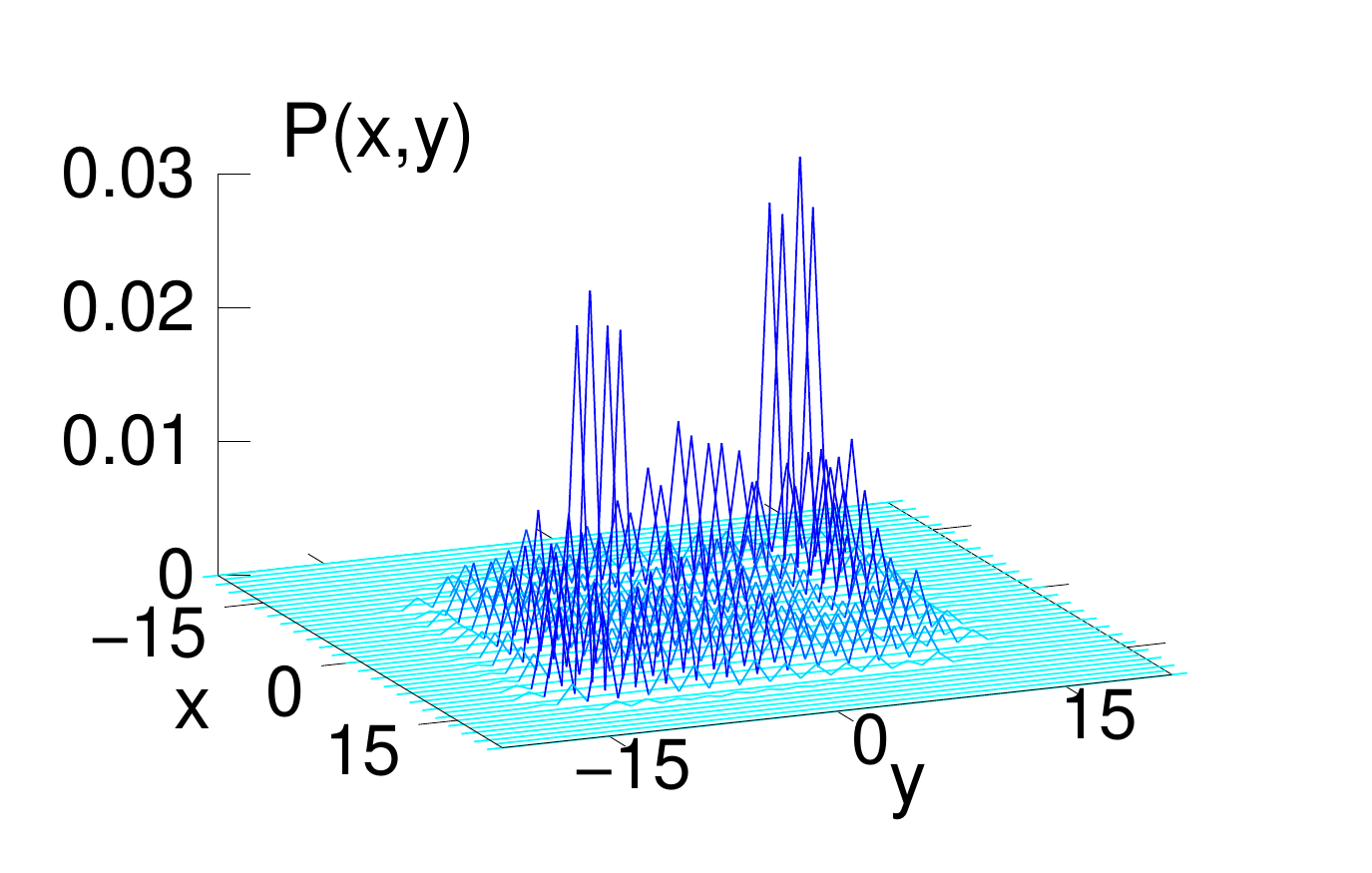}}\hspace*{-.35cm}
\subfigure[\label{fig:3c} ]{\includegraphics[scale=0.25]{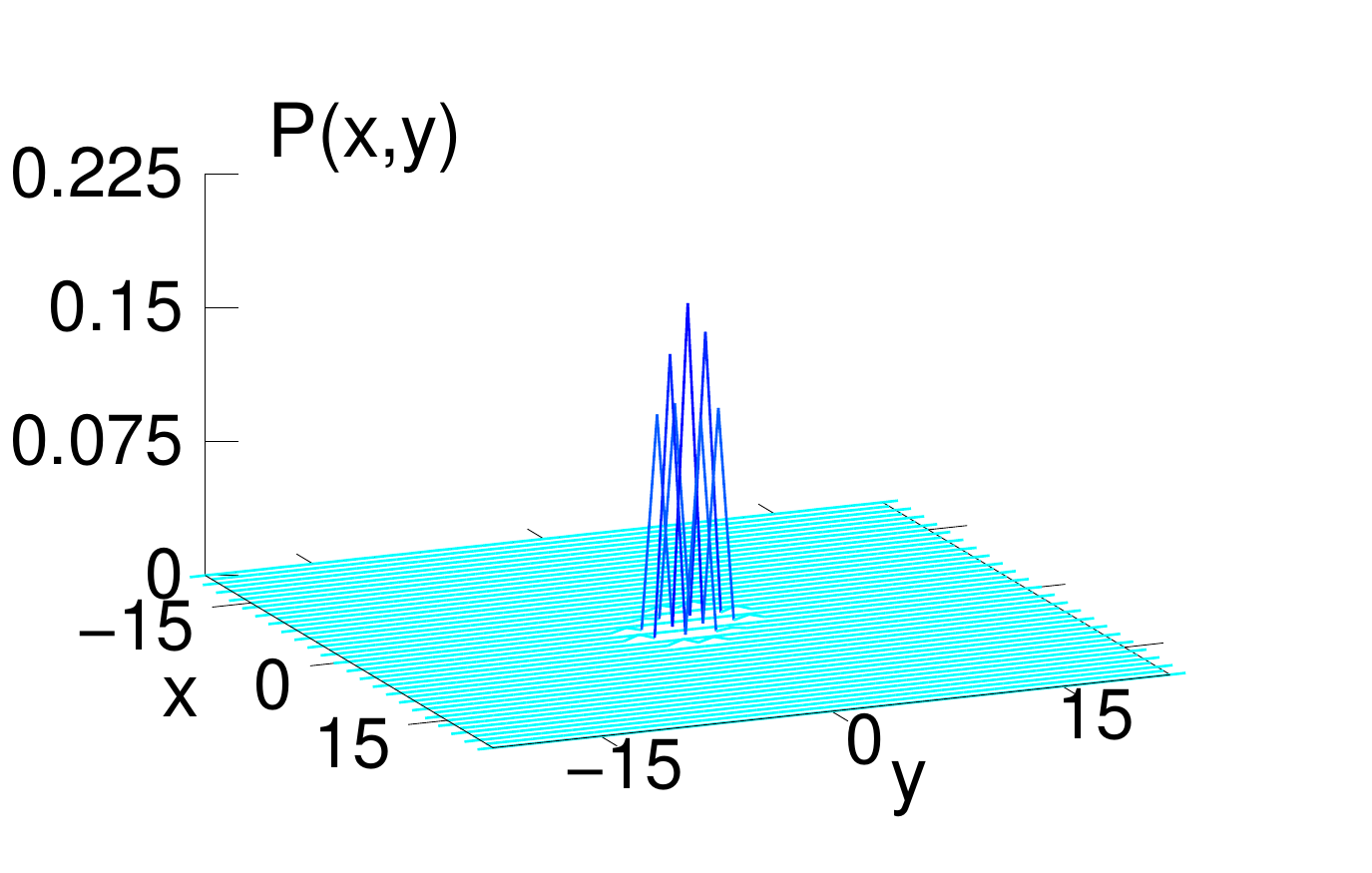}}\hspace*{-.35cm}
\subfigure[\label{fig:3d} ]{\includegraphics[scale=0.25]{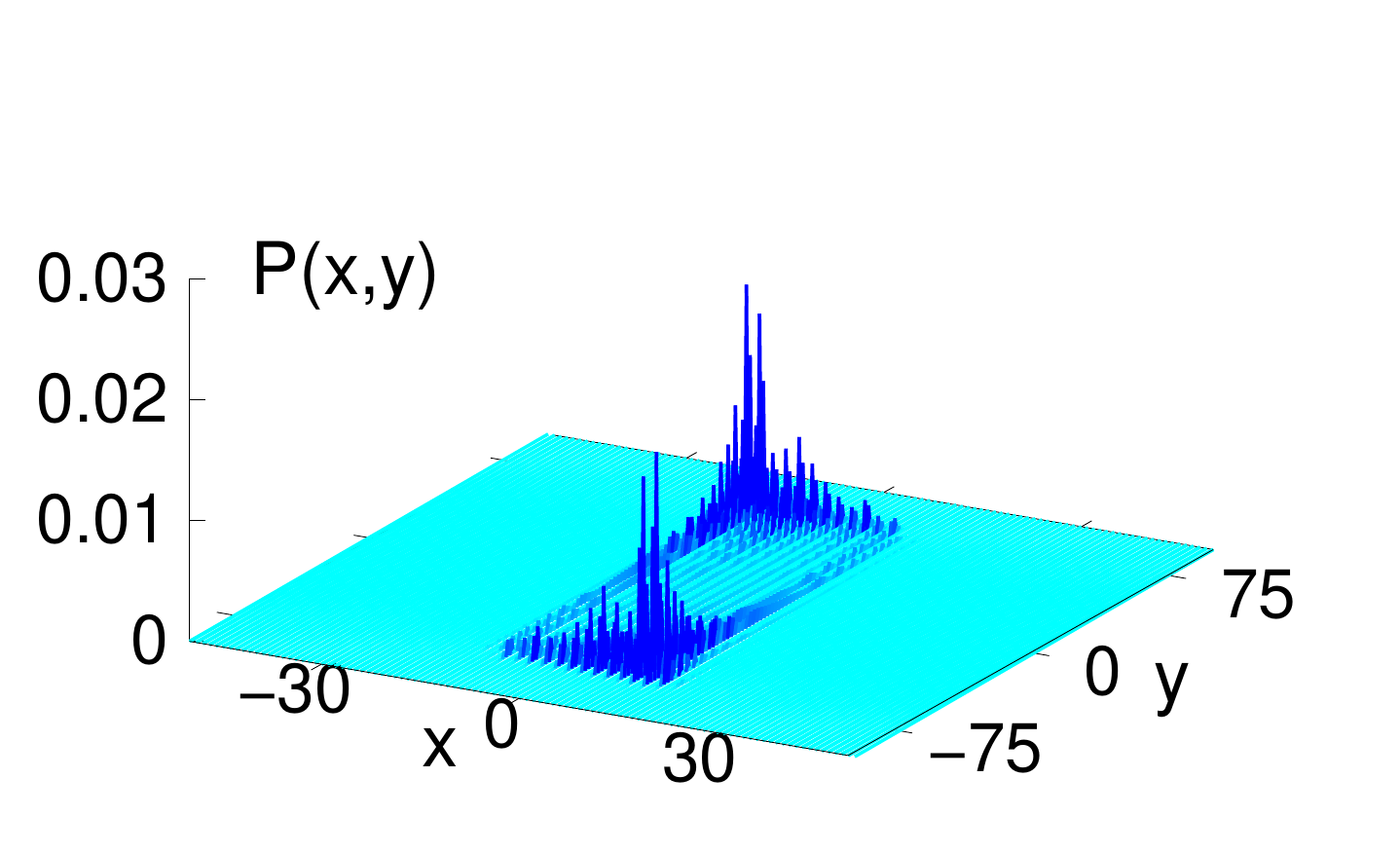}}\hspace*{-.15cm}
\subfigure[\label{fig:3e} ]{\includegraphics[scale=0.25]{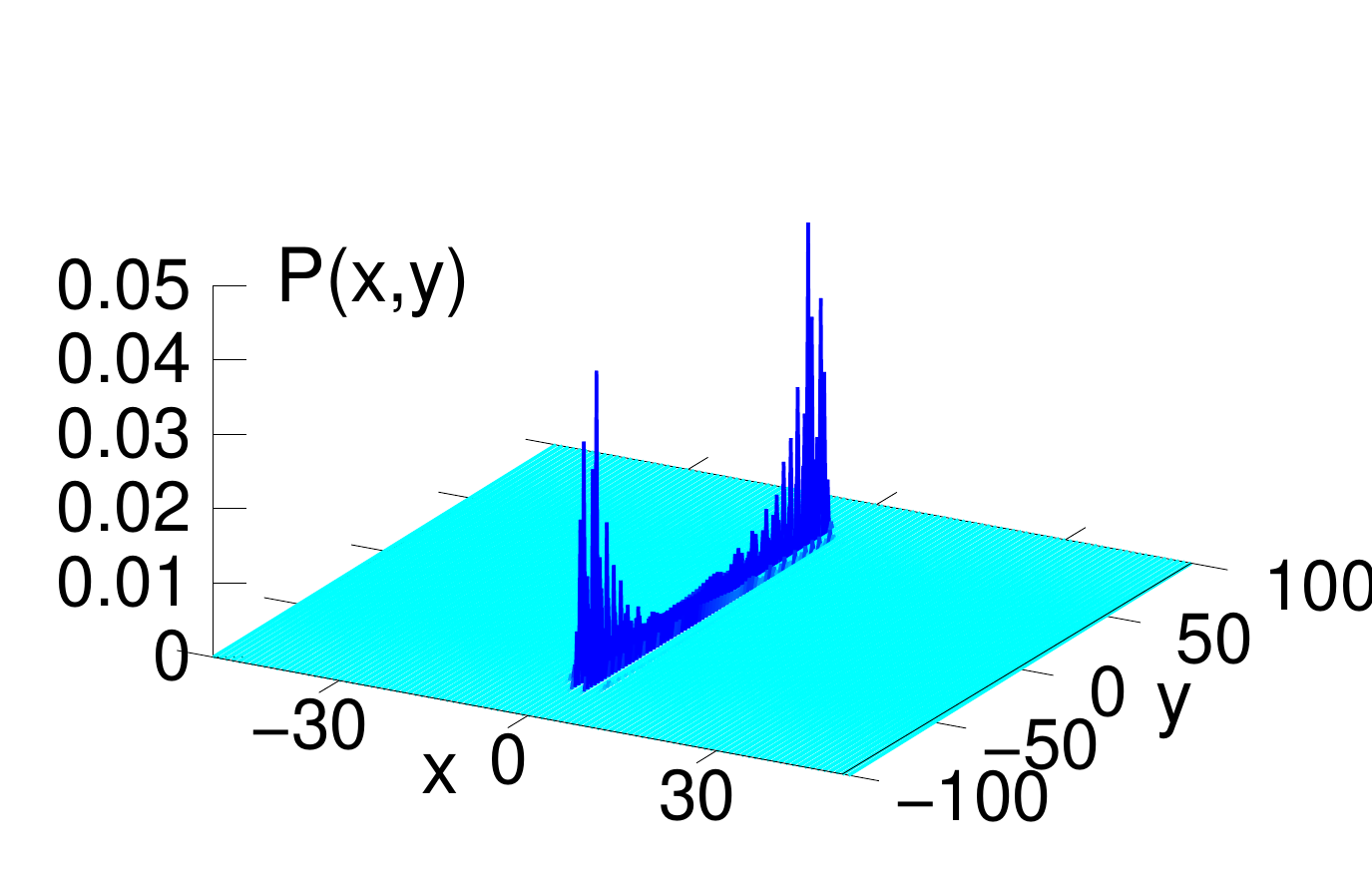}}
\subfigure[\label{fig:2f} ]{\includegraphics[scale=0.40]{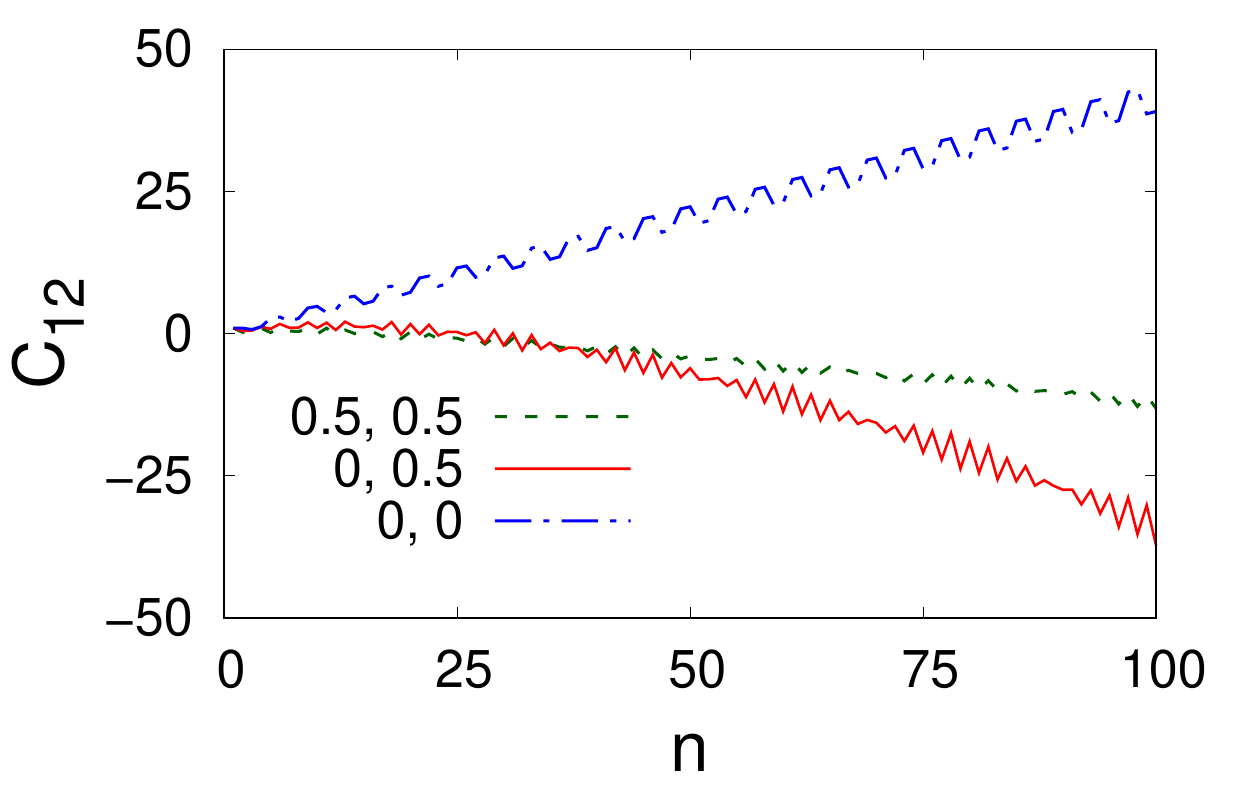}}
\subfigure[\label{fig:2g1} ]{\includegraphics[scale=0.40]{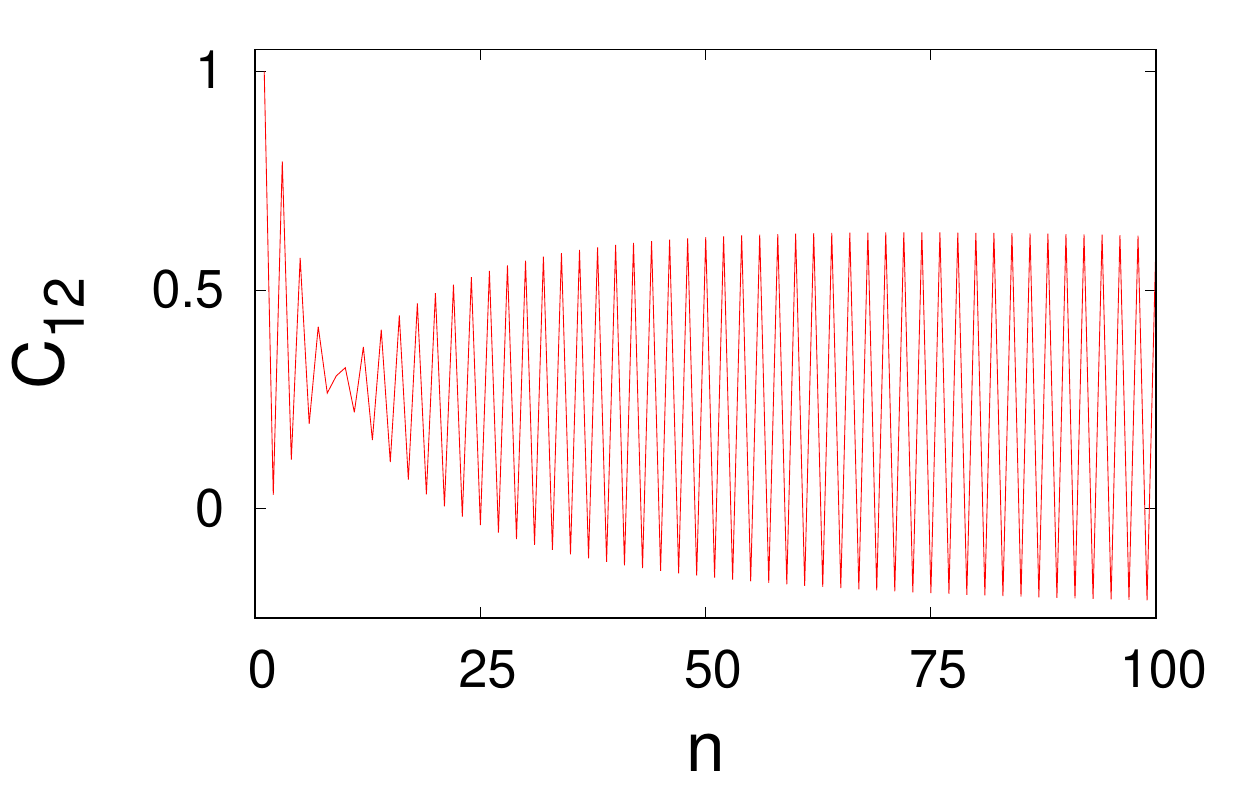}}
\subfigure[\label{fig:2g2} ]{\includegraphics[scale=0.40]{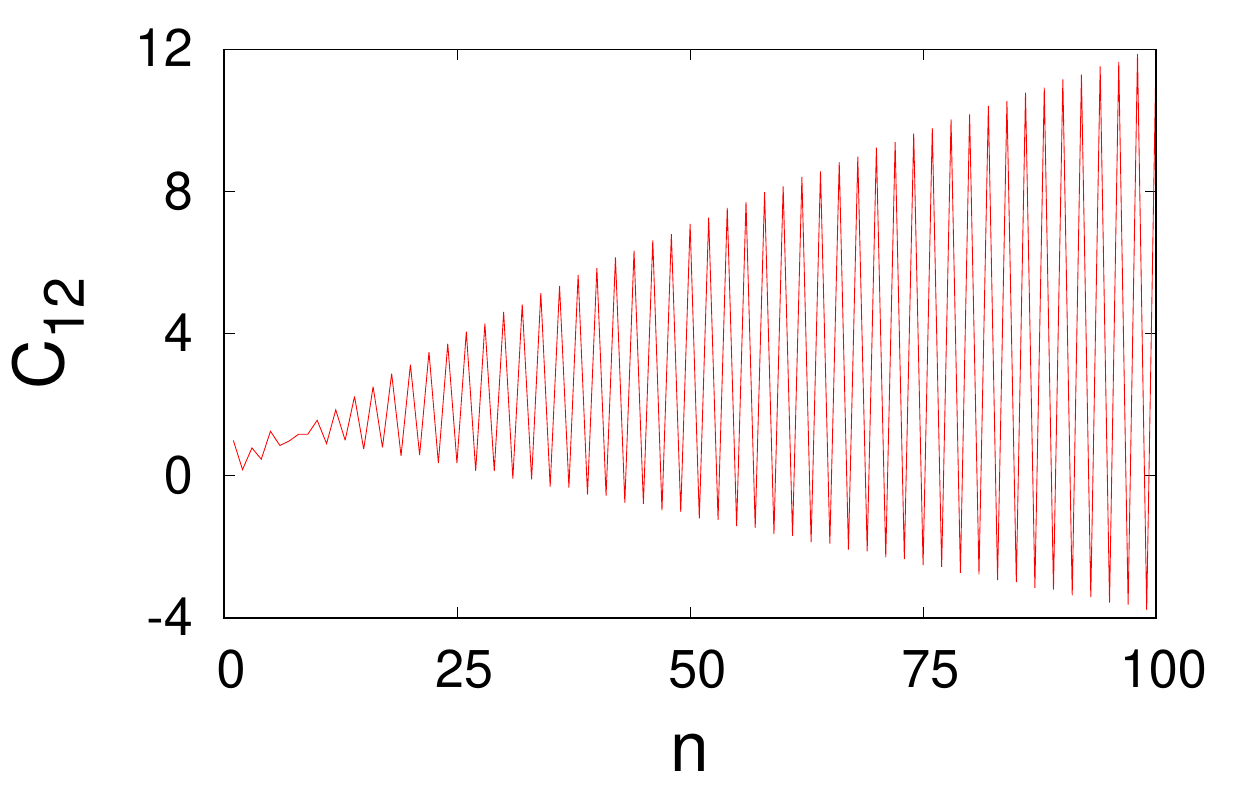}}
\subfigure[\label{fig:3f} ]{\includegraphics[scale=0.40]{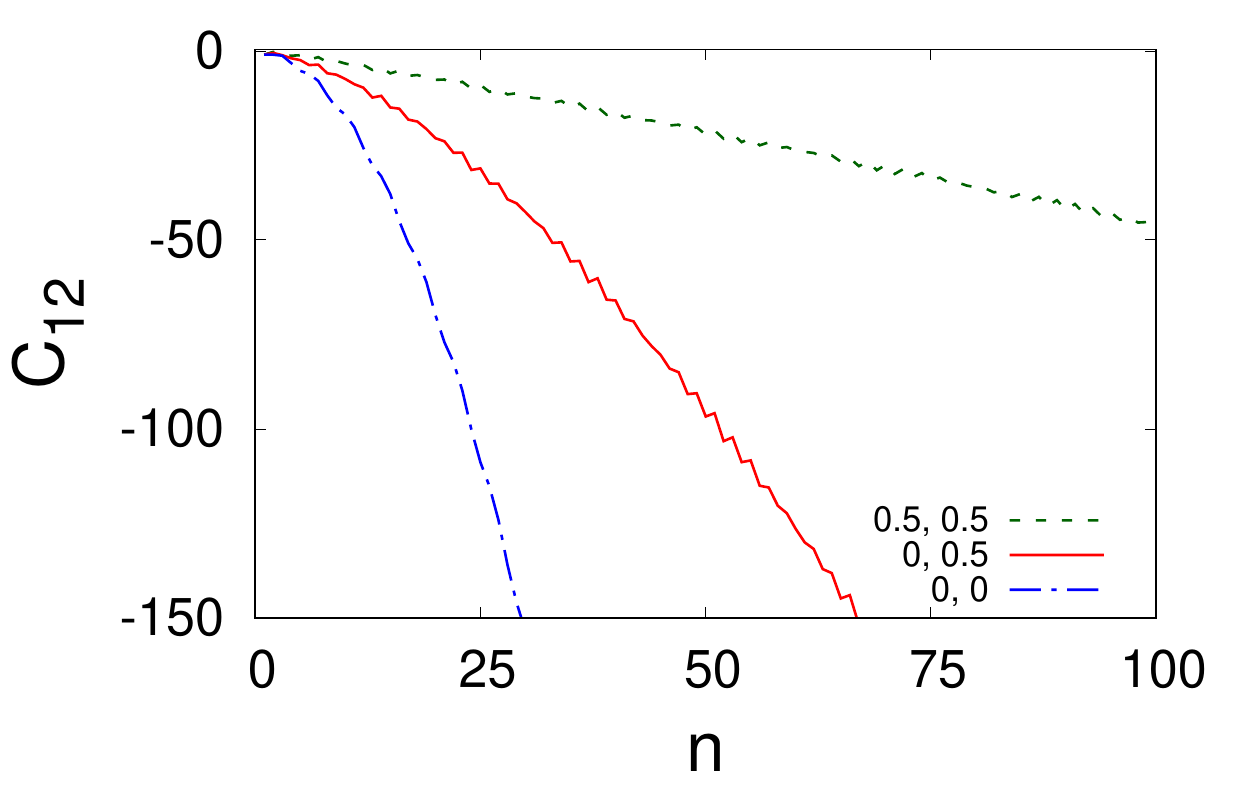}}
\subfigure[\label{fig:3g1} ]{\includegraphics[scale=0.40]{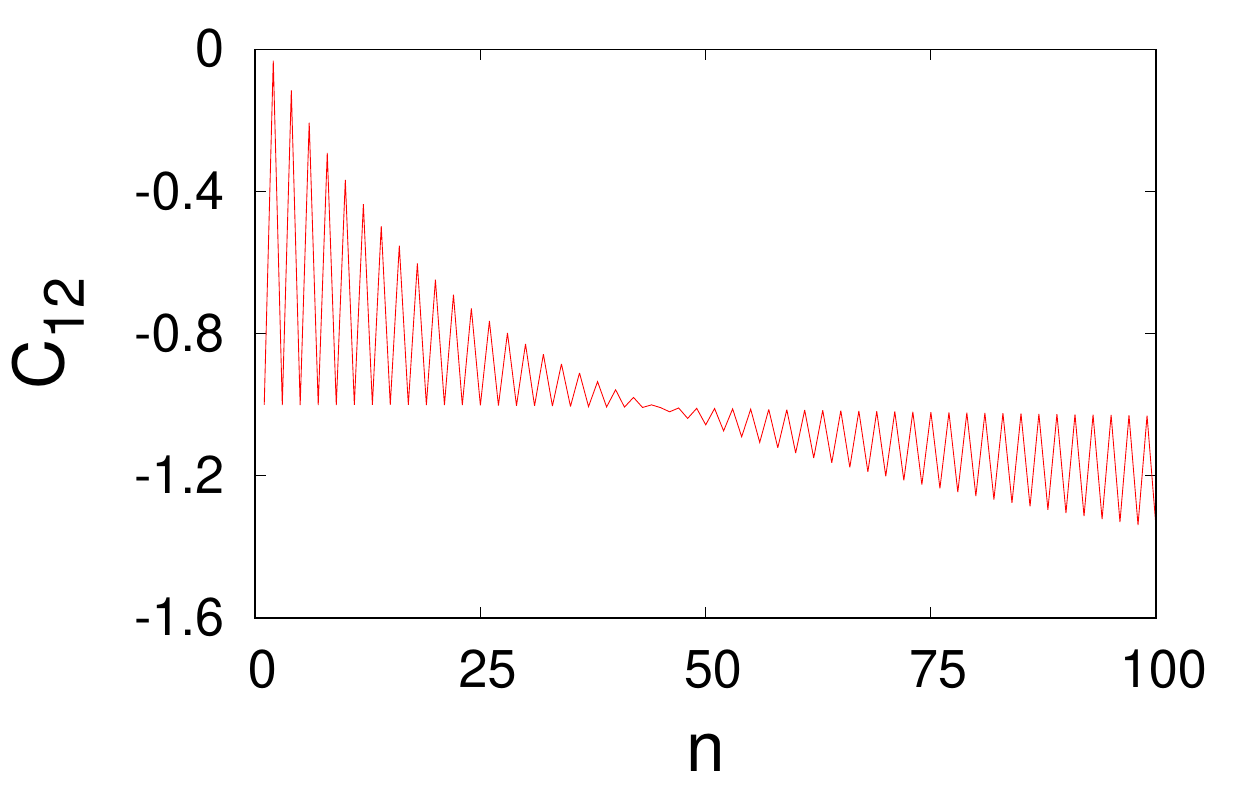}}
\subfigure[\label{fig:3g2} ]{\includegraphics[scale=0.40]{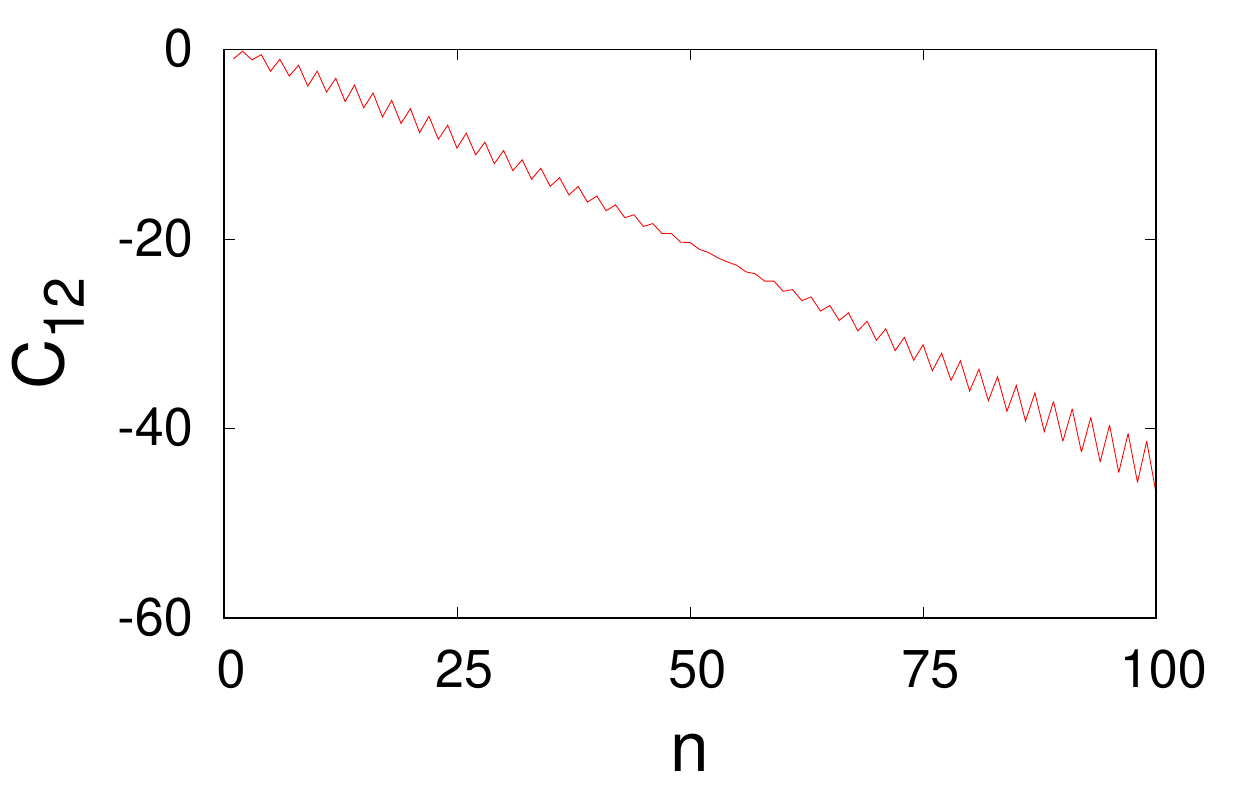}}

\caption{\label{fig:1}{ Here figures (a)-(o) show two-particle probability distributions \(P(x,y)\) after 100 time steps for two non interacting walkers evolving under the influences of the coin \(\hat{C}_{\alpha_{1},\alpha_{2}}(t)\). \(P(x,y)\) for \(|Sep\rangle\) initial state :(a) \(\alpha_{1}\) = \(\alpha_{2}\) = 0; (b) \(\alpha_{1}\) = \(\alpha_{2}\)=0.5; (c) \(\alpha_{1}\) = \(\alpha_{2}\) = 1.25, (d) \(\alpha_{1}\) = 0, \(\alpha_{2}\) = 0.5; (e) \(\alpha_{1}\) = 0, \(\alpha_{2}\) = 1.25. \(P(x,y)\) for \(|\psi^{+}\rangle\) initial state : (f) \(\alpha_{1}\) = \(\alpha_{2}\) = 0; (g) \(\alpha_{1}\) = \(\alpha_{2}\) = 0.5; (h) \(\alpha_{1}\) = \(\alpha_{2}\) = 1.25, (i) \(\alpha_{1}\) = 0, \(\alpha_{2}\) = 0.5; (j) \(\alpha_{1}\) = 0, \(\alpha_{2}\) = 1.25. \(P(x,y)\) for \(|\psi^{-}\rangle\) initial state : (k) \(\alpha_{1}\) = \(\alpha_{2}\) = 0; (l) \(\alpha_{1}\) = \(\alpha_{2}\) = 0.5; (m) \(\alpha_{1}\) = \(\alpha_{2}\) = 1.25, (n) \(\alpha_{1}\) = 0, \(\alpha_{2}\) = 0.5; (o) \(\alpha_{1}\) = 0, \(\alpha_{2}\) = 1.25. Figures (p)-(u) show the variations of the correlation function \(C_{12}\) against dimension less time \(n\) for two non-interacting walkers evolving under the influences of \(\hat{C}_{\alpha_{1},\alpha_{2}}(t)\). Variations of \(C_{12}\) for \(|\psi^{+}\rangle\) initial state : (p) \(\alpha_{1}\) = \(\alpha_{2}\) = 0; \(\alpha_{1}\) = \(\alpha_{2}\) = 0.5; \(\alpha_{1}\) = 0, \(\alpha_{2}\) = 0.5 (q) \(\alpha_{1}\) = \(\alpha_{2}\) = 1.25 (r) \(\alpha_{1}\) = 0, \(\alpha_{2}\) = 1.25. Variations of \(C_{12}\) for \(|\psi^{-}\rangle\) initial state : (s) \(\alpha_{1}\) = \(\alpha_{2}\) = 0; \(\alpha_{1}\) = \(\alpha_{2}\) = 0.5; \(\alpha_{1}\) = 0, \(\alpha_{2}\) = 0.5 (t) \(\alpha_{1}\) = \(\alpha_{2}\) = 1.25 (u) \(\alpha_{1}\) = 0, \(\alpha_{2}\) = 1.25
 }
 }
\end{figure*}

\begin{figure}[th]
\centering
\subfigure[\label{fig:4a} ]{\includegraphics[scale=0.30]{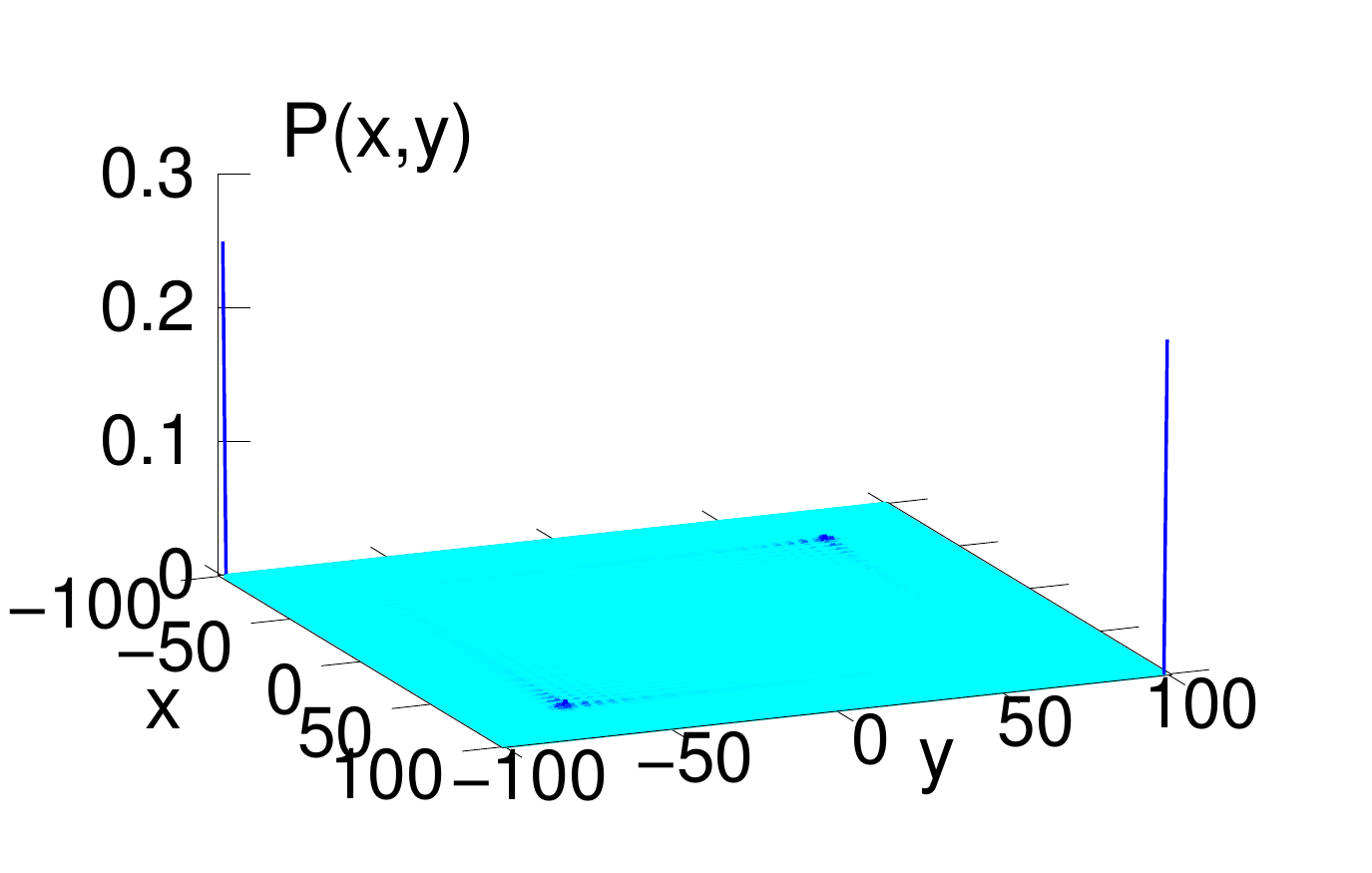}}\hspace*{-.15cm}
\subfigure[\label{fig:4f} ]{\includegraphics[scale=0.30]{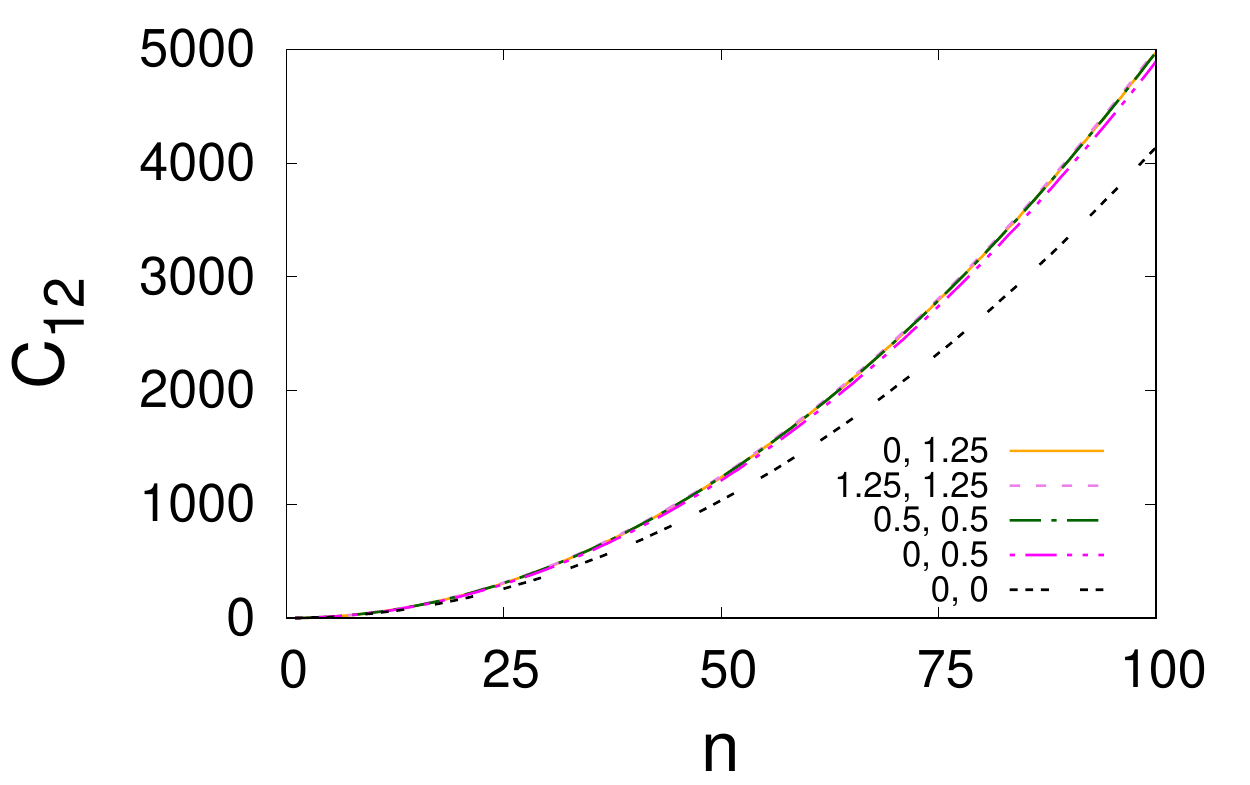}}
\caption{\label{fig:2}{ (a) Two-particle probability distribution \(P(x,y)\) after 100 time steps of evolution of two \(\mathbb{1}\)-interacting walkers starting from a \(|Sep\rangle\) initial state with the following coin parameters : \(\alpha_{1}\)=\(\alpha_{2}\)=0; (b) Variations of the correlation function against time for evolutions starting from a \(|Sep\rangle\) initial state with different coin parameter combinations. Some of the curves for different combinations of the two coin parameters nearly overlap each other. }
 }
 
\end{figure}

\begin{figure*}[th]
\centering

\subfigure[\label{fig:5a} ]{\includegraphics[scale=0.25]{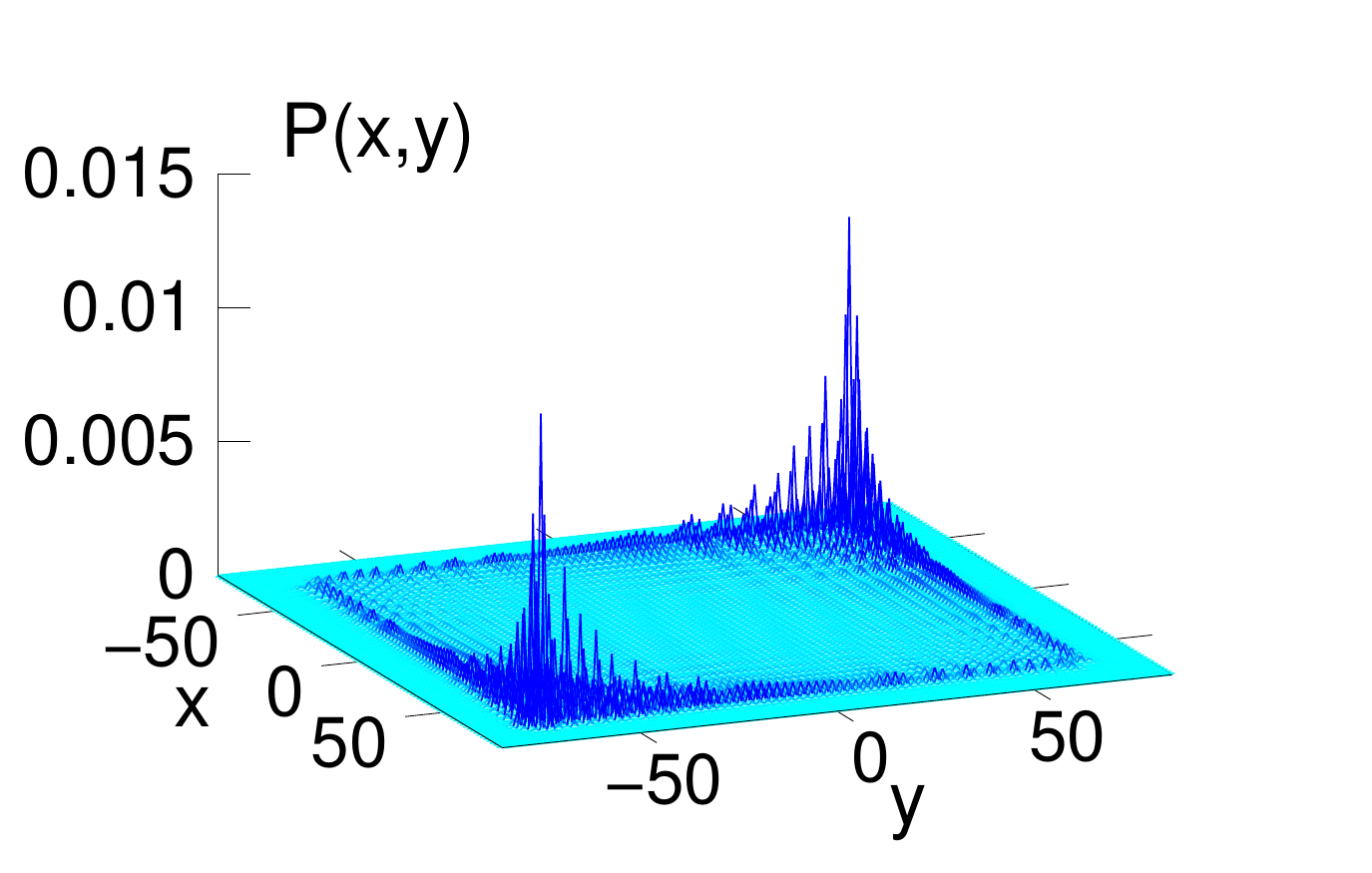}}\hspace*{-.5cm}
\subfigure[\label{fig:5b} ]{\includegraphics[scale=0.25]{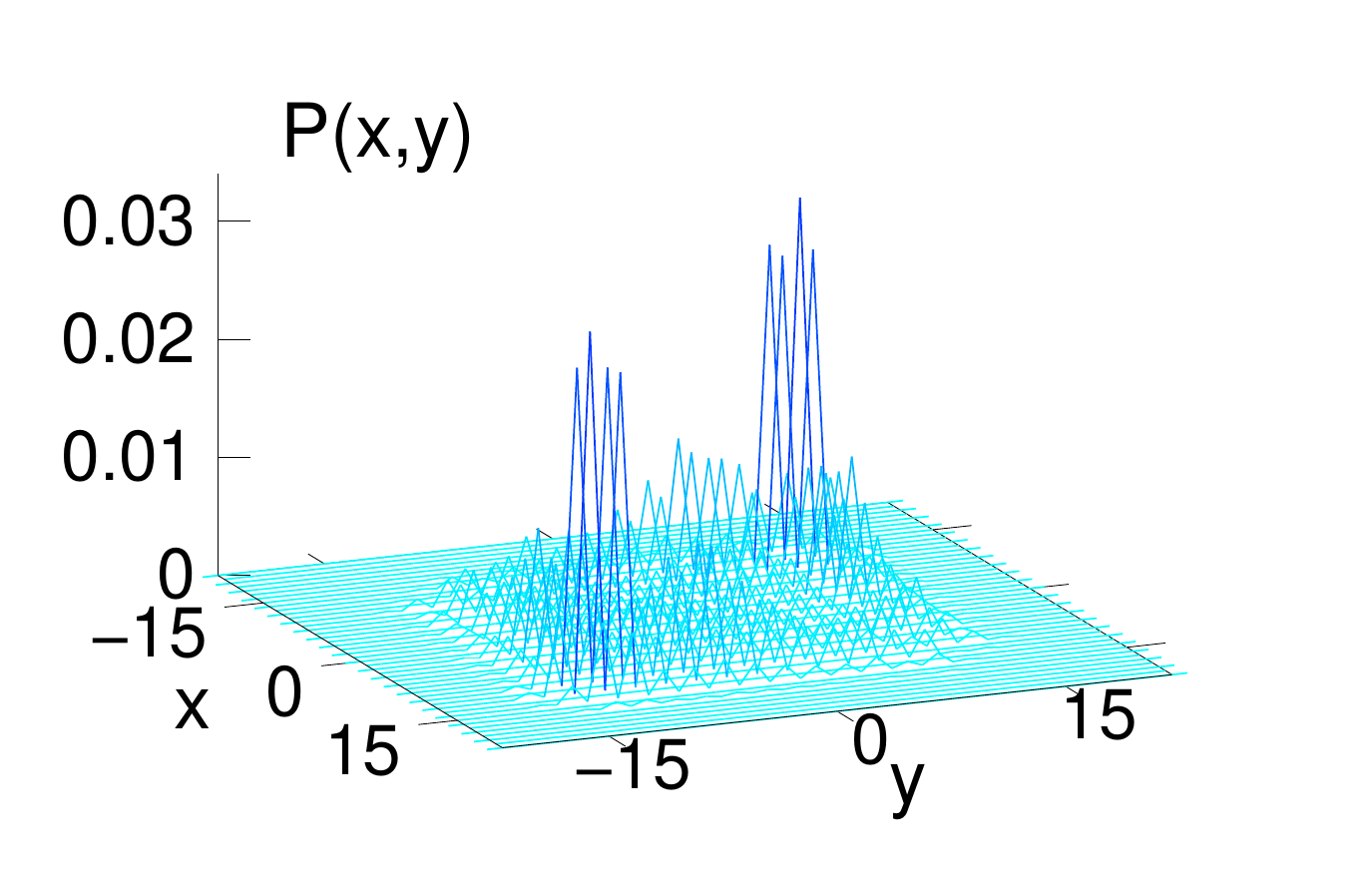}}\hspace*{-.5cm}
\subfigure[\label{fig:5c} ]{\includegraphics[scale=0.25]{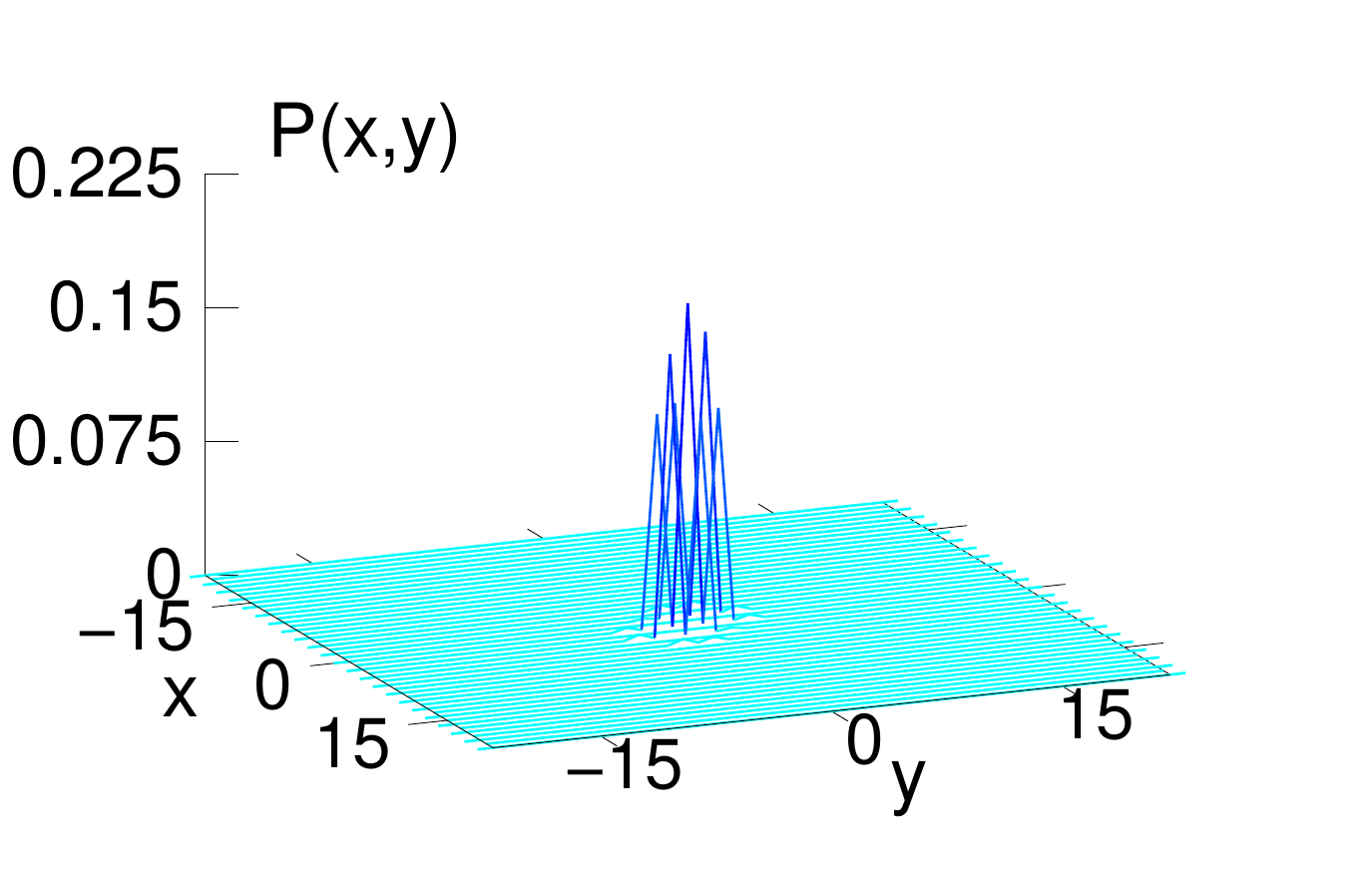}}\hspace*{-.5cm}
\subfigure[\label{fig:5d} ]{\includegraphics[scale=0.25]{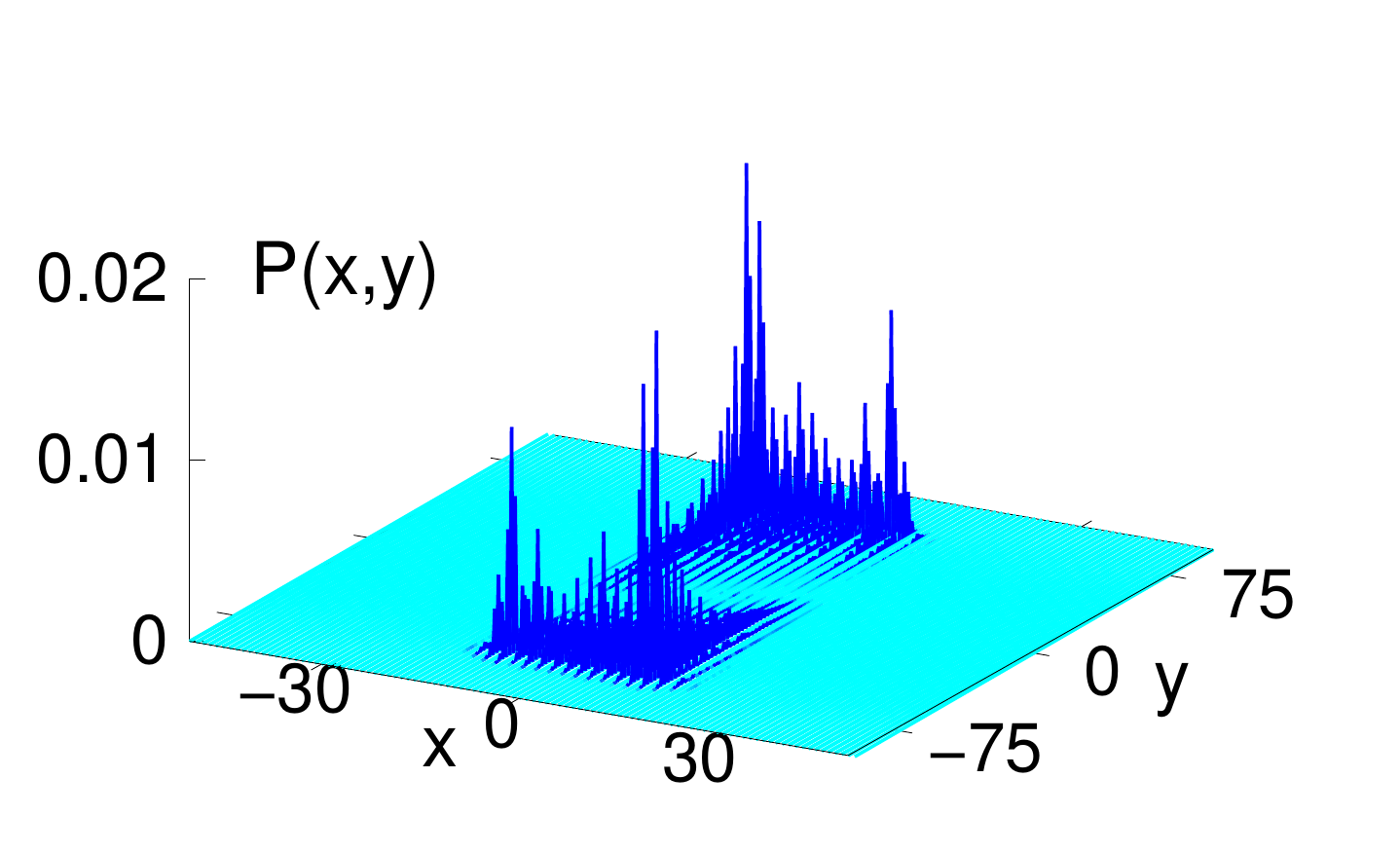}}\hspace*{-.35cm}
\subfigure[\label{fig:5e} ]{\includegraphics[scale=0.25]{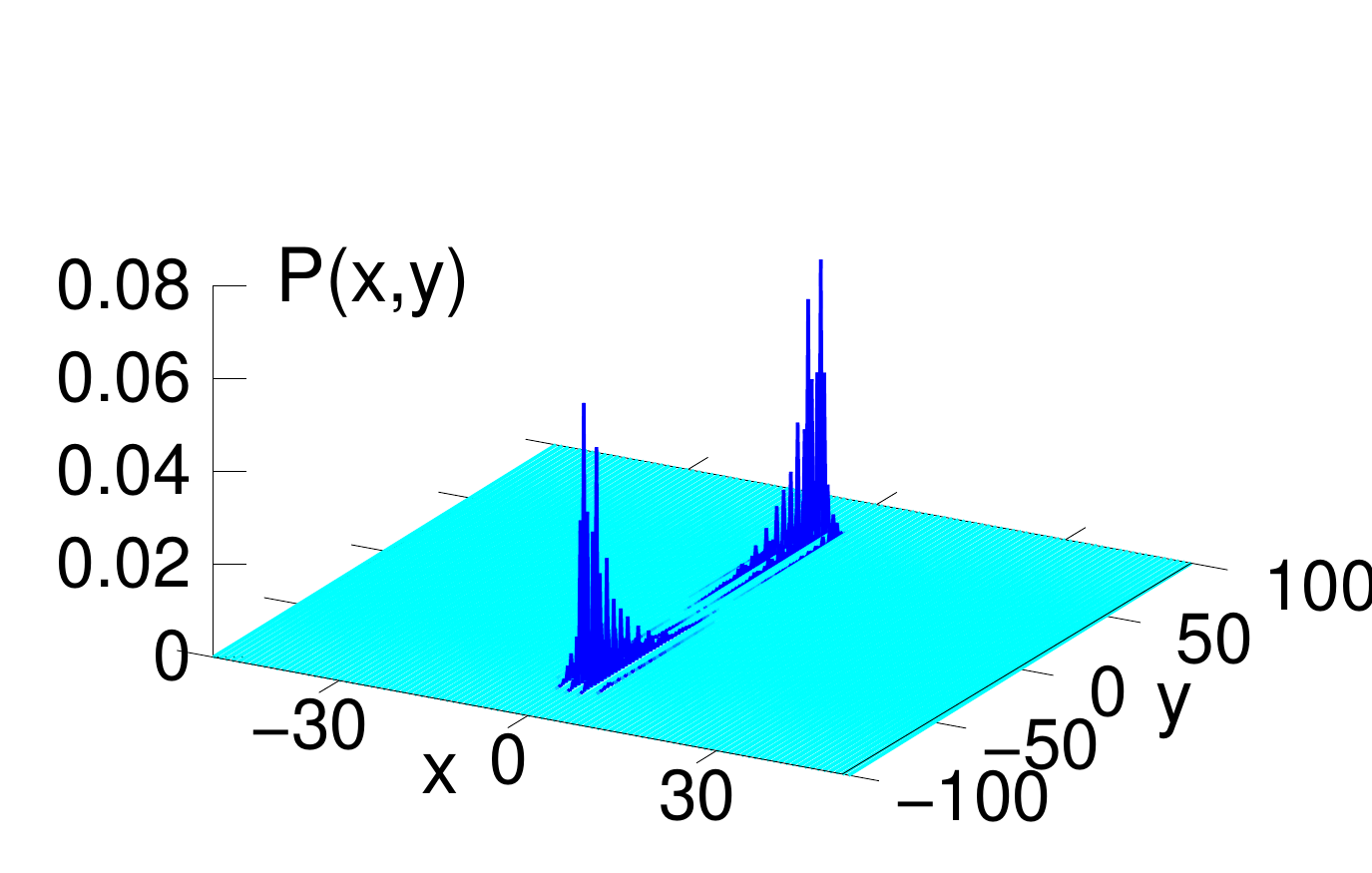}}
\subfigure[\label{fig:6a} ]{\includegraphics[scale=0.25]{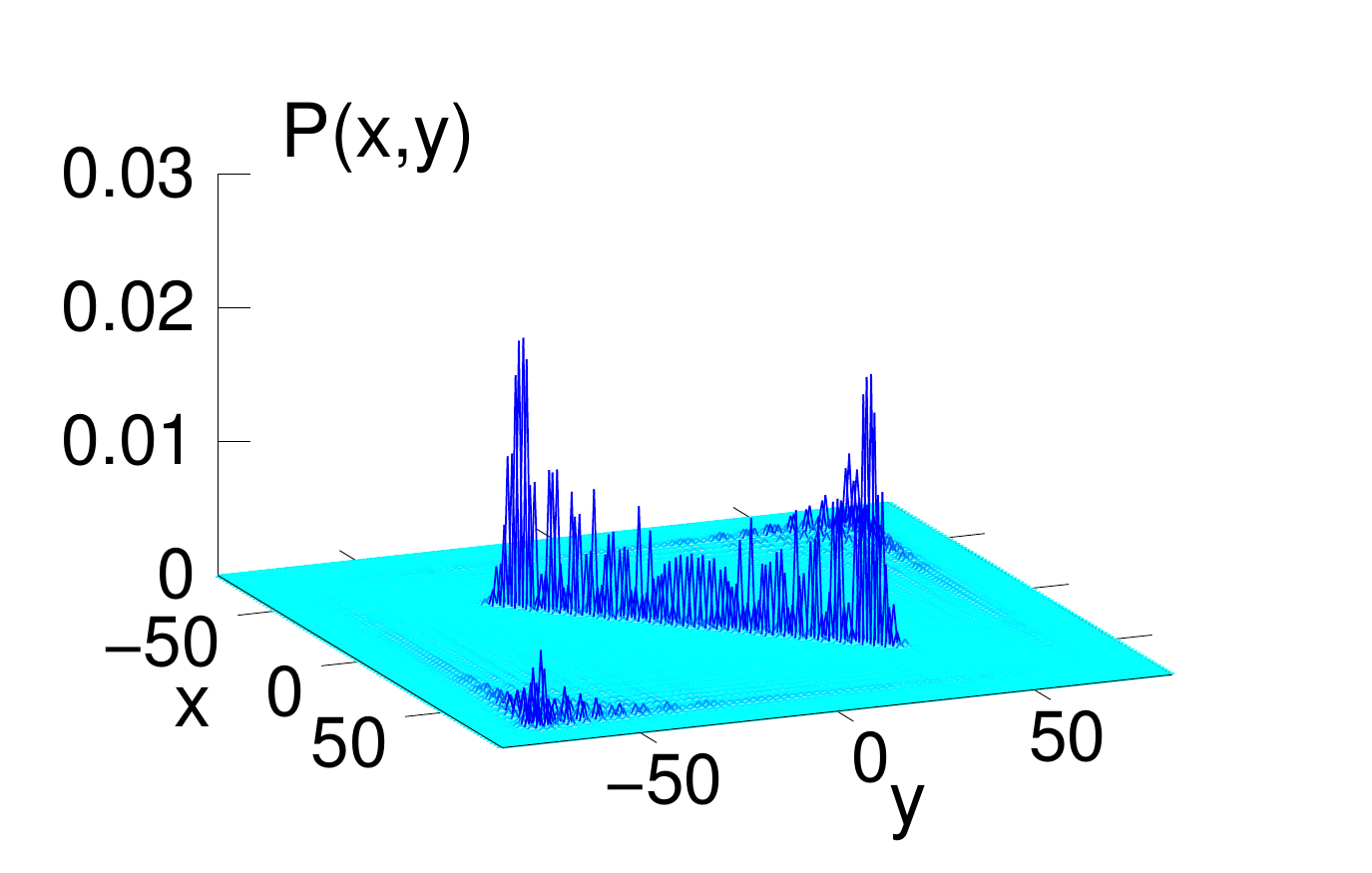}}\hspace*{-.5cm}
\subfigure[\label{fig:6b} ]{\includegraphics[scale=0.25]{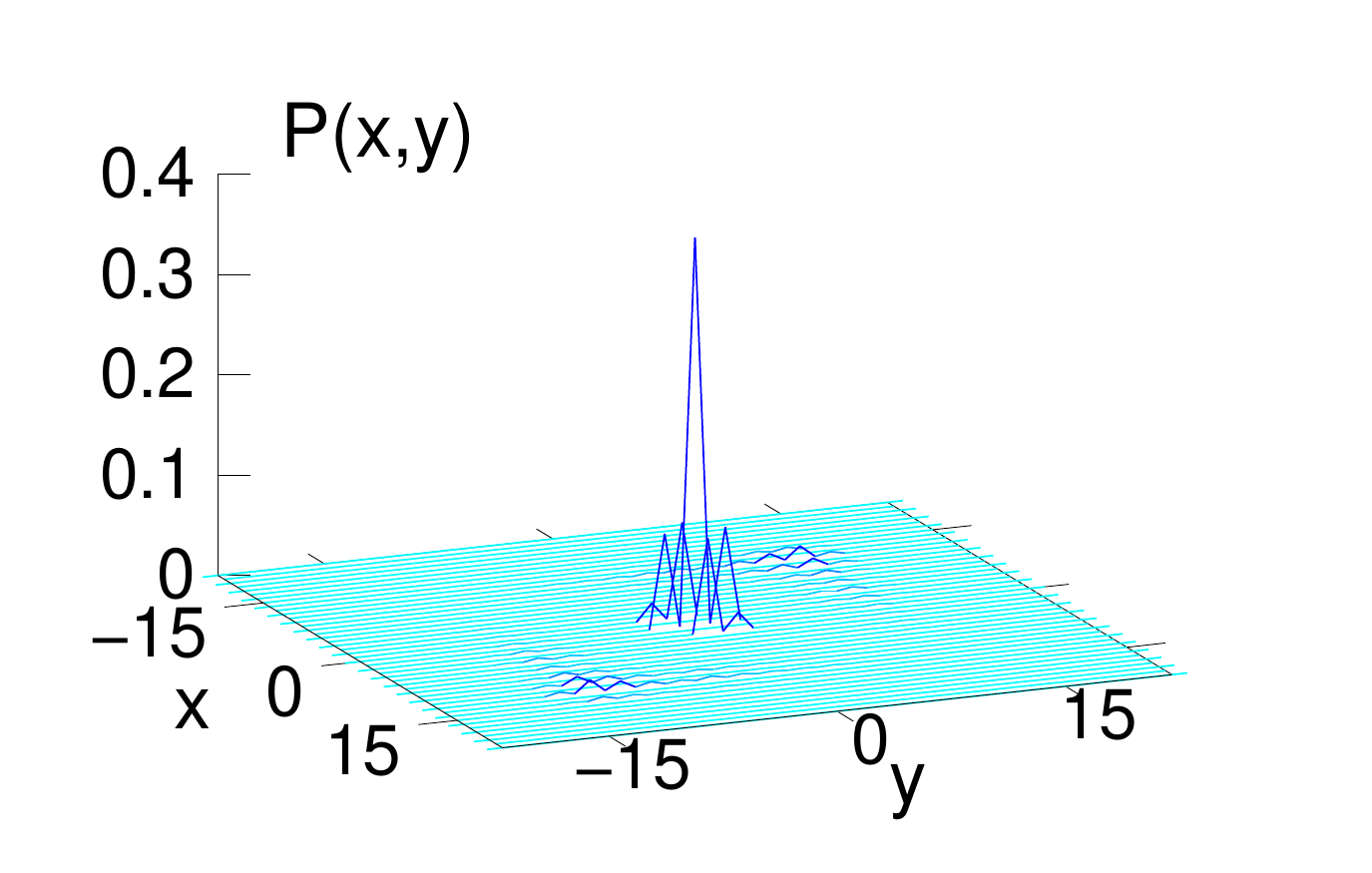}}\hspace*{-.5cm}
\subfigure[\label{fig:6c} ]{\includegraphics[scale=0.25]{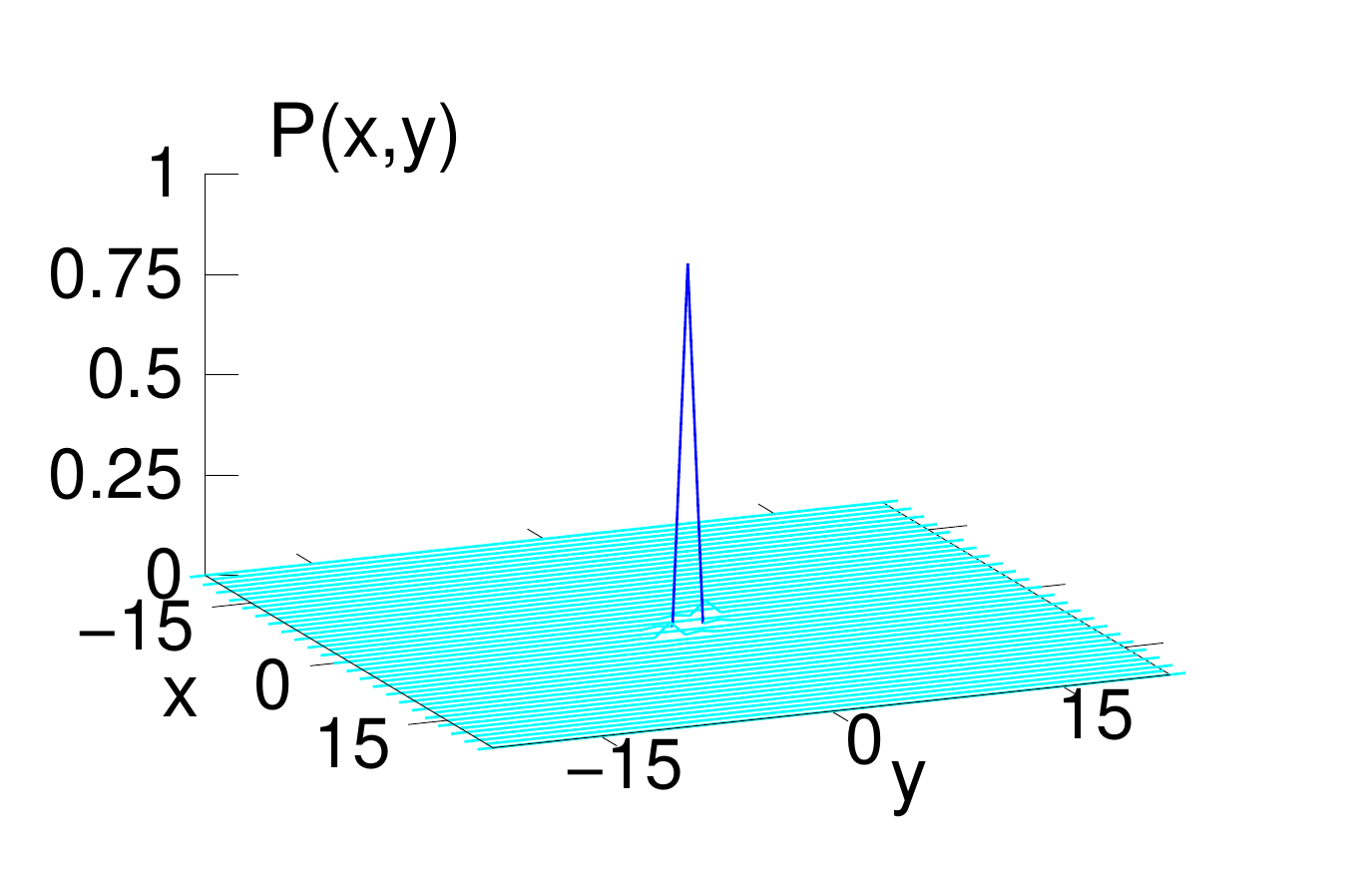}}\hspace*{-.5cm}
\subfigure[\label{fig:6d} ]{\includegraphics[scale=0.25]{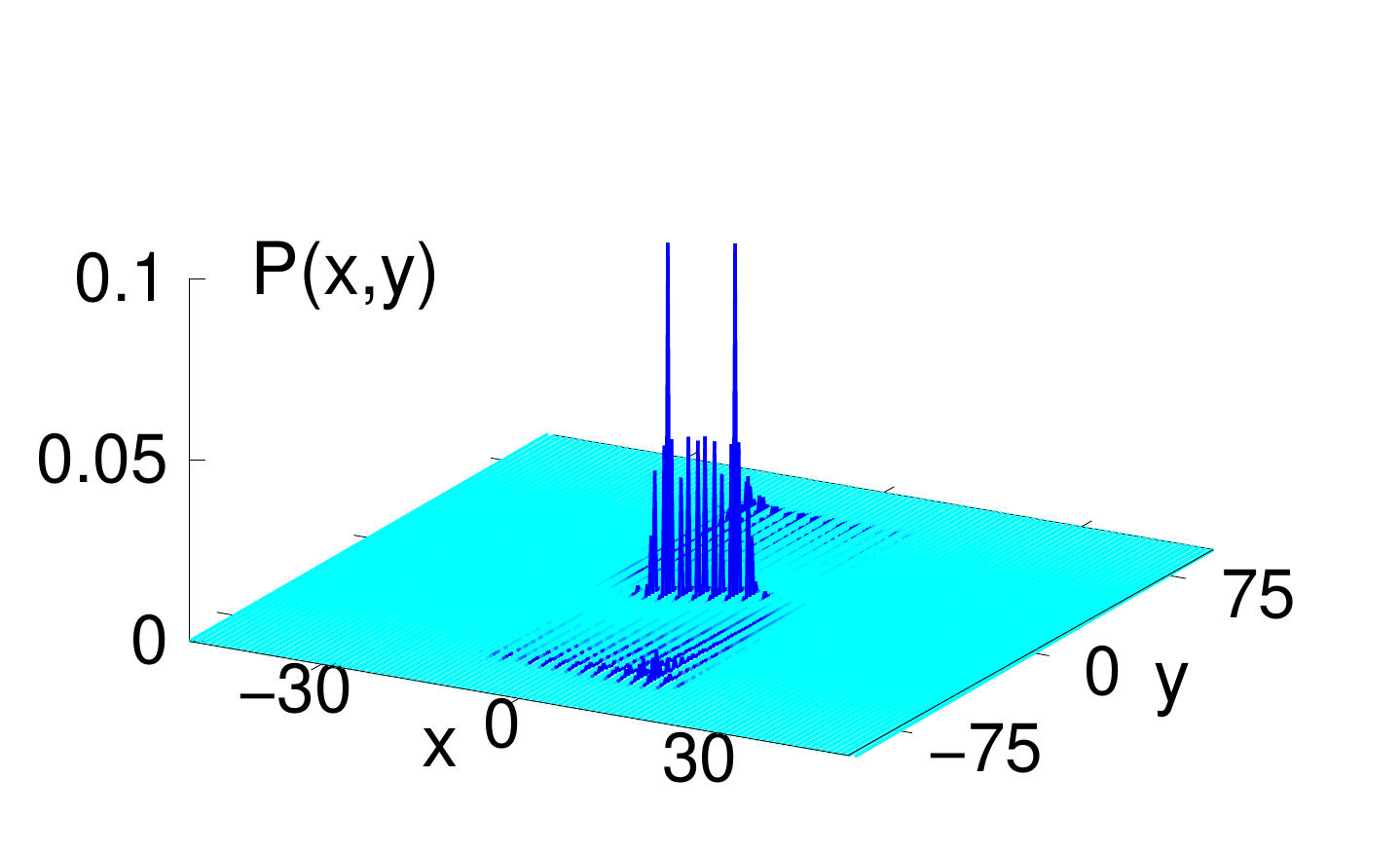}}\hspace*{-.35cm}
\subfigure[\label{fig:6e} ]{\includegraphics[scale=0.25]{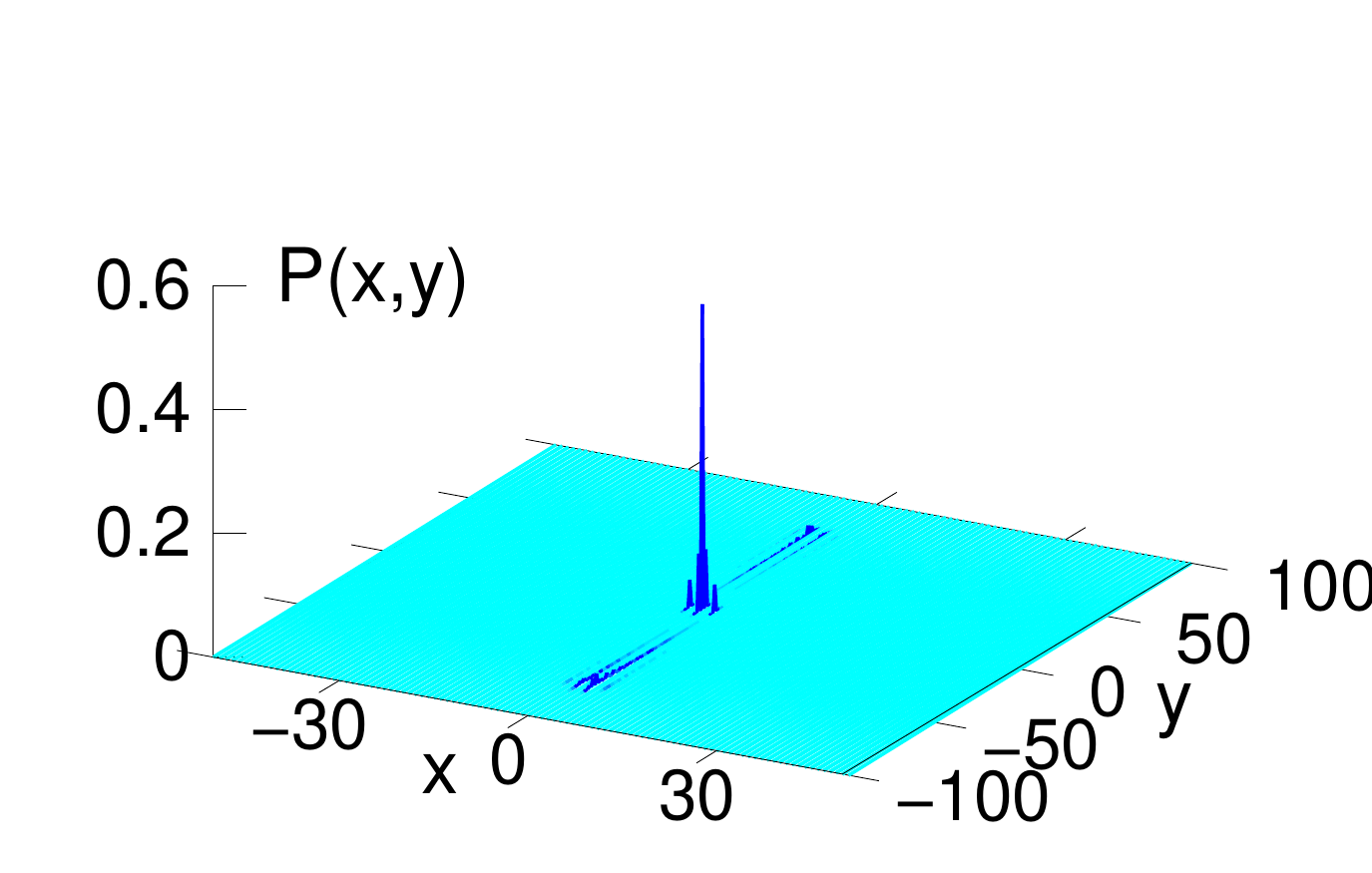}}
\subfigure[\label{fig:5f} ]{\includegraphics[scale=0.40]{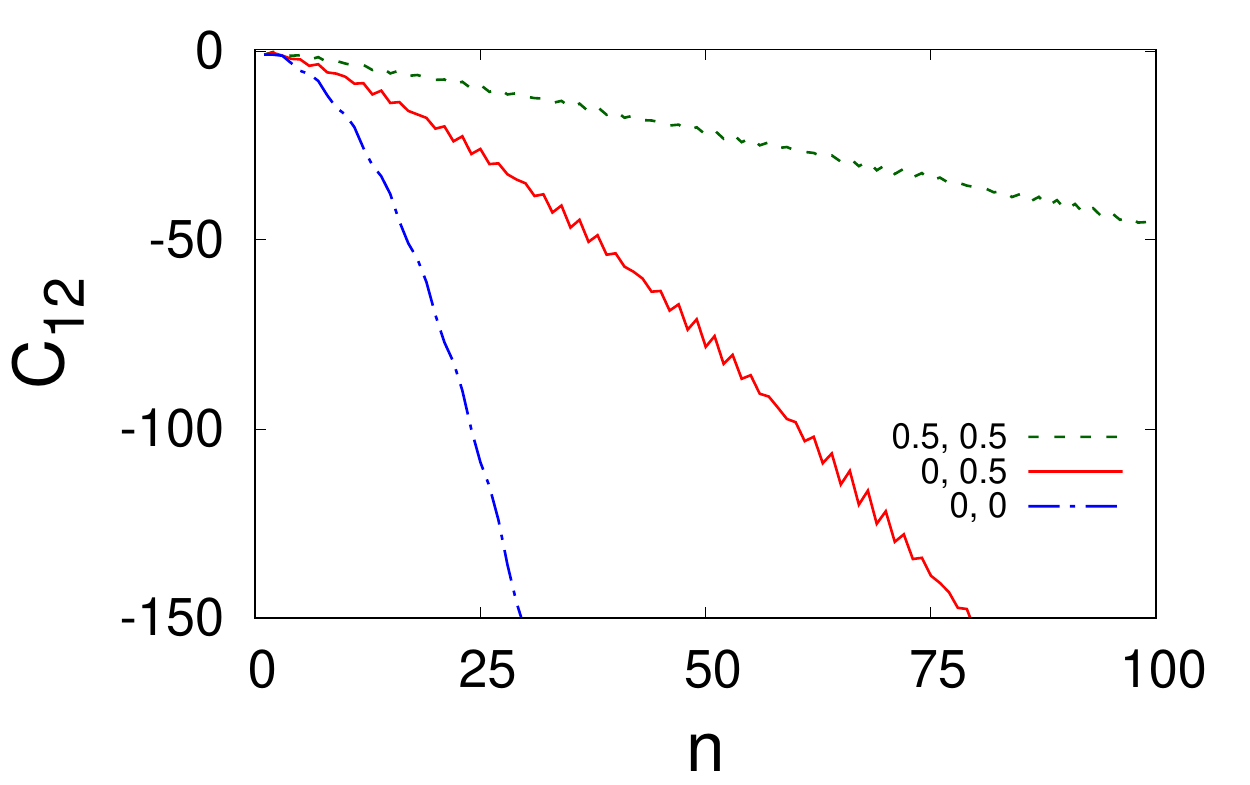}}
\subfigure[\label{fig:5g} ]{\includegraphics[scale=0.40]{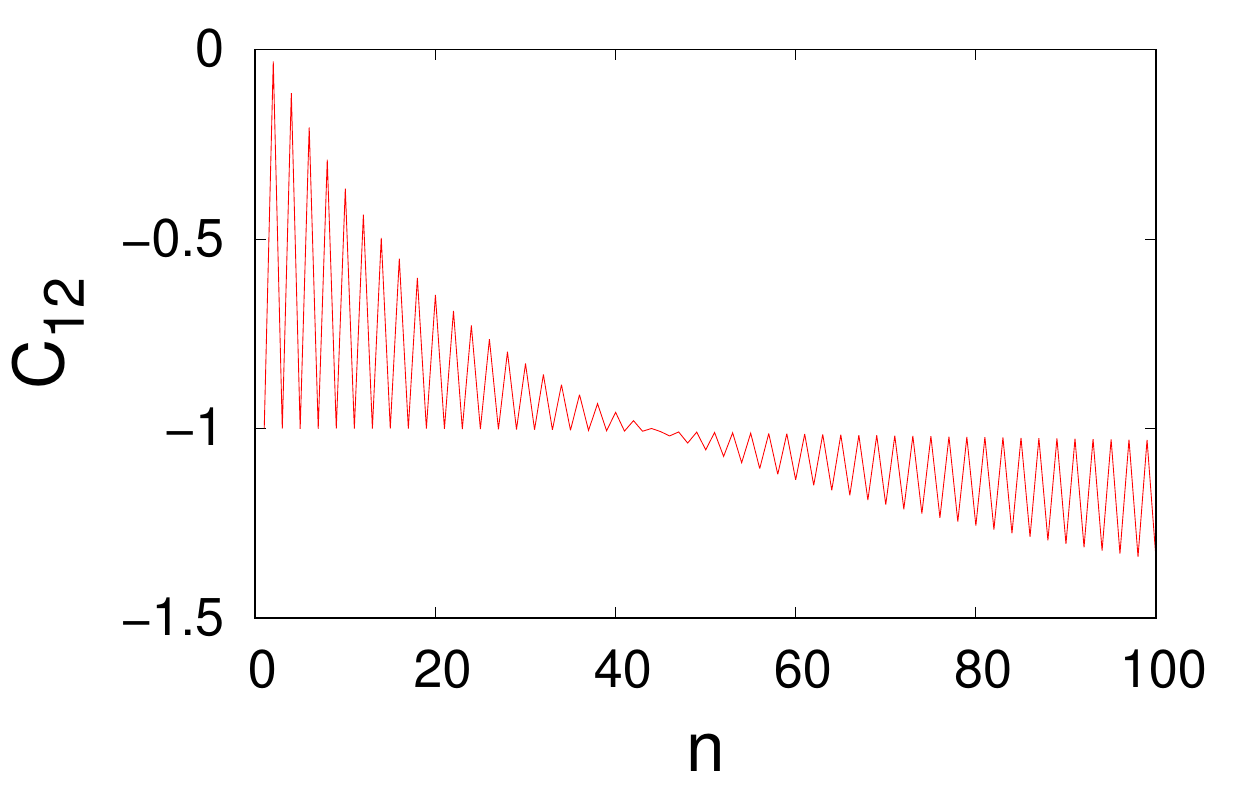}}
\subfigure[\label{fig:5g1} ]{\includegraphics[scale=0.40]{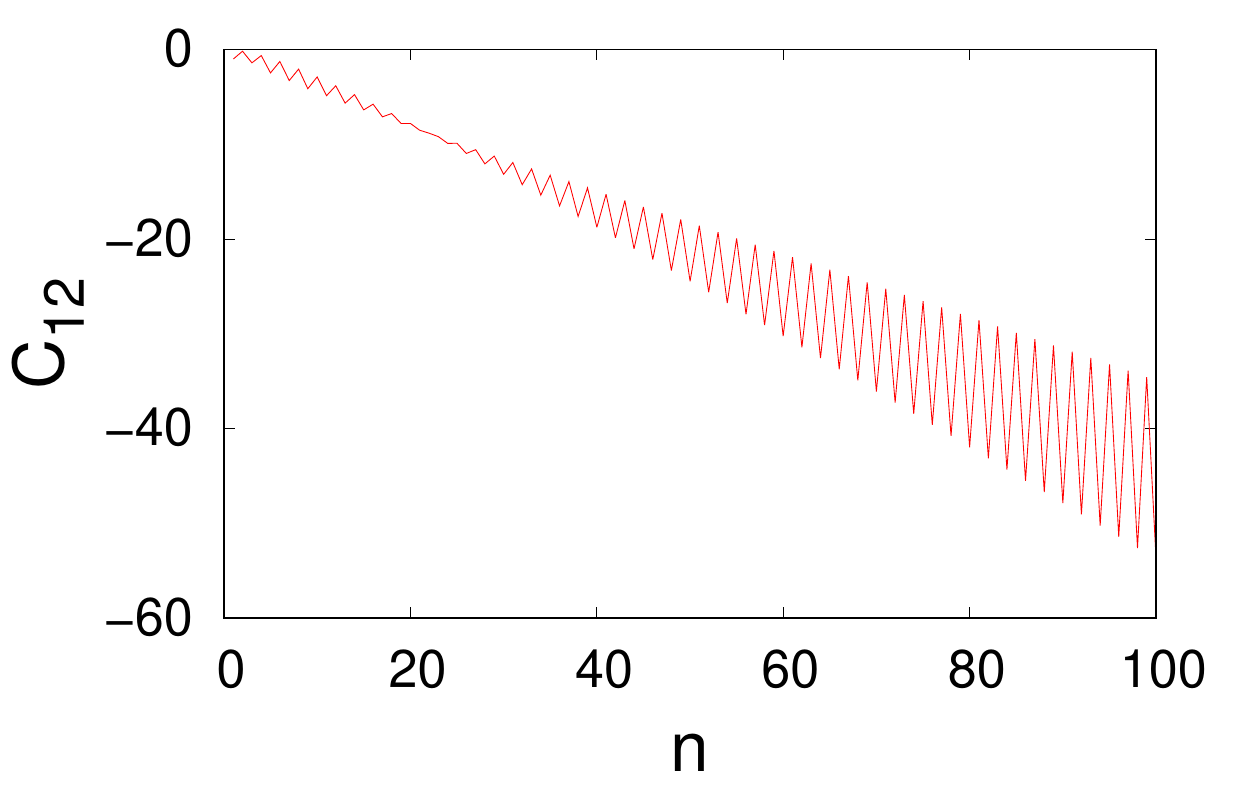}}
\subfigure[\label{fig:6f} ]{\includegraphics[scale=0.40]{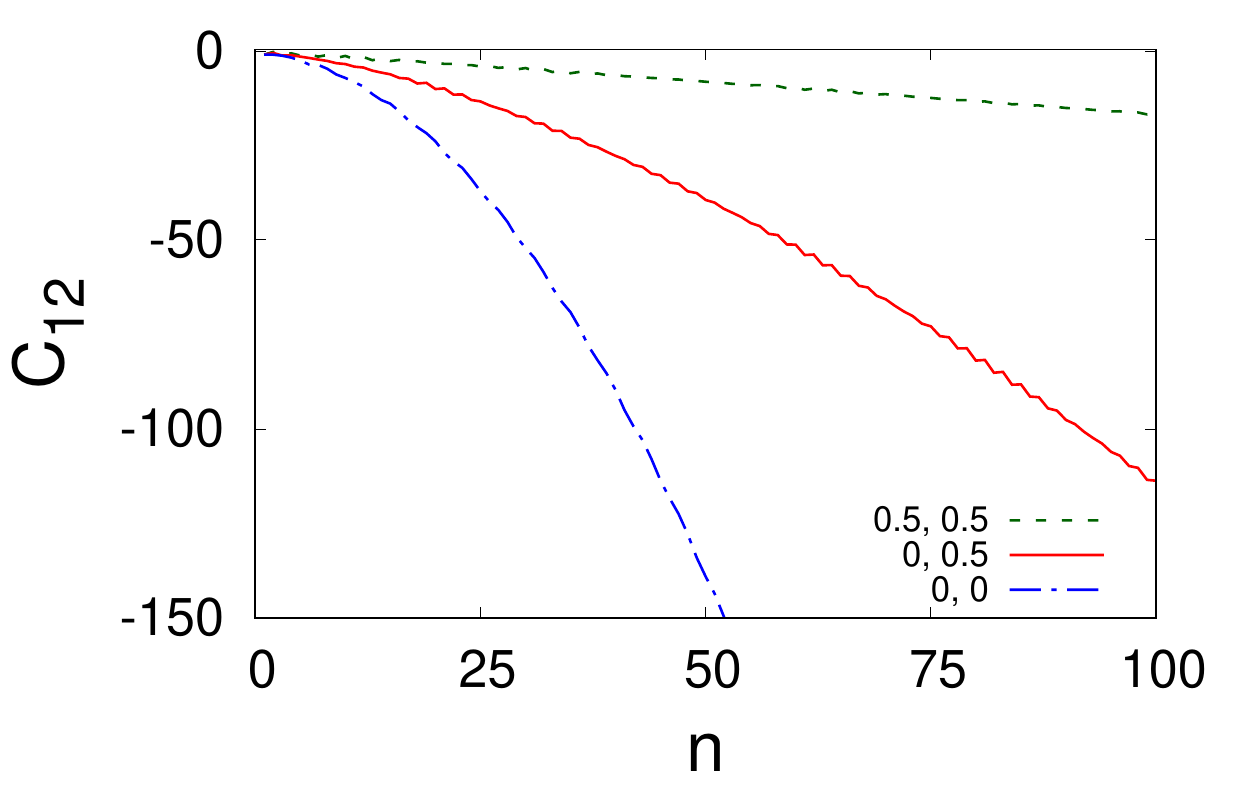}}
\subfigure[\label{fig:6g} ]{\includegraphics[scale=0.40]{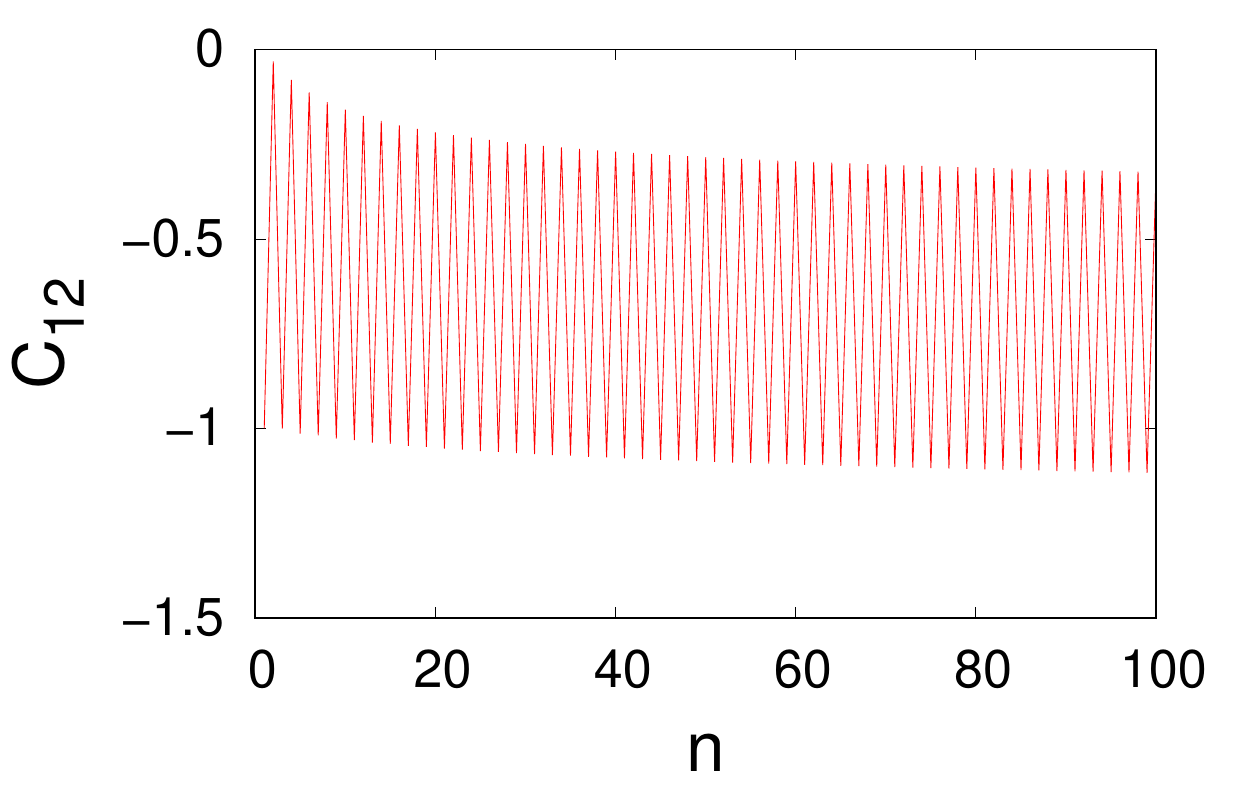}}
\subfigure[\label{fig:6h} ]{\includegraphics[scale=0.40]{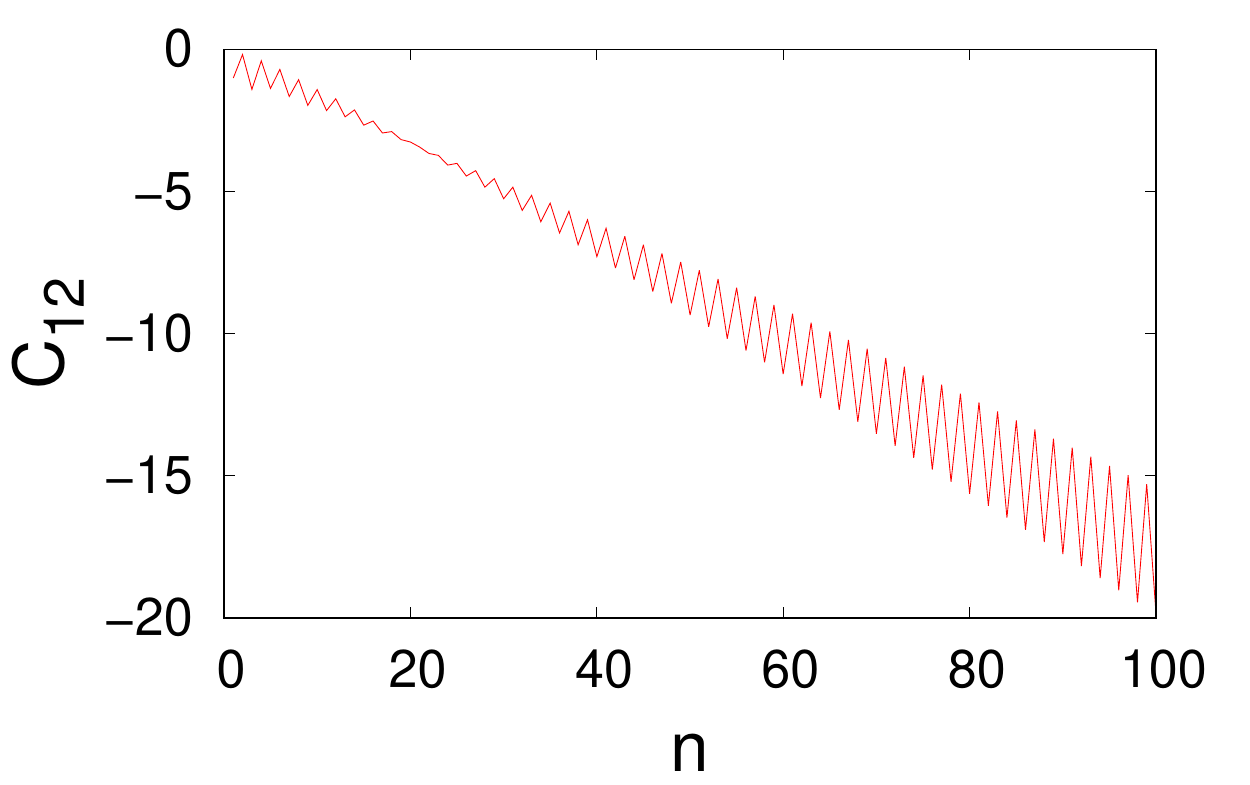}}

\caption{\label{fig:2}{ Here figures (a)-(j) show two-particle probability distributions \(P(x,y)\) after 100 time steps for two \(\mathbb{1}\)- interacting walkers evolving under the influences of the coin \(\hat{C}_{\alpha_{1},\alpha_{2}}(t)\). \(P(x,y)\) for \(|\psi^{+}\rangle\) initial state : (a) \(\alpha_{1}\) = \(\alpha_{2}\) = 0; (b) \(\alpha_{1}\) = \(\alpha_{2}\) = 0.5; (c) \(\alpha_{1}\) = \(\alpha_{2}\) = 1.25, (d) \(\alpha_{1}\) = 0, \(\alpha_{2}\) = 0.5; (e) \(\alpha_{1}\) = 0, \(\alpha_{2}\) = 1.25. \(P(x,y)\) for \(|\psi^{-}\rangle\) initial state : (f) \(\alpha_{1}\) = \(\alpha_{2}\) = 0; (g) \(\alpha_{1}\) = \(\alpha_{2}\) = 0.5; (h) \(\alpha_{1}\) = \(\alpha_{2}\) = 1.25, (i) \(\alpha_{1}\) = 0, \(\alpha_{2}\) = 0.5, (j) \(\alpha_{1}\) = 0, \(\alpha_{2}\) = 1.25. Figures (k)-(p) show the variations of the correlation function \(C_{12}\) against dimension less time \(n\) for two \(\mathbb{1}\) interacting walkers evolving under the influences of \(\hat{C}_{\alpha_{1},\alpha_{2}}(t)\). Variations of \(C_{12}\) for \(|\psi^{+}\rangle\) initial state : (k) \(\alpha_{1}\) = \(\alpha_{2}\) = 0; \(\alpha_{1}\) = \(\alpha_{2}\) = 0.5; \(\alpha_{1}\) = 0, \(\alpha_{2}\) = 0.5 (l) \(\alpha_{1}\) = \(\alpha_{2}\) = 1.25 (m) \(\alpha_{1}\) = 0, \(\alpha_{2}\) = 1.25. Variations of \(C_{12}\) for \(|\psi^{-}\rangle\) initial state : (n) \(\alpha_{1}\) = \(\alpha_{2}\) = 0; \(\alpha_{1}\) = \(\alpha_{2}\) = 0.5; \(\alpha_{1}\) = 0, \(\alpha_{2}\) = 0.5 (o) \(\alpha_{1}\) = \(\alpha_{2}\) = 1.25 (p) \(\alpha_{1}\) = 0, \(\alpha_{2}\) = 1.25.} }

\end{figure*}

\begin{figure*}[th]
\centering

\subfigure[\label{fig:7a} ]{\includegraphics[scale=0.25]{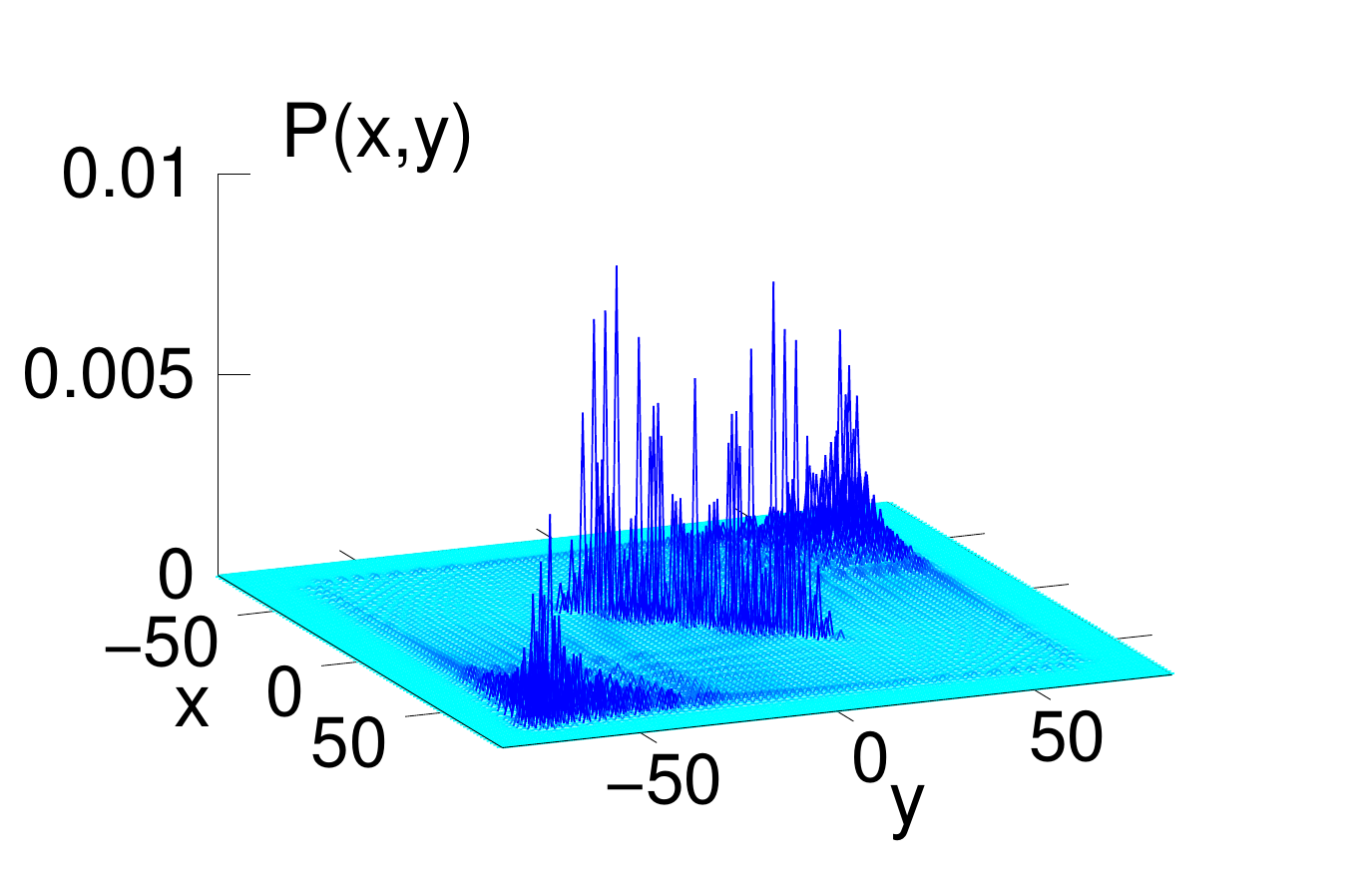}}\hspace*{-.5cm}
\subfigure[\label{fig:7b} ]{\includegraphics[scale=0.25]{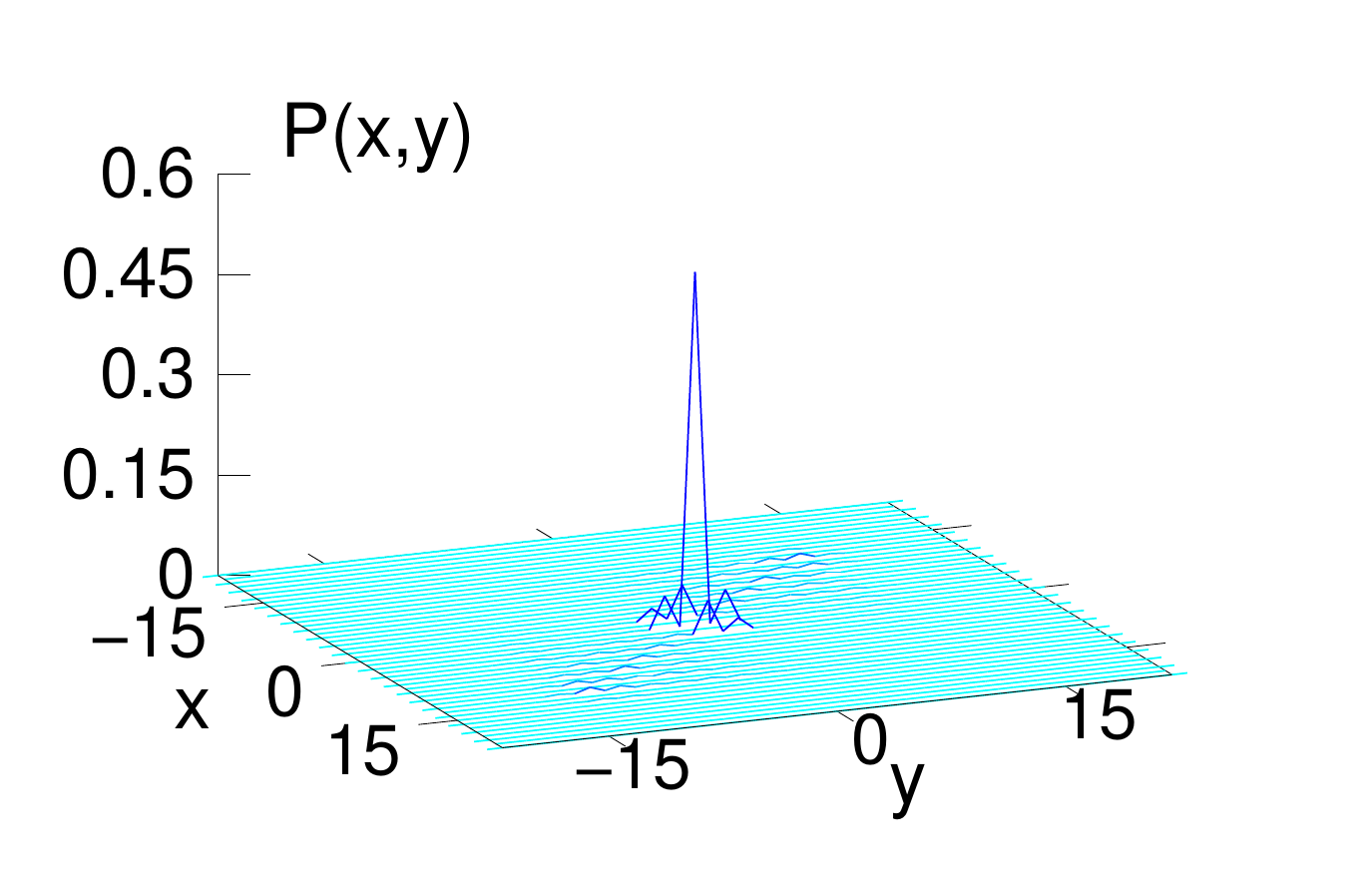}}\hspace*{-.5cm}
\subfigure[\label{fig:7c} ]{\includegraphics[scale=0.25]{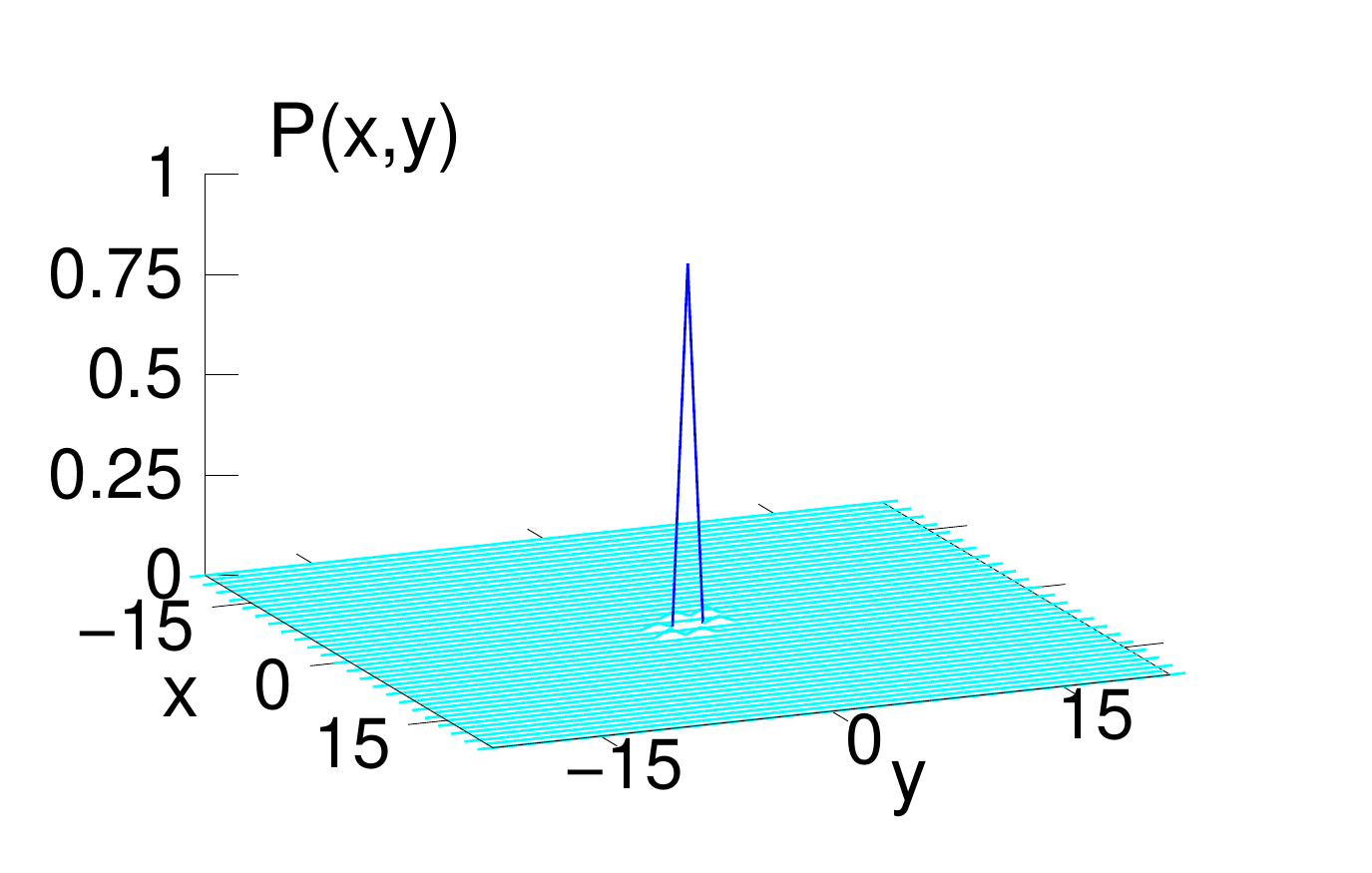}}\hspace*{-.55cm}
\subfigure[\label{fig:7d} ]{\includegraphics[scale=0.25]{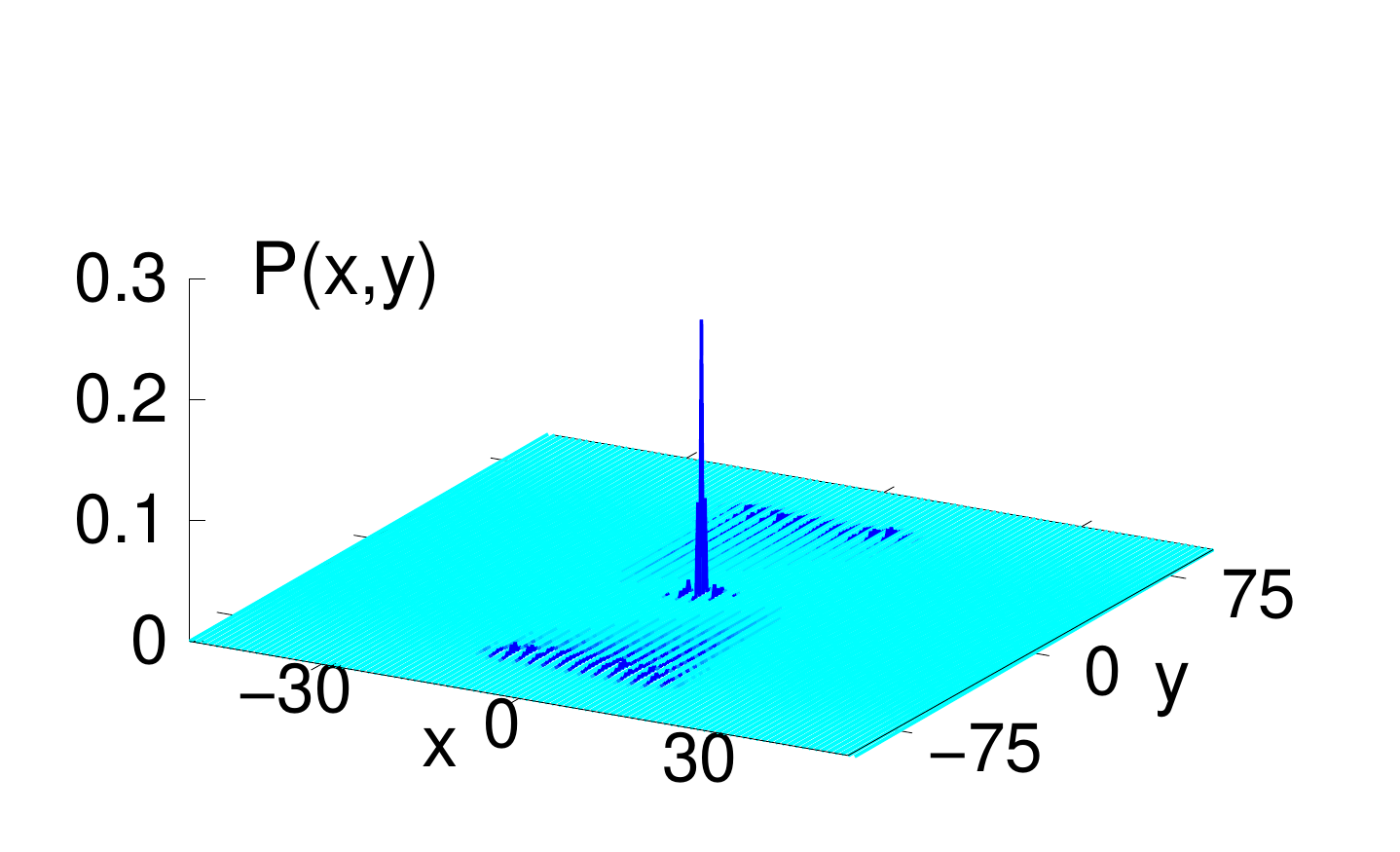}}\hspace*{-.35cm}
\subfigure[\label{fig:7e} ]{\includegraphics[scale=0.25]{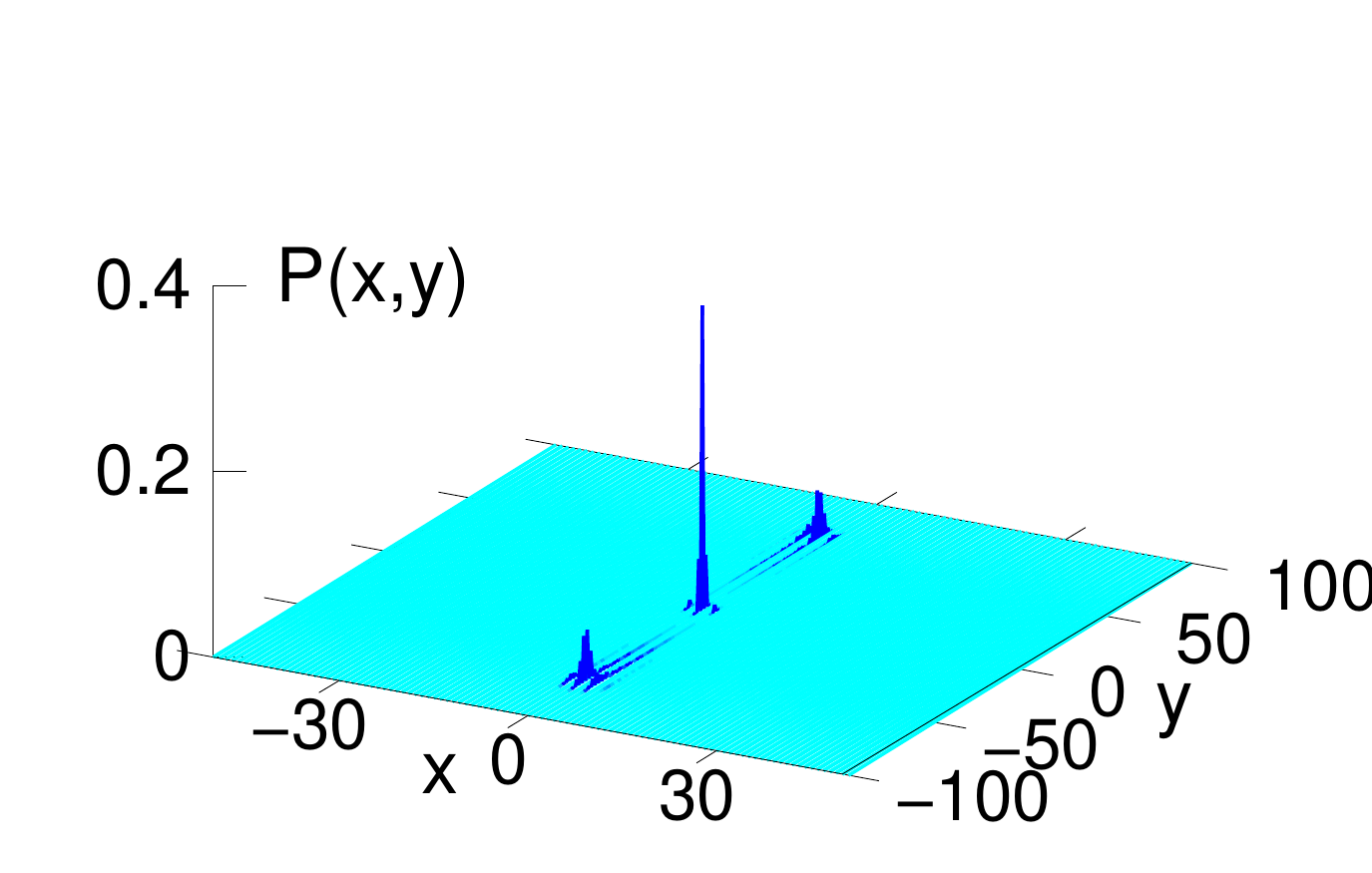}}
\subfigure[\label{fig:8a} ]{\includegraphics[scale=0.25]{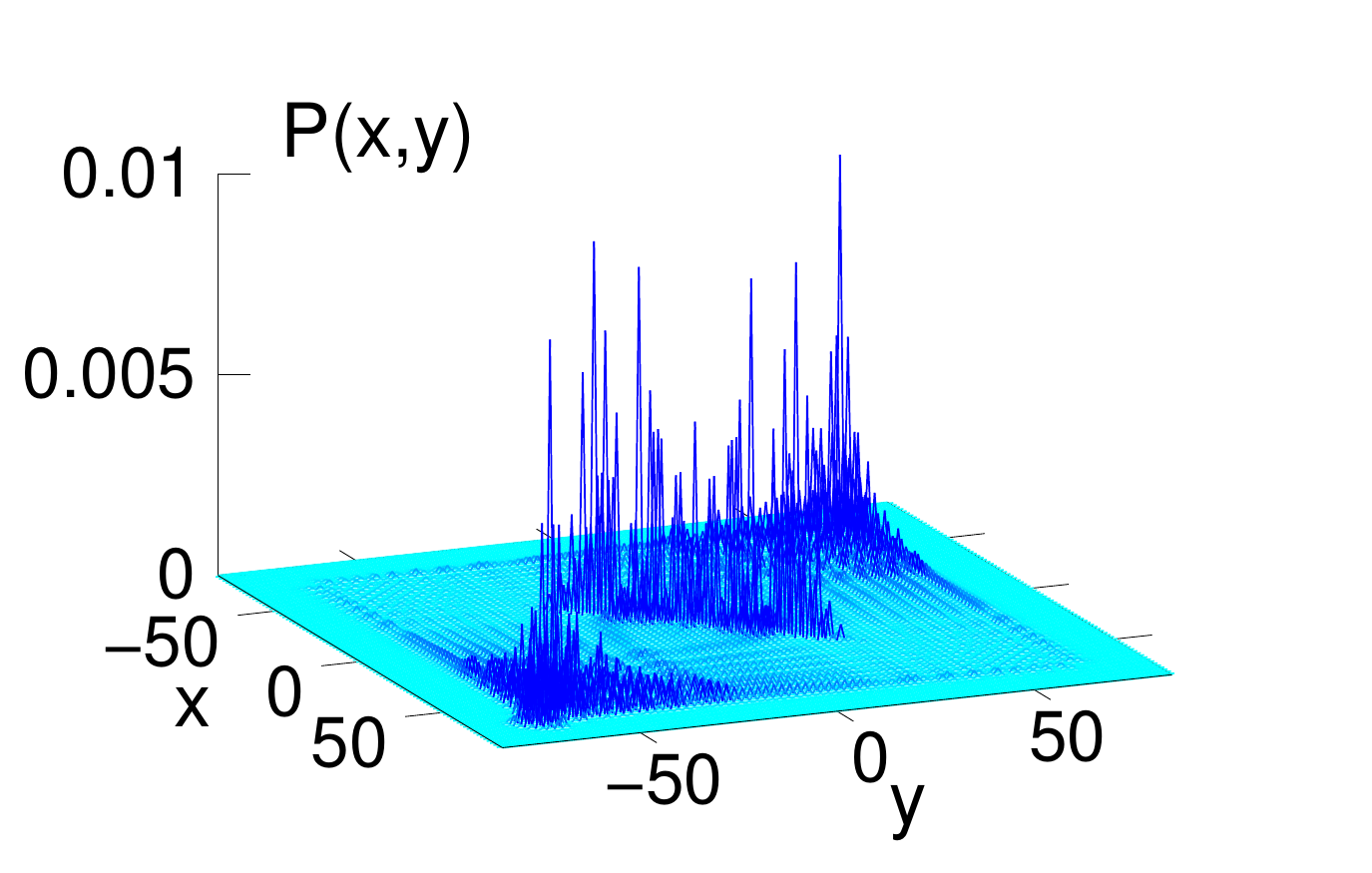}}\hspace*{-.55cm}
\subfigure[\label{fig:8b} ]{\includegraphics[scale=0.25]{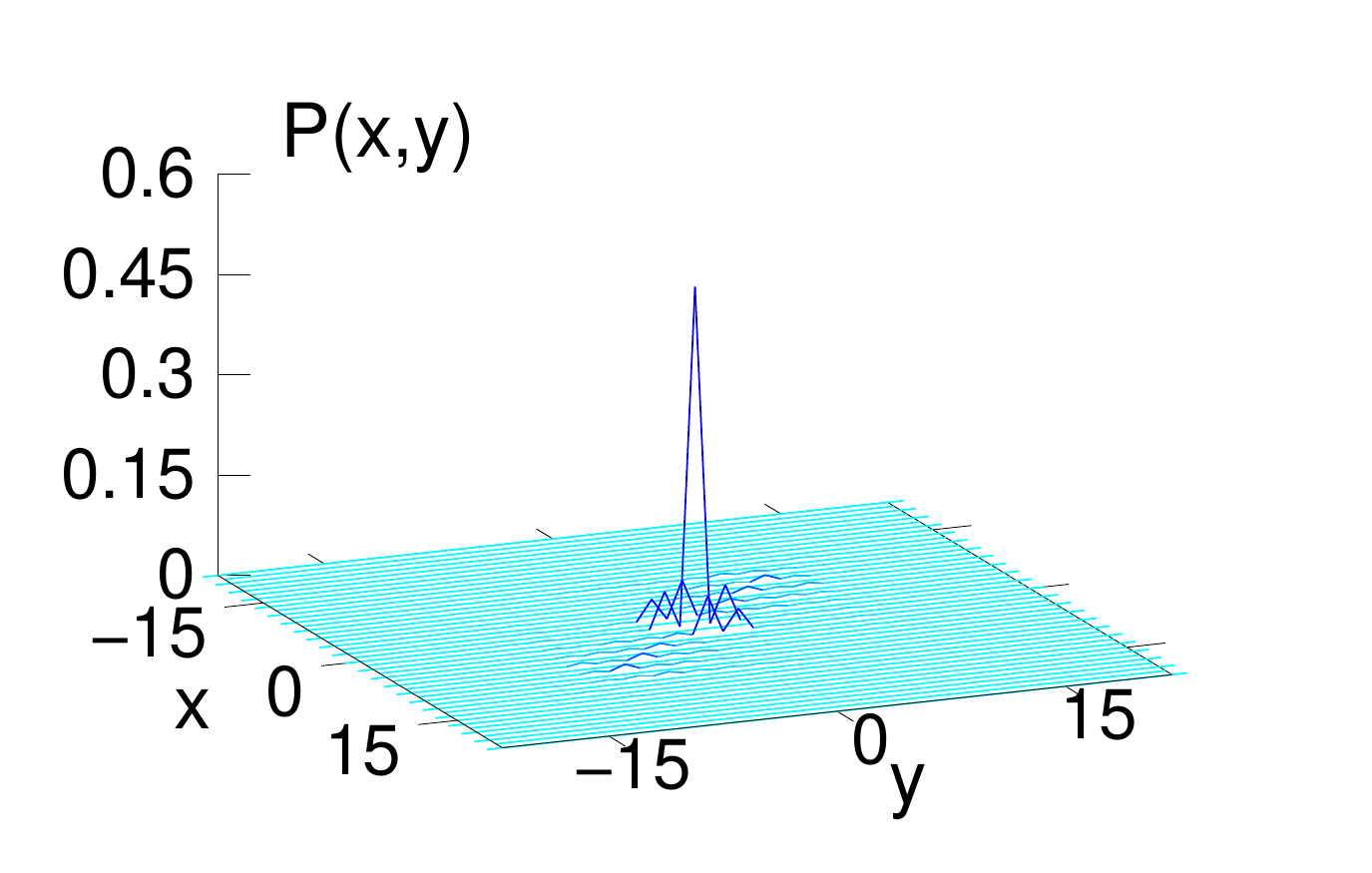}}\hspace*{-.55cm}
\subfigure[\label{fig:8c} ]{\includegraphics[scale=0.25]{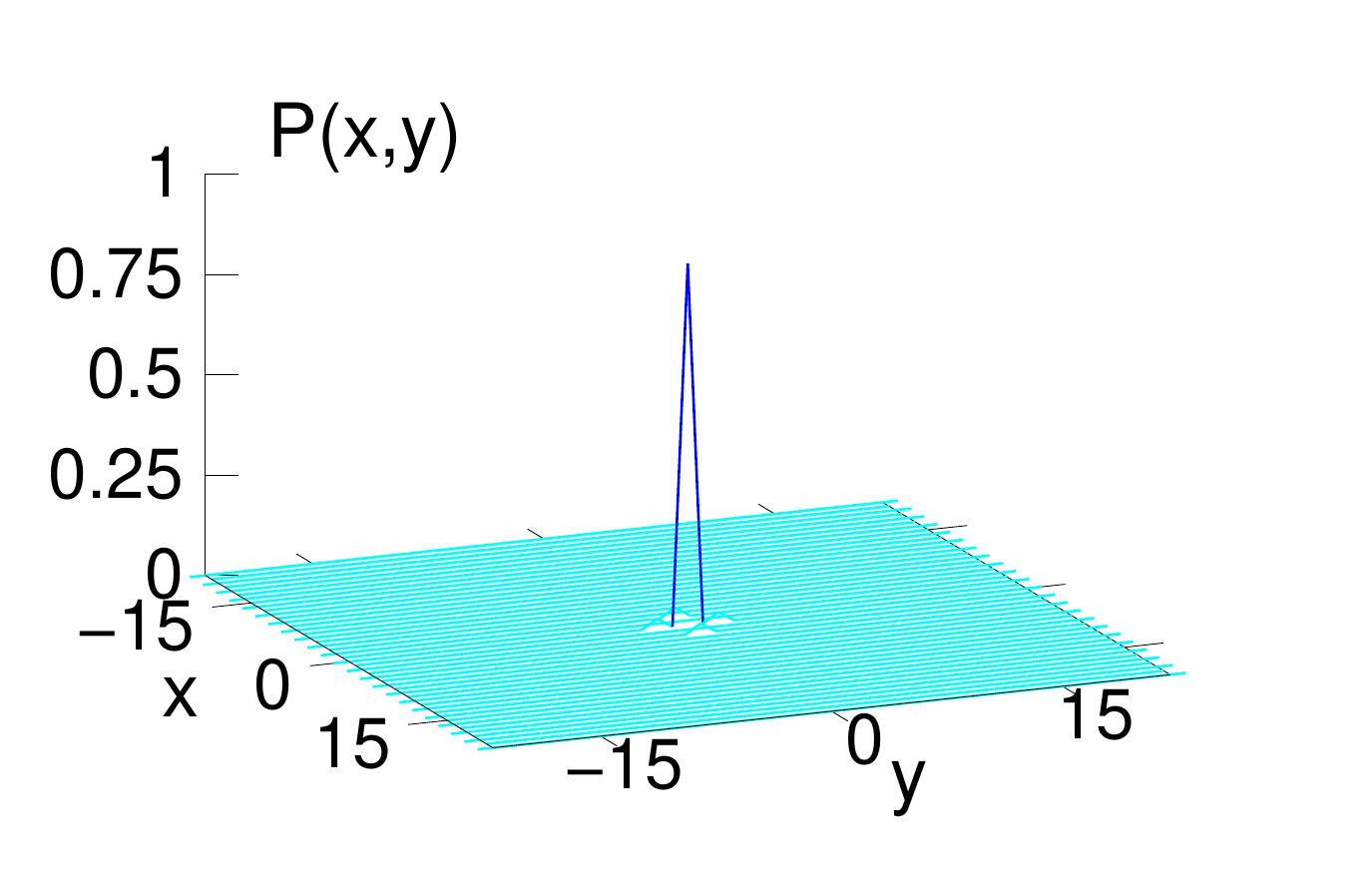}}\hspace*{-.55cm}
\subfigure[\label{fig:8d} ]{\includegraphics[scale=0.25]{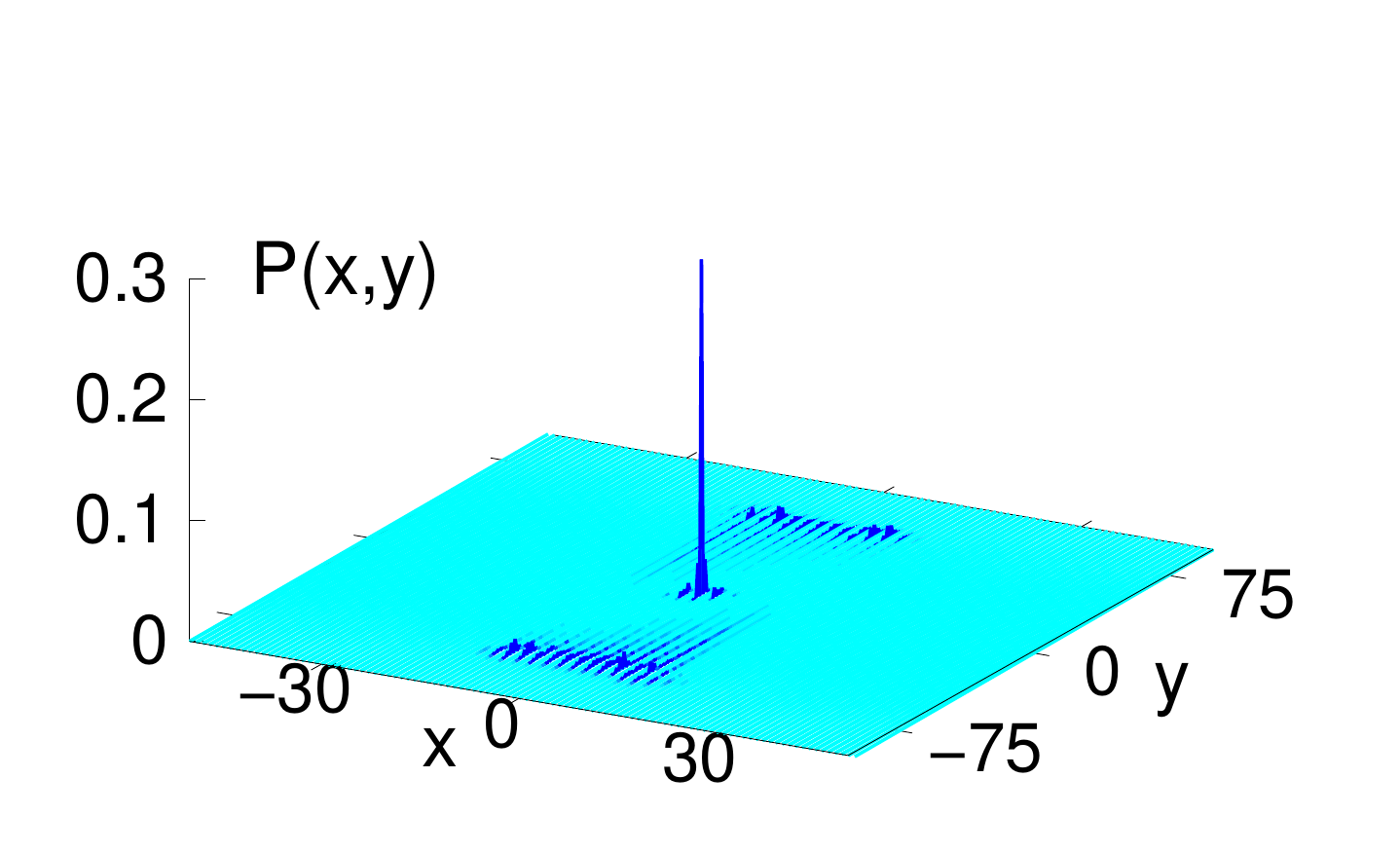}}\hspace*{-.35cm}
\subfigure[\label{fig:8e} ]{\includegraphics[scale=0.25]{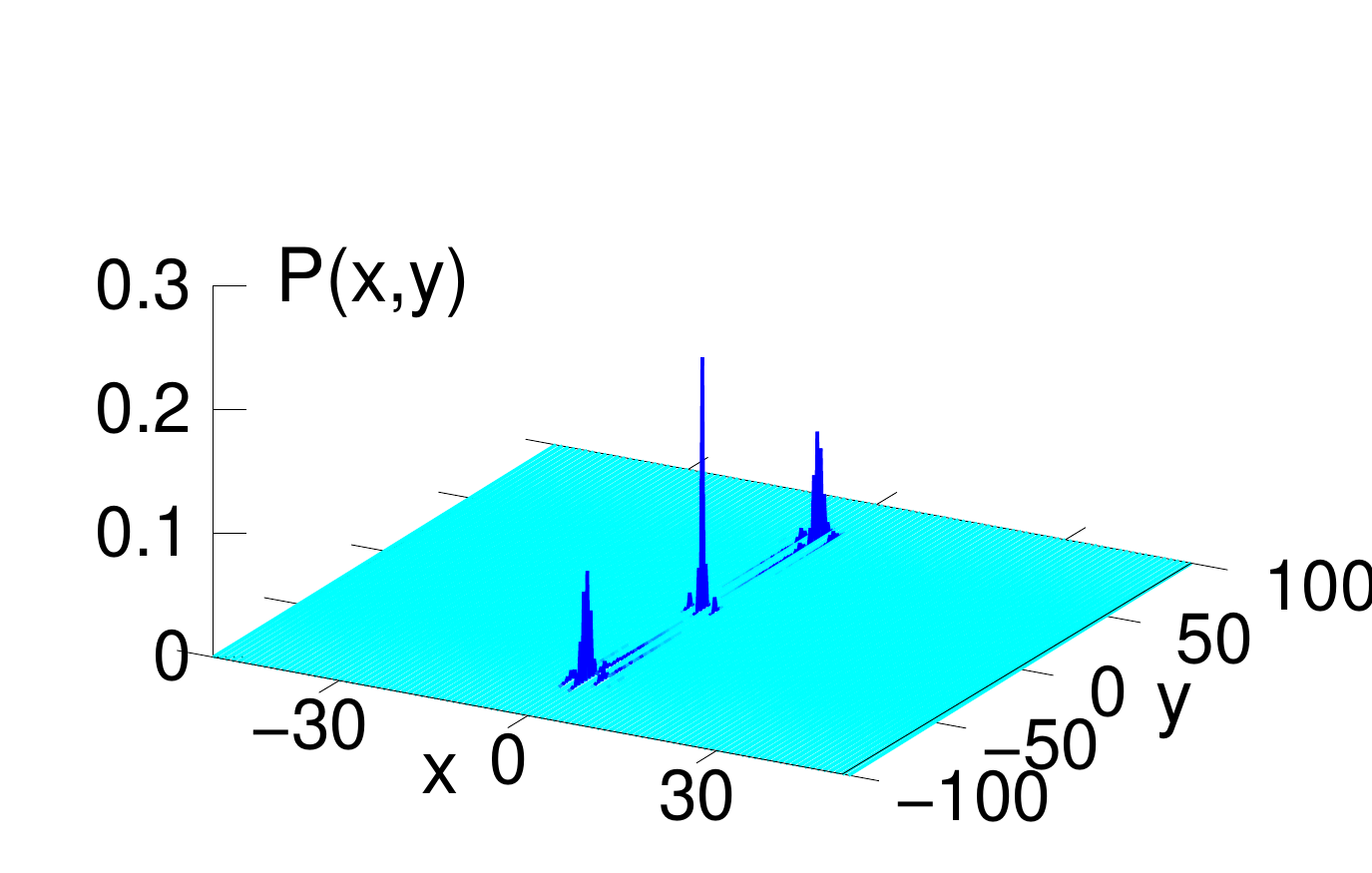}}
\subfigure[\label{fig:9a} ]{\includegraphics[scale=0.25]{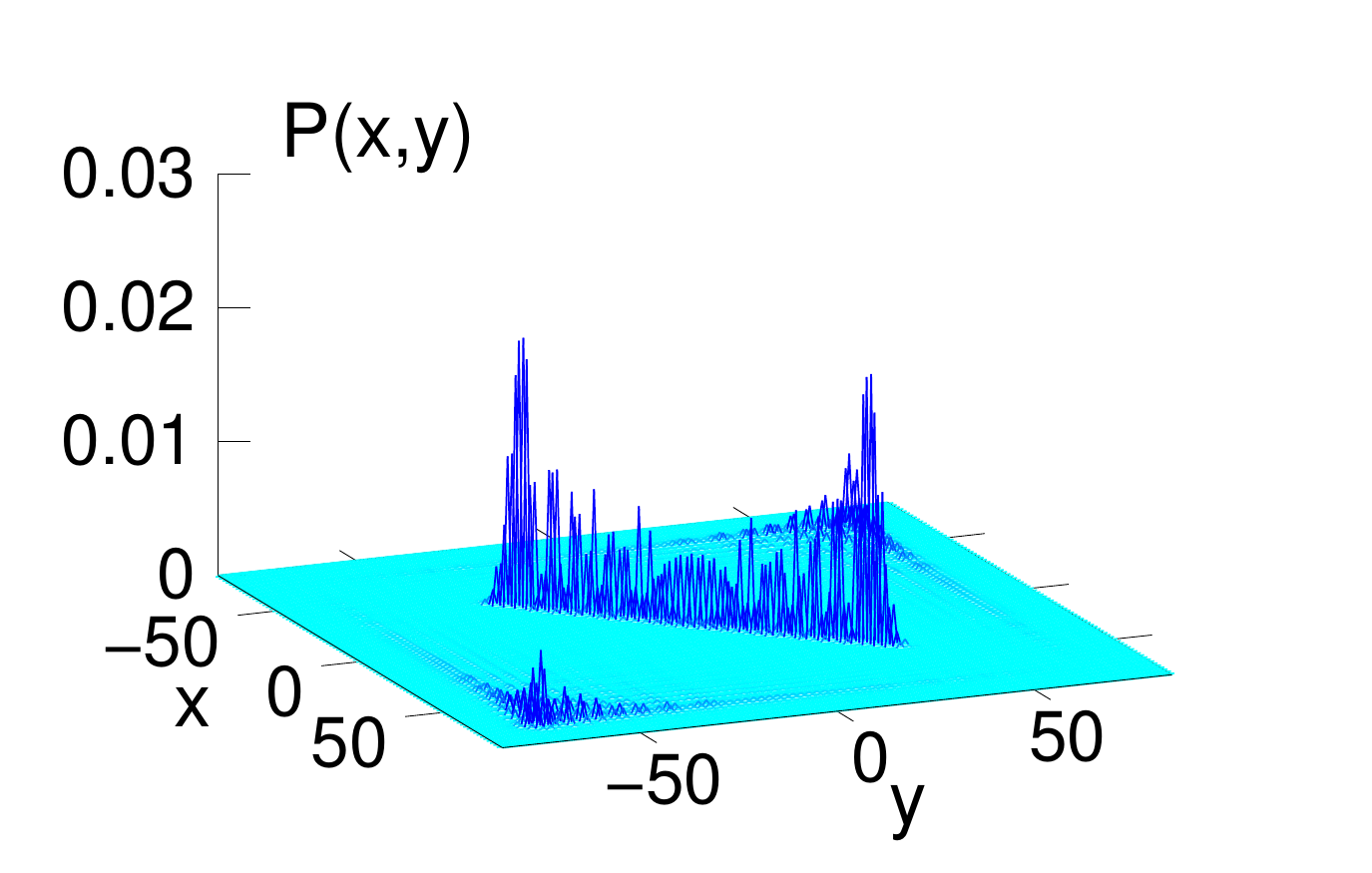}}\hspace*{-.55cm}
\subfigure[\label{fig:9b} ]{\includegraphics[scale=0.25]{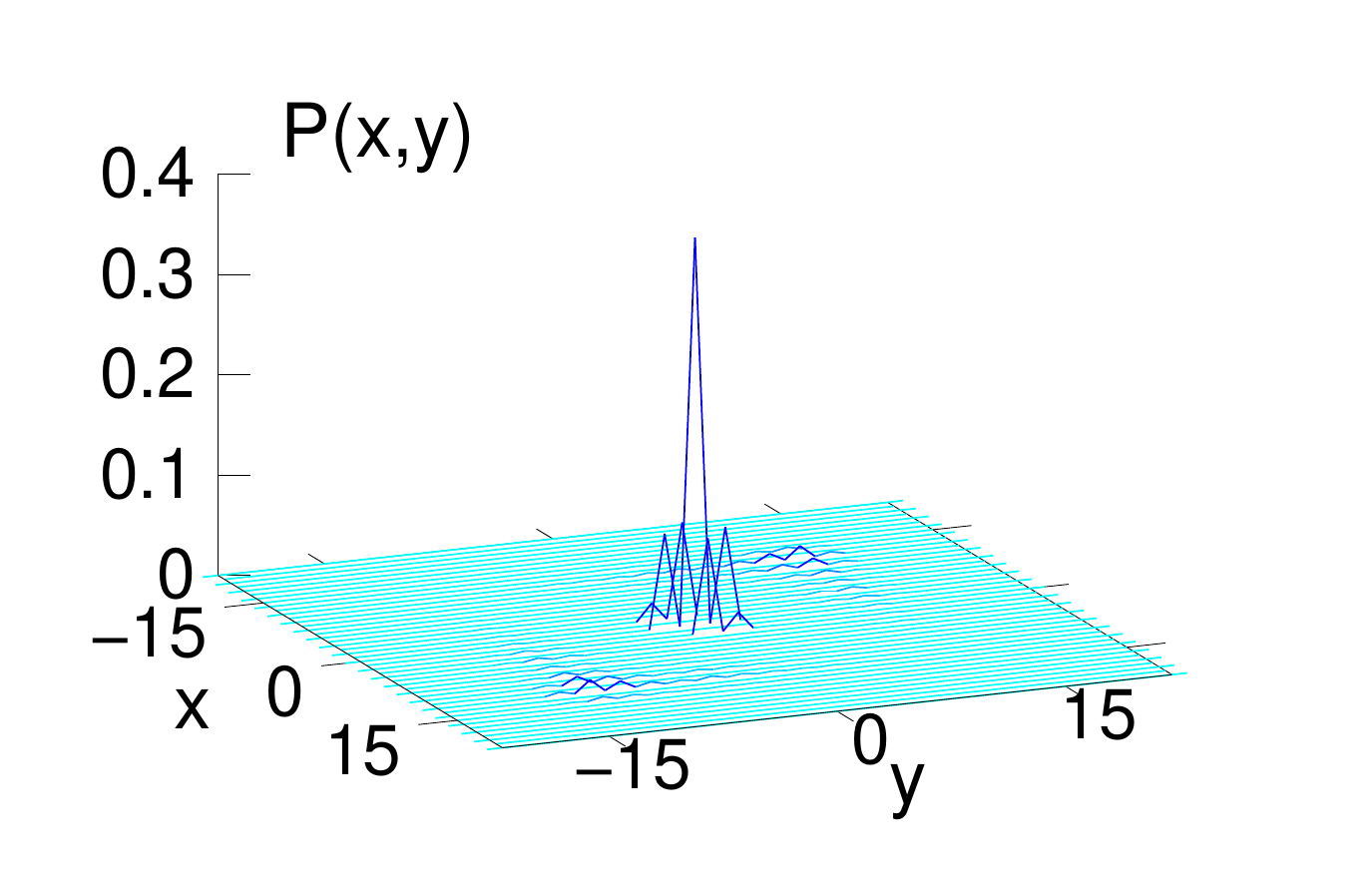}}\hspace*{-.55cm}
\subfigure[\label{fig:9c} ]{\includegraphics[scale=0.25]{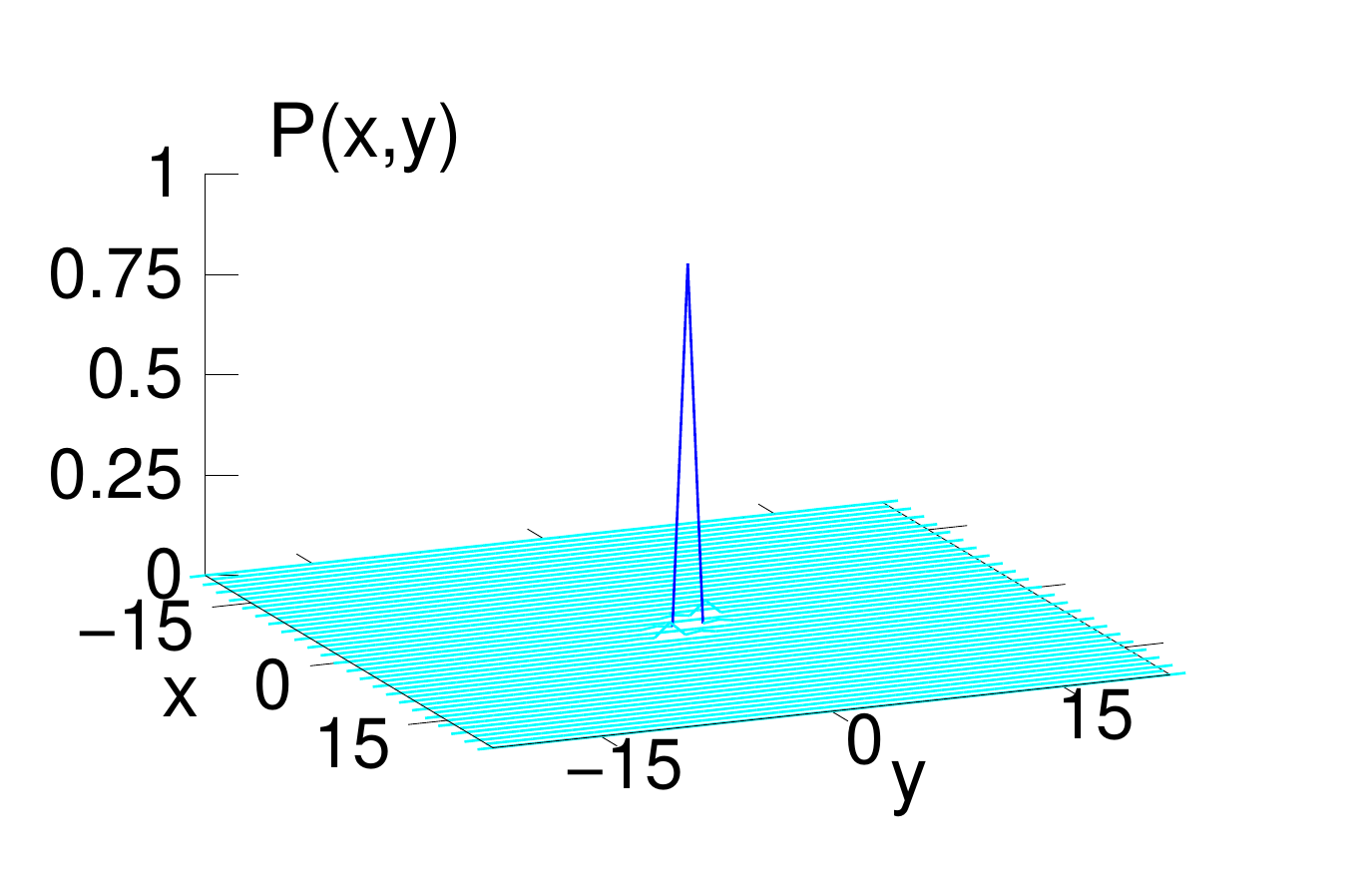}}\hspace*{-.55cm}
\subfigure[\label{fig:9d} ]{\includegraphics[scale=0.25]{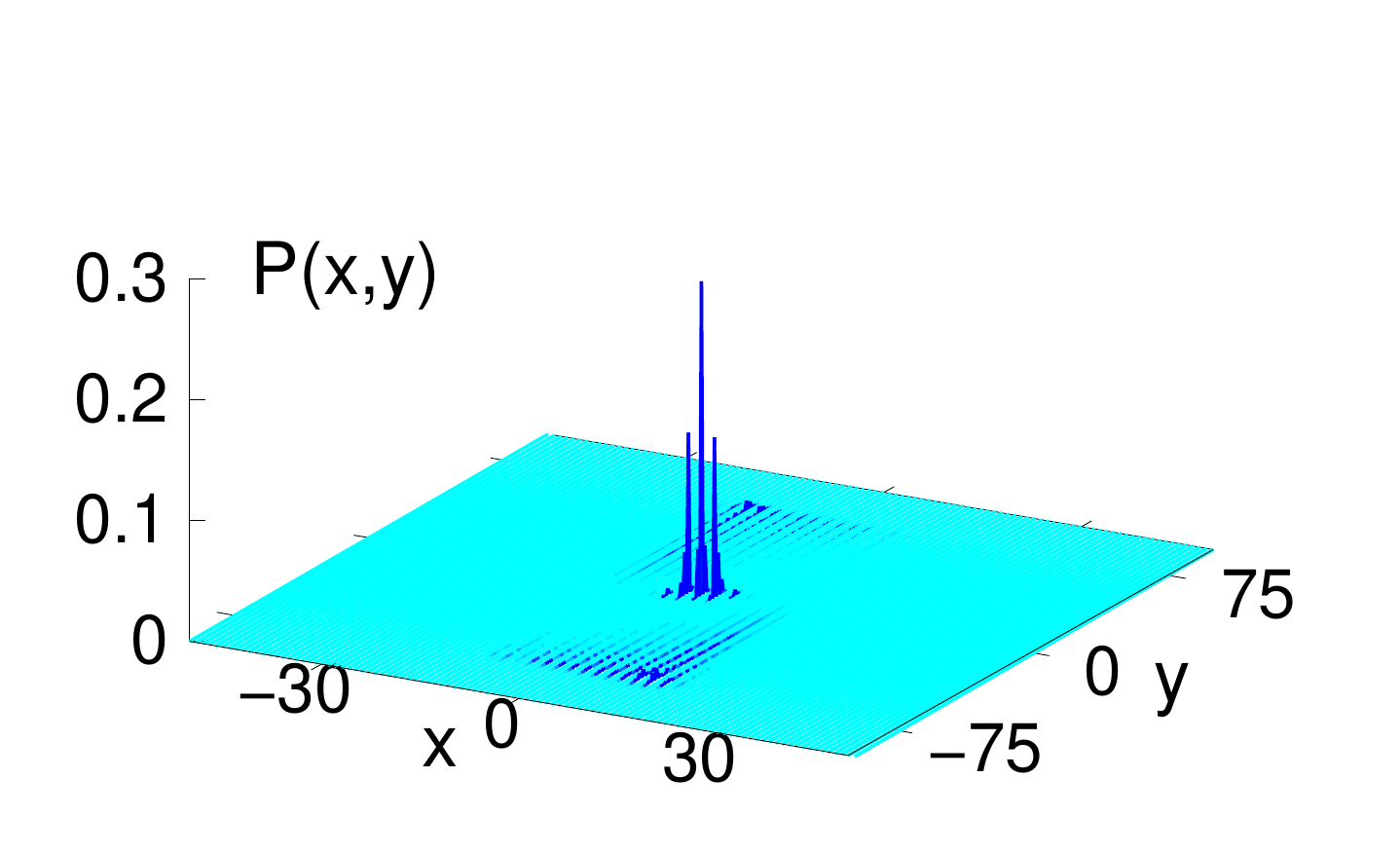}}\hspace*{-.35cm}
\subfigure[\label{fig:9e} ]{\includegraphics[scale=0.25]{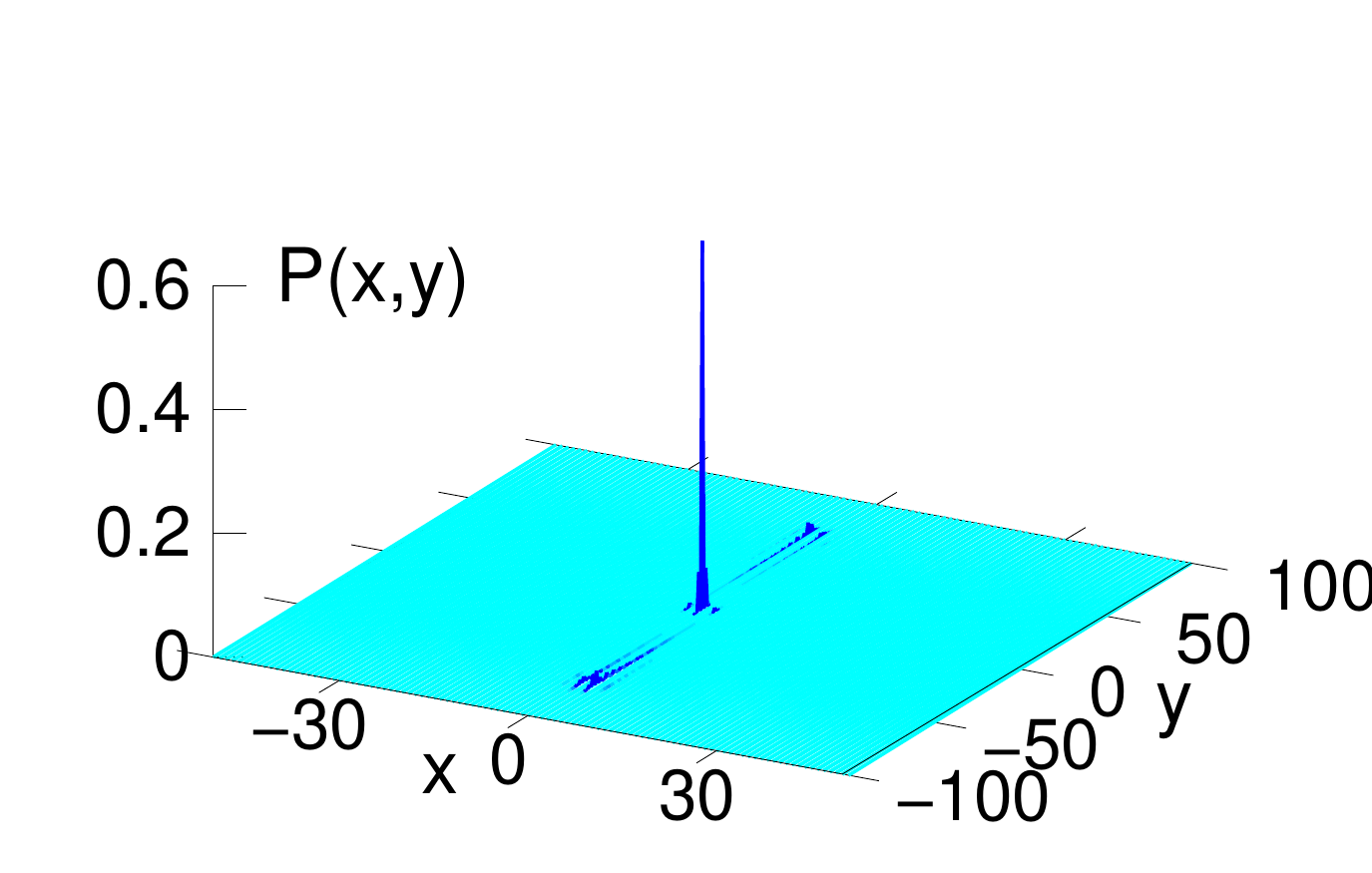}}
\subfigure[\label{fig:7f1} ]{\includegraphics[scale=0.35]{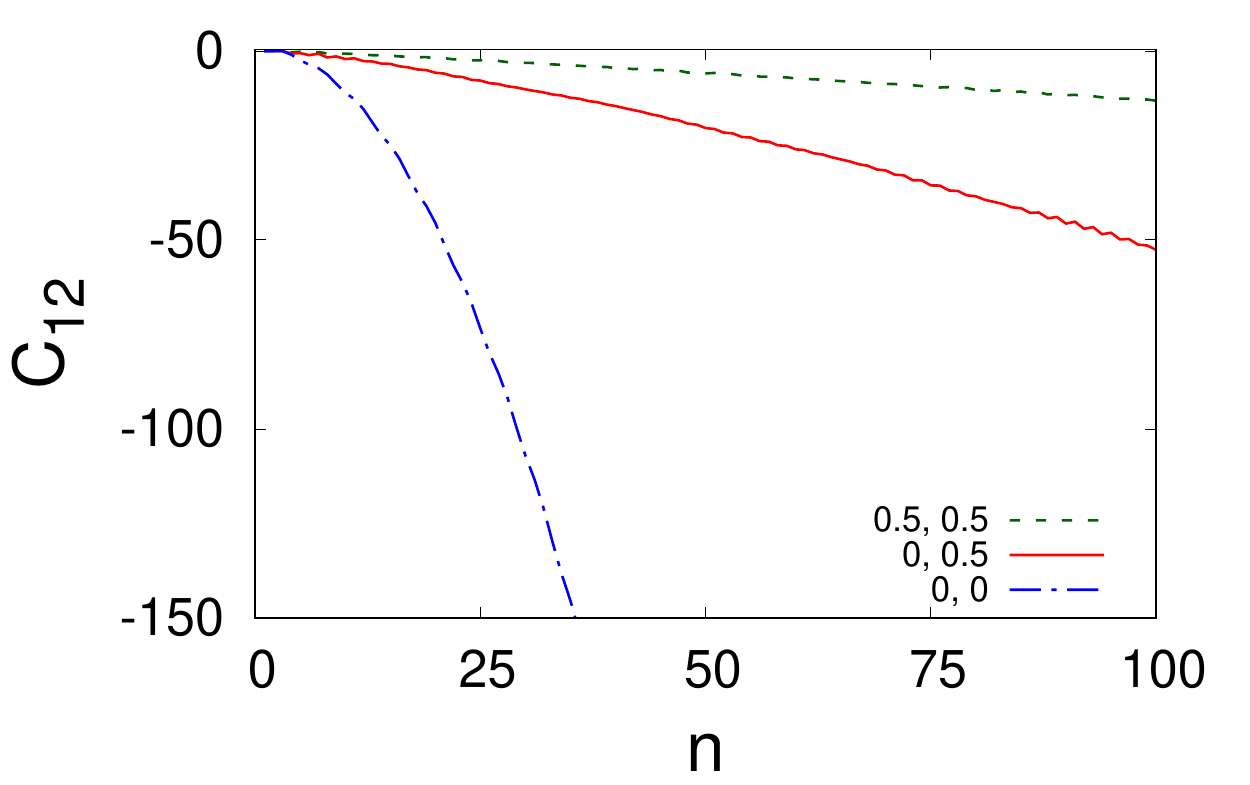}}
\subfigure[\label{fig:7g1} ]{\includegraphics[scale=0.35]{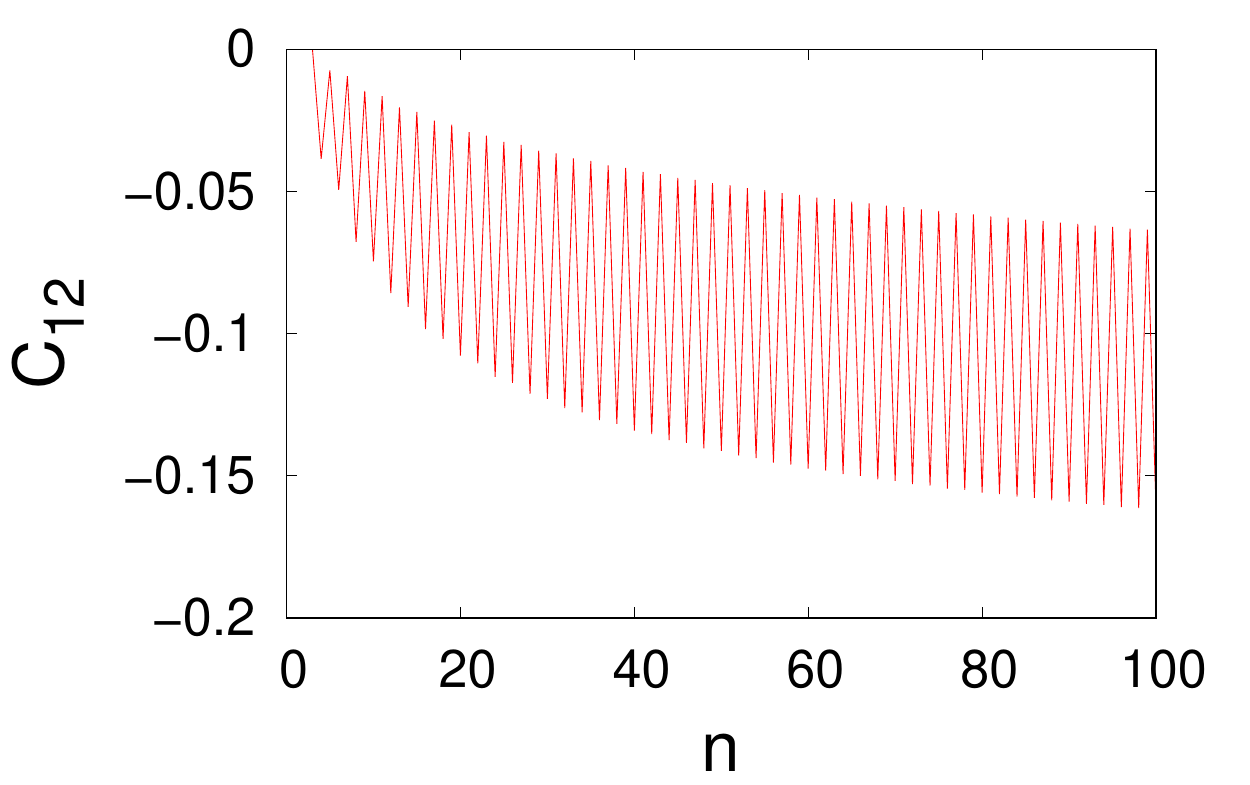}}
\subfigure[\label{fig:7h1} ]{\includegraphics[scale=0.35]{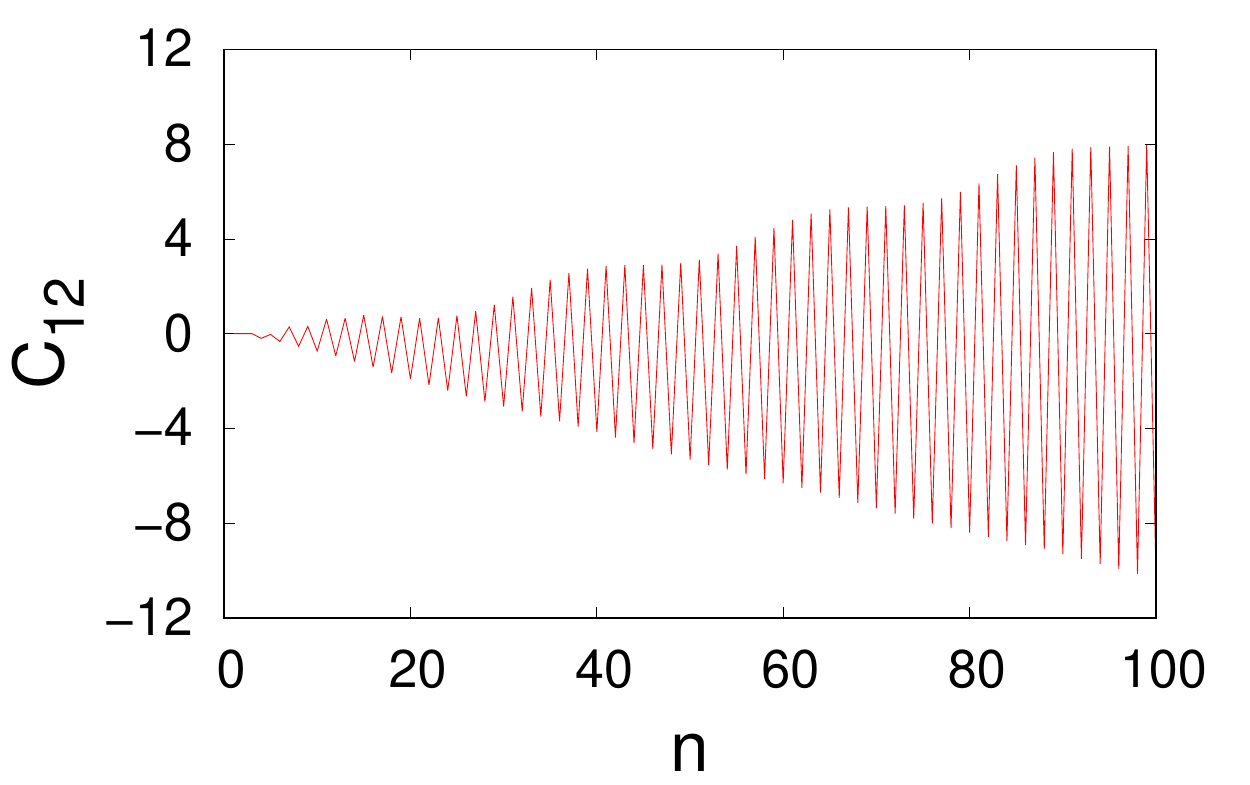}}\\
\subfigure[\label{fig:8f1} ]{\includegraphics[scale=0.35]{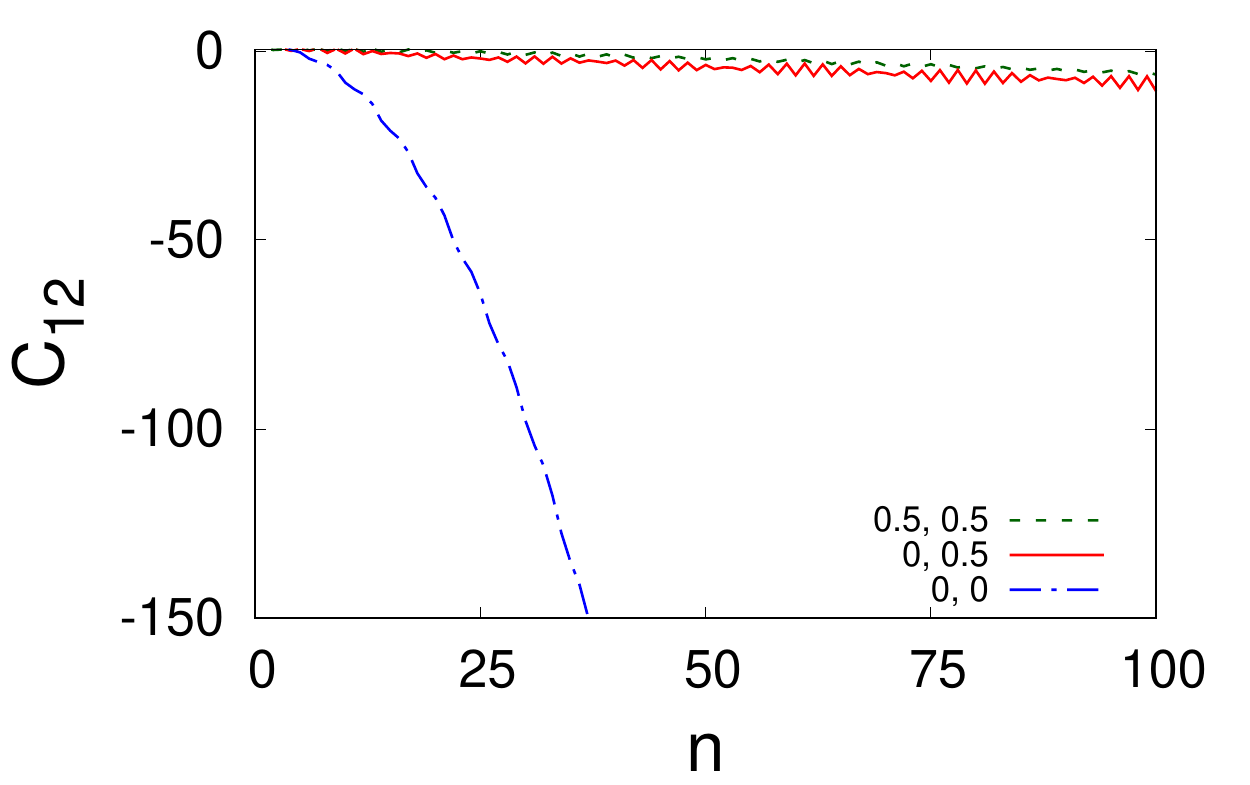}}
\subfigure[\label{fig:8g1} ]{\includegraphics[scale=0.35]{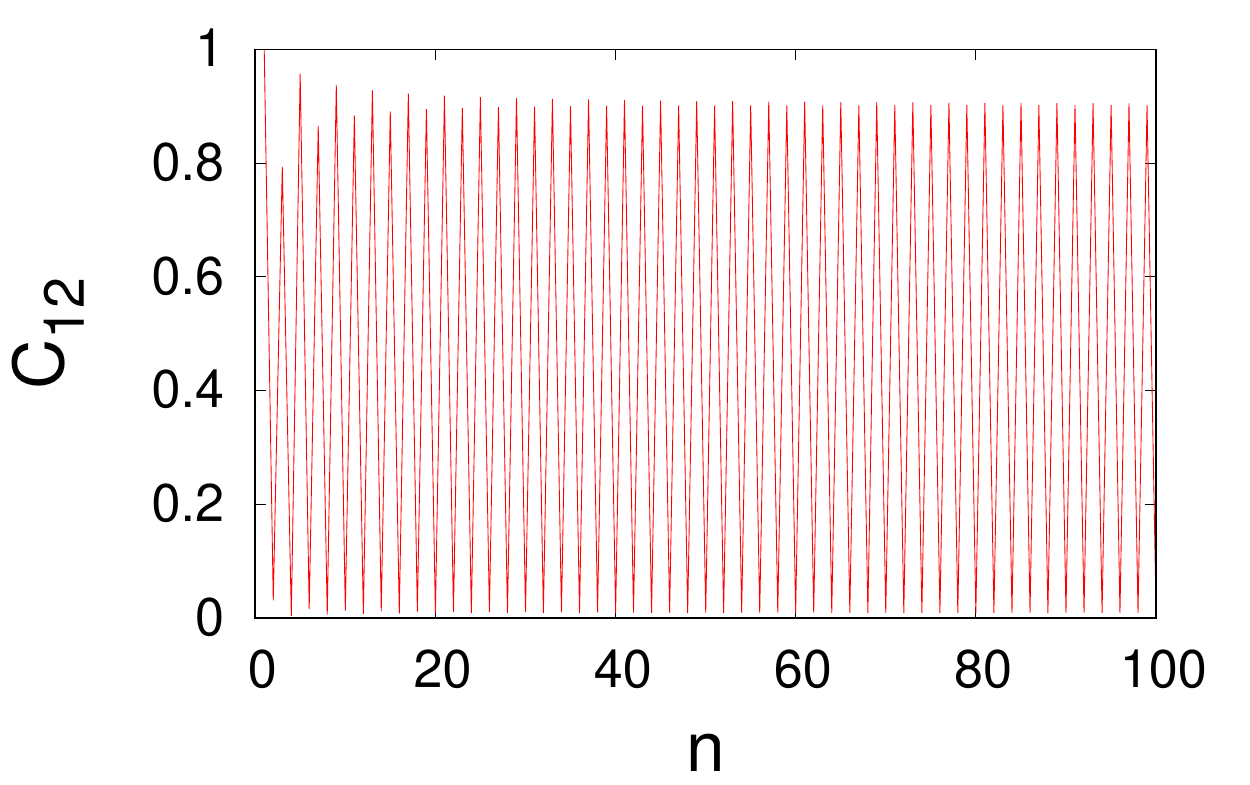}}
\subfigure[\label{fig:8h1} ]{\includegraphics[scale=0.35]{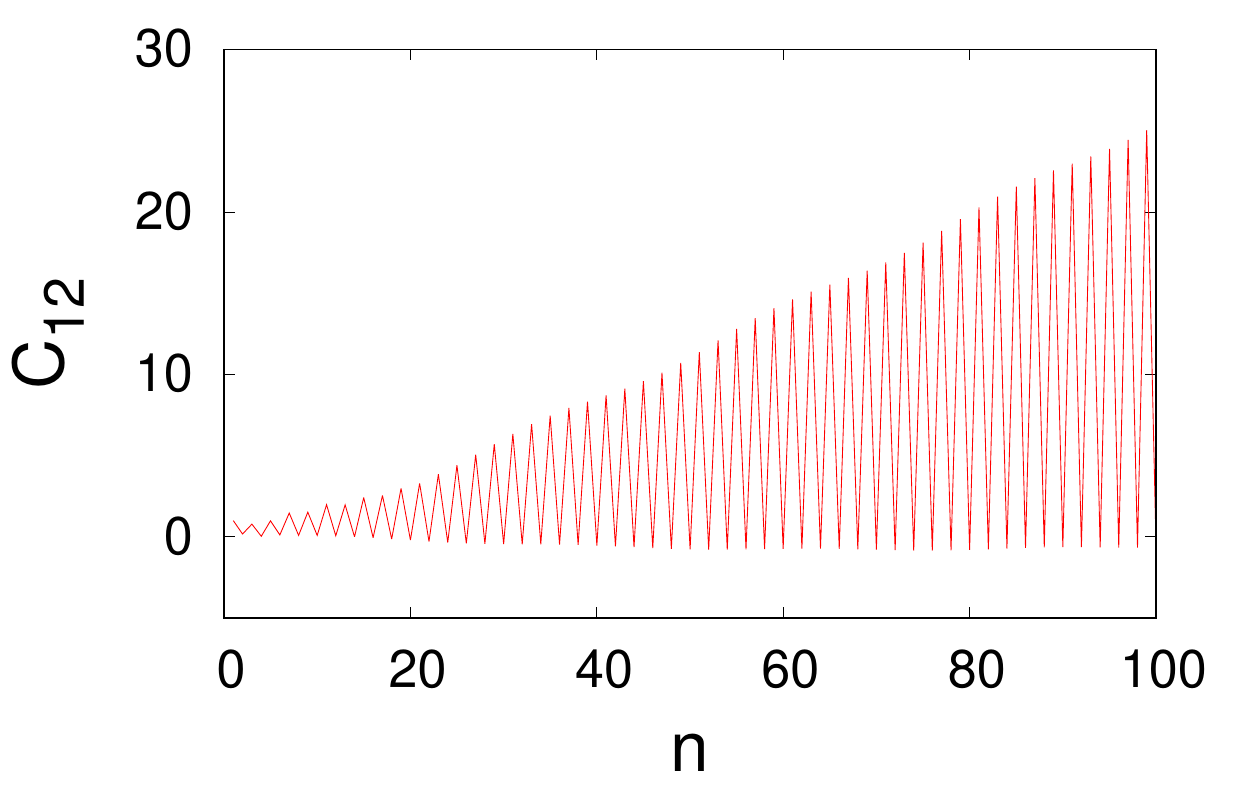}}\\
\subfigure[\label{fig:9f1} ]{\includegraphics[scale=0.35]{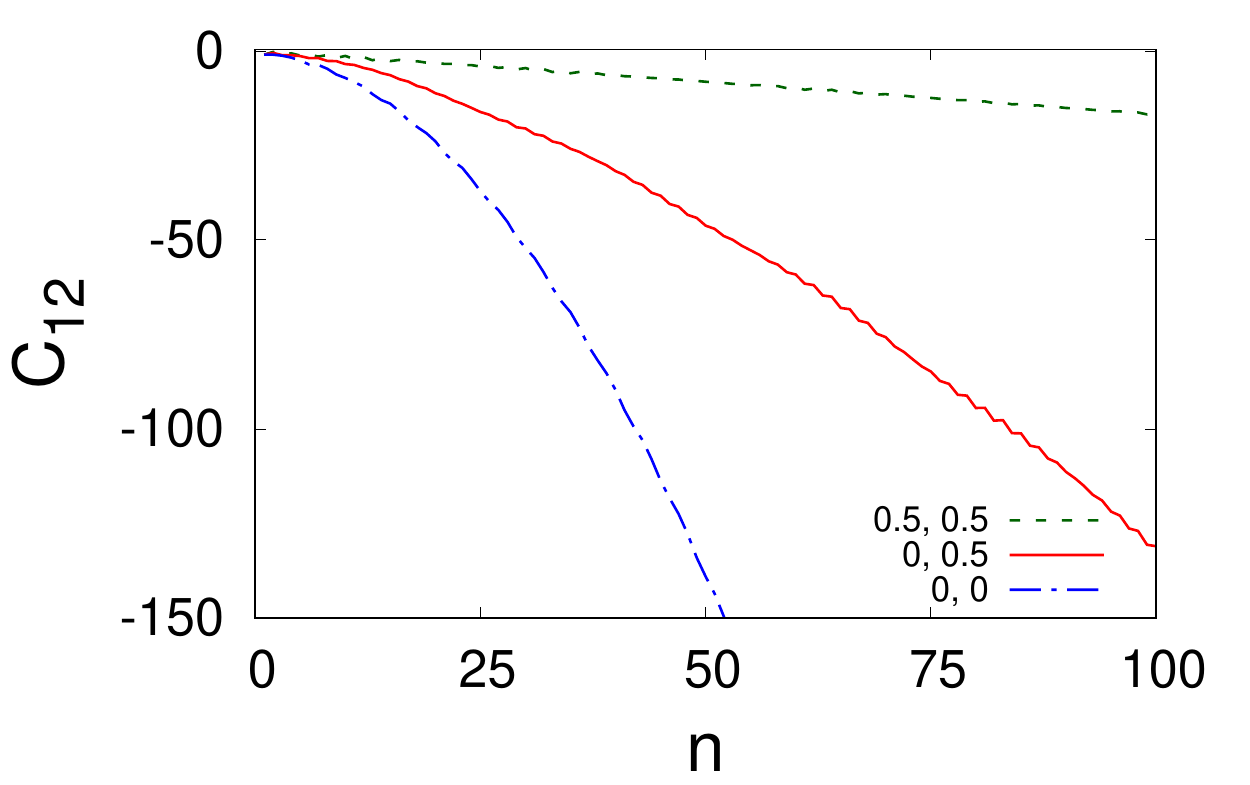}}
\subfigure[\label{fig:9g1} ]{\includegraphics[scale=0.35]{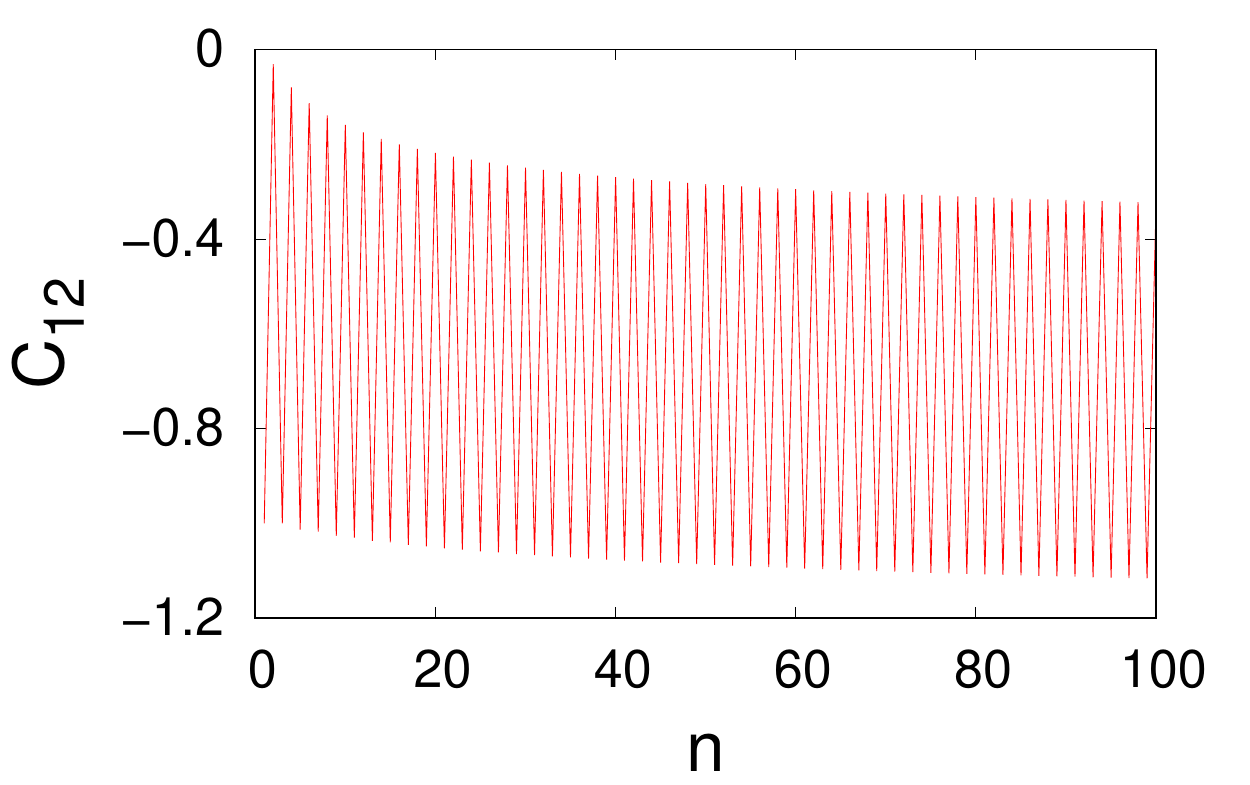}}
\subfigure[\label{fig:9h1} ]{\includegraphics[scale=0.35]{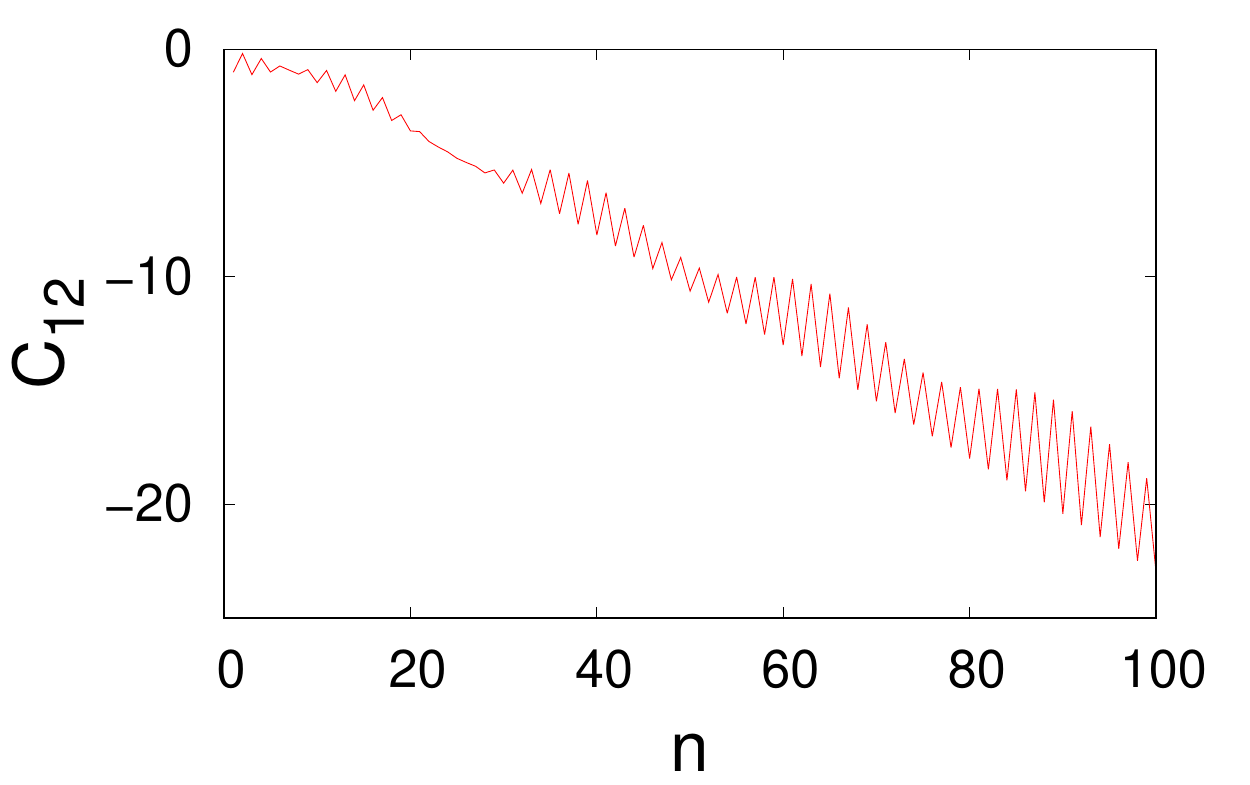}}

\caption{\label{fig:3}{ Here figures (a)-(o) show two-particle probability distributions \(P(x,y)\) after 100 time steps for two \(\pi\)-phase interacting walkers evolving under the influences of the coin \(\hat{C}_{\alpha_{1},\alpha_{2}}(t)\). \(P(x,y)\) for \(|Sep\rangle\) initial state : (a) \(\alpha_{1}\) = \(\alpha_{2}\) = 0; (b) \(\alpha_{1}\) = \(\alpha_{2}\) = 0.5; (c) \(\alpha_{1}\) = \(\alpha_{2}\) = 1.25, (d) \(\alpha_{1}\) = 0, \(\alpha_{2}\) = 0.5; (e) \(\alpha_{1}\) = 0, \(\alpha_{2}\) = 1.25. \(P(x,y)\) for \(|\psi^{}+\rangle\) initial state : (f) \(\alpha_{1}\) = \(\alpha_{2}\) = 0; (g) \(\alpha_{1}\) = \(\alpha_{2}\) = 0.5; (h) \(\alpha_{1}\) = \(\alpha_{2}\) = 1.25, (i) \(\alpha_{1}\) = 0, \(\alpha_{2}\) = 0.5, (j) \(\alpha_{1}\) = 0, \(\alpha_{2}\) = 1.25. \(P(x,y)\) for \(|\psi^{-}\rangle\) initial state : (k) \(\alpha_{1}\) = \(\alpha_{2}\) = 0; (l) \(\alpha_{1}\) = \(\alpha_{2}\) = 0.5; (m) \(\alpha_{1}\) = \(\alpha_{2}\) = 1.25, (n) \(\alpha_{1}\) = 0, \(\alpha_{2}\) = 0.5; (o) \(\alpha_{1}\) = 0, \(\alpha_{2}\) = 1.25. Figures (p)-(x) show the variations of the correlation function \(C_{12}\) against dimensionless time \(n\) for two \(\pi\)-phase interacting walkers evolving under the influences of \(\hat{C}_{\alpha_{1},\alpha_{2}}(t)\). Variations of \(C_{12}\) for \(|Sep\rangle\) initial state : (p) \(\alpha_{1}\) = \(\alpha_{2}\) = 0; \(\alpha_{1}\) = \(\alpha_{2}\) = 0.5; \(\alpha_{1}\) = 0, \(\alpha_{2}\) = 0.5 (q) \(\alpha_{1}\) = \(\alpha_{2}\) = 1.25 (r) \(\alpha_{1}\) = 0, \(\alpha_{2}\) = 1.25. Variations of \(C_{12}\) for \(|\psi^{+}\rangle\) initial state : (s) \(\alpha_{1}\) = \(\alpha_{2}\) = 0; \(\alpha_{1}\) = \(\alpha_{2}\) = 0.5; \(\alpha_{1}\) = 0, \(\alpha_{2}\) = 0.5 (t) \(\alpha_{1}\) = \(\alpha_{2}\) = 1.25 (u) \(\alpha_{1}\) = 0, \(\alpha_{2}\) = 1.25. Variations of \(C_{12}\) for \(|\psi^{-}\rangle\) initial state : (v) \(\alpha_{1}\) = \(\alpha_{2}\) = 0; \(\alpha_{1}\) = \(\alpha_{2}\) = 0.5; \(\alpha_{1}\) = 0, \(\alpha_{2}\) = 0.5 (w) \(\alpha_{1}\) = \(\alpha_{2}\) = 1.25 (x) \(\alpha_{1}\) = 0, \(\alpha_{2}\) = 1.25.
} }
 
\end{figure*}

Here we describe the results obtained for time dependent coins of kind \(\hat{C}_{\alpha}(t)\).  
In general, the coin parameter \(\alpha\) can be different for the two different coins used in two-particle QW. 
We have studied the QW evolutions for five different combinations of these two parameters : (1) \(\alpha_{1}=\alpha_{2}=0\), (2) \(\alpha_{1}=\alpha_{2}=0.50\), (3) \(\alpha_{1}=\alpha_{2}=1.25\), (4) \(\alpha_{1}=0,\alpha_{2}=0.5\) and (5) \(\alpha_{1}=0,\alpha_{2}=1.25\). The first three combinations represent cases where the walkers are driven by identical coins. The other two combinations represent cases where two different walkers are controlled by two different coins of contrasting nature, e.g., \(\alpha=0\) generates ballistic evolution whereas \(\alpha=1.25\) generates localization.\\

When \(\alpha_{1}=\alpha_{2}=0\), the coins transform to time-independent Hadamard coins. We discuss the results for time-independent coins here so that we can compare and contrast these results with all other cases studied here with time-dependent coins. The two particle probability distributions for time independent Hadamard coins were studied by Berry et al.\cite{Berry} for different initial states.  Our results for \(\alpha_{1}=\alpha_{2}=0\) agree with the results obtained by Berry et al.\cite{Berry}. The second(third) combination is used to explore the evolution of two quantum walkers which would have generated diffusive(localized) evolutions at the individual level had there been no quantum entanglement and interactions between the two walkers. \\

\subsubsection{Dynamics of two non-interacting walkers under the influence of \(\hat{C}_{\alpha_{1},\alpha_{2}}(t)\)}
 
The joint probability distribution \(P(x,y)\) for two non-interacting walkers starting from the state \(|Sep\rangle\) is simply equal to the product of two single particle probability distributions. For \(\alpha_{1}=\alpha_{2}=0\), a snapshot of the resultant distribution \(P(x,y)\) is shown in Fig.\ref{fig:1a}. Formation of multiple high peaks at four different corners of the \(xy\) plane indicates that if, upon measurement on the system, one particle is found to be placed relatively far from the origin then the other particle is likely to be found either near the first particle (bunching) or at the opposite end of the line (anti-bunching). The average separation \(\Delta_{12}\) increases gradually with time \cite{supl}. For \(\alpha_{1}=\alpha_{2}=0.5\), \(P(x,y)\) spreads much more slowly (see figure \ref{fig:1b}), as expected. \(\Delta_{12}\) also increases with time in a much slower fashion. For \(\alpha_{1}=\alpha_{2}=1.25\), the probability distribution remains localized near origin (see Fig.\ref{fig:1c}). The weak time-variation of \(\Delta_{12} \) and its saturation to a small value \( \sim 1.4\) also indicate localization of the walkers \cite{supl}. The nature of QW evolution changes when the two walkers are driven by two quantum coins of contrasting nature. For \(\alpha_{1}=0,\alpha_{2}=1.25\), \(P(x,y)\) spreads along the y-axis of the related plot shown in figure \ref{fig:1e} indicating that one of the walkers remains close to origin throughout the evolution while the other one moves away from the origin. The higher peaks are at certain non-zero values of \(y\). The width of the distribution increases in case of \(\alpha_{1}=0,\alpha_{2}=0.5\)(see Fig.\ref{fig:1d}). These results simply indicate that one can tune the value of \(\alpha_{2}\) to change the width of \(P(x,y)\) while keeping \(\alpha_{1}\) at a fixed value. On the other hand, the position of the peaks can be controlled by changing the value of \(\alpha_{1}\).\\

Two non-interacting walkers remains uncorrelated and un-entangled when they start evolving from the \(|Sep\rangle\) state. On the other hand, entangled initial states generates two-particles correlation even in the absence of pair interactions. The figures \ref{fig:2a}-\ref{fig:2e} show the joint probability distributions \(P(x,y)\) for the bosonic \(|\psi^{+}\rangle\) initial state. The probability distributions look somewhat similar to those obtained in case of \(|Sep\rangle\) initial state. This occurs due to presence of two similar terms in \(|\psi^{+}\rangle\) and \(|Sep\rangle\) states. The temporal variation of \(C_{12}\) for different combinations of the coin parameters are shown in the figures \ref{fig:2f}-\ref{fig:2g2}. When \(\alpha_{1}=\alpha_{2}=0\), \(C_{12}\) increases with time (from zero) indicating that the walkers become more and more correlated with time. \(C_{12}\) also exhibits some periodic oscillations where the oscillation amplitude increases with time (see Fig.\ref{fig:2f}). It indicates a competition between bunching and anti-bunching behaviors. However, the relative dominance of bunching behavior is quite clear here. When both the coins become time-dependent (\(\alpha_{1}=\alpha_{2}=0.5\)), the particles become anti-correlated as \(C_{12}\) remains negative and its value gradually decreases with time (see Fig.\ref{fig:2f}). So, for \(|\psi^{+}\rangle\), changing \(\alpha\) from 0 to 0.5 not only slowers the spreading of \(P(x,y)\) but also changes the character of the dynamics. For \(\alpha_{1}=\alpha_{2}=1.25\), the particles remains localized near the origin (see Fig.\ref{fig:2c}) as expected but there are distinct periodic oscillations of \(C_{12}\)(see Fig.\ref{fig:2g1}). The amplitude of oscillation is although small. For \(\alpha_{1}=0,\alpha_{2}=1.25\), \(C_{12}\) exhibit oscillations and the oscillation amplitude increases with time (see Fig.\ref{fig:2g2}). This occurs as one of the particle is moving away from the origin. So, the amplitude of such oscillation can be controlled by varying the parameter \(\alpha_{1}\). The particles become periodically correlated and anti-correlated with time. For \(\alpha_{1}=0,\alpha_{2}=0.5\), the particles remain anti-correlated. There are packets of oscillation in the temporal variation of \(C_{12}\) (see Fig\ref{fig:2f}).

For \(|\psi^{-}\rangle\) initial state, the evolution is completely different from that found in cases of \(|Sep\rangle\) and \(|\psi^{+}\rangle\). The particles remain anti-correlated and anti-bunching behavior dominates. When \(\alpha_{1}=\alpha_{2}=0\), the particles exhibit nearly pure anti-bunching behavior as shown in Fig. \ref{fig:3a}. It implies that upon measurement on the system, if one particle is found to be placed near one end of the line then the other particle is likely to be found at the opposite end of the line. The correlation function rapidly decays with time as shown in Fig. \ref{fig:3f}. For \(\alpha_{1}=\alpha_{2}=0.5\), \(P(x,y)\) spreads in a slower fashion but the presence of anti-bunching behavior can still be seen in the related plot of \(P(x,y)\) (see Fig.\ref{fig:3b}). \(C_{12}\) also decrease in a slower fashion. For \(\alpha_{1}=\alpha_{2}=1.25\), the probability distribution remains localized near origin (see Fig.\ref{fig:3c}). The particles remain anti-correlated but \(C_{12}\) also shows oscillatory behavior indicating competition between bunching and anti-bunching (see Fig.\ref{fig:3g1}). When \(\alpha_{1}=0, \alpha_{2}=1.25\) the probability distribution looks quite similar to that obtained with \(|\psi^{+}\rangle\) (see Fig.\ref{fig:3e}). However, the correlation function rapidly decreases with time (see Fig.\ref{fig:3g2}). For \(\alpha_{1}=0,\alpha_{2}=0.5\), the plots of \(P(x,y)\) indicate that upon measurement, it is highly probable that both the particles are to be found on two opposite sides of the origin, one near and the other far from the origin. So, the particles remain strongly anti-correlated (see Fig.\ref{fig:3d}).\\
It is interesting to find that the system can generate localization behavior of significantly different character depending on the nature of the initial states in case of \(\alpha_{1}=\alpha_{2}=1.25\), even in the absence of interactions.\\
\subsubsection{Dynamics of two \(\mathbb{1}\) interacting walkers under the influence of \(\hat{C}_{\alpha_{1},\alpha_{2}}(t)\)}
 
In the presence of interactions, the particles become entangled \cite{supl}. Let us first describe the results obtained for \(|Sep\rangle\) initial state. In this interacting walk, the identity operator acts on the \(|0, \uparrow ; 0, \uparrow \rangle\) and \(|0, \downarrow ; 0, \downarrow \rangle\) terms of the \(|Sep\rangle\) state. Since both particles are in the same position and coin states and the identity operator does not mix the coin states of each particle, they are translated together and move in the same direction at each step of the walk. The related probability distribution \(P(x,y)\) and time variation of \(C_{12}\) are shown in the Figures \ref{fig:4a} and \ref{fig:4f} respectively. \(C_{12}\) rapidly increases with time. Since this strong bunching behavior is independent of the coin parameters, all the curves for different parameter sets nearly overlap each-other. The contribution of the other two terms present in \(|Sep\rangle\) is relatively weaker. The influence of the other terms is visible in Fig. \ref{fig:4f}, as the curve for \(\alpha_{1}=\alpha_{2}=0\) does not exactly overlap the other curves. The temporal variations of \(\Delta_{12}\) and \(E{|\psi}\rangle\) are quite similar to that observed in the previously studied case of two non-interacting walkers starting from \(|Sep\rangle\) initial state \cite{supl}. This is also due to the contribution of the other two terms present in \(|Sep\rangle\). The entanglement entropy saturates to values \(\sim 2\) for different coin parameter combinations \cite{supl}. \\ 

When the walker pairs start from bosonic \(|\psi^{+}\rangle\) initial state with \(\alpha_{1}=\alpha_{2}=0\), the particles exhibit fermionic anti-bunching behavior as is clearly shown in Fig. \ref{fig:5a}. This is quite interesting as fermionic anti-bunching is obtained from bosonic initial state. The correlation function \(C_{12}\) rapidly decays with time indicating that the walkers become more and more anti-correlated with time (see Fig.\ref{fig:5f}). \(P(x,y)\) spreads much more slowly as the values of the coin parameters are increased to \(\alpha_{1}=\alpha_{2}=0.50\). The probability distributions for \(\alpha_{1}=\alpha_{2}=0.50\) and \(\alpha_{1}=\alpha_{2}=1.25\) are shown in the figures \ref{fig:5b} and \ref{fig:5c} respectively. The walkers exhibit anti-correlated evolution in all the three cases. However, the rate of decay of \(C_{12}\) decreases as the values of the coin parameters are increased (see Fig.\ref{fig:5f}, \ref{fig:5g}). For \(\alpha_{1}=0, \alpha_{2}\neq 0\), the probability distributions are found to exhibit behaviors qualitatively similar to the previously studied cases (see figures \ref{fig:5d} and \ref{fig:5e}). A difference is that the probability of finding both the particles together at the origin is smaller than that found in the previously studied cases with \(\alpha_{1}=0, \alpha_{2}=1.25\).\\
The scenario becomes more interesting when the walker pairs start from \(|\psi^{-}\rangle\) initial state with \(\alpha_{1}=\alpha_{2}=0\). The probability distribution \(P(x,y)\) consists of two parts. One part spreads along one diagonal of the \(xy\) plane in Fig. \ref{fig:6a} indicating the presence of bunching behavior and the other part corresponds to anti-bunching. So, there are finite probabilities that upon measurement, two particles can either be found bunched together or they can also be found far apart from each-other situated on the opposite ends of the line. The correlation function \(C_{12}\) rapidly decays with time indicating that the walkers remain anticorrelated. This is because the anti-bunching peaks are more distant from the origin in comparison to the bunching peaks. As the values of \(\alpha_{1},\alpha_{2}\) are simultaneously increased from \(0\) to \(0.5\), the dynamics becomes very slow. The average separation reaches a value \(\sim 4.5\) after 100 steps. Fig. \ref{fig:6b} shows that a significant part of \(P(x,y)\) is localized near origin. Both \(P(x,y)\) and \(C_{12}\) evolves quite slowly with time (see Figs.\ref{fig:6b} and \ref{fig:6f}). The system exhibits dynamics which is much more slower than what is naively expected in case of two such walkers. For \(\alpha_{1}=\alpha_{2}=1.25\), although the probability distribution remains localized near origin(see Fig.\ref{fig:6c}), the dynamical behavior is quite different from our expectations. Here both the correlation function and the average separation \(\Delta_{12}\) exhibits periodic oscillatory behavior (see Fig.\ref{fig:6g}) \cite{supl}. This is a ``dynamical" kind of localization. When \(\alpha_{1}=0, \alpha_{2}= 1.25\), the probability distribution shown in Fig. \ref{fig:6e} indicates that it is highly probable that upon measurement both the particles would be found at origin. Therefore, for \(\alpha_{2}=1.25\), the system exhibits very slow evolution even if \(\alpha_{1}=0\) (see Fig.\ref{fig:6h}). If one walker has slow dynamics then the other walker also evolves slowly. Even for \(\alpha_{1}=0, \alpha_{2}= 0.5\), \(P(x,y)\) has peaks near origin(see Fig.\ref{fig:6d}). The particles become more and more anticorrelated with time but in a much more slower fashion in comparison to the previous cases (except the non-interacting case with \(|\psi^{+}\rangle\))with \(\alpha_{1}=0, \alpha_{2}= 0.5\) (see Fig.\ref{fig:6f}).
\\ 
One interesting observation is that for \(|\psi^{-}\rangle\) initial state, two interacting walkers will exhibit very slow dynamics whenever one of the coin parameters is \(\geq0.5\).\\
\subsubsection{Dynamics of two \(\pi\)-phase interacting walkers under the influence of \(\hat{C}_{\alpha_{1},\alpha_{2}}(t)\)}
Let us now describe the results for \(\pi\)-phase interaction. When the walkers start from the \(|Sep\rangle\) initial state with \(\alpha_{1}=\alpha_{2}=0\), the probability distribution \(P(x,y)\) exhibits a diagonally spreading part alongwith an anti-bunching part (see Fig.\ref{fig:7a}), similar to the previously described case of \(\mathbb{1}\)-interaction and \(|\psi^{-}\rangle\) state. However, the plot of \(P(x,y)\) in Fig.\ref{fig:7a} also shows that the contribution to the anti-bunching part is relatively stronger here. As a result, the correlation function \(C_{12}\) decays relatively faster (see Fig.\ref{fig:7f1}) 
For \(\alpha_{1}=\alpha_{2}=0.5\), a major part of \(P(x,y)\) remains localized near origin (see Fig.\ref{fig:7b}). The other part generates anti-bunching. For \(\alpha_{1}=\alpha_{2}=1.25\), the system becomes localized near origin (see Fig. \ref{fig:7c})  and the particles remain anticorrelated (see Fig. \ref{fig:7g1}). For \(\alpha_{1}=0,\alpha_{2}\ne 0\), the situation becomes more interesting as the probability distribution \(P(x,y)\) is quite different from the previously described cases. For \(\alpha_{1}=0,\alpha_{2}=1.25\), \(P(x,y)\) has three sharp peaks in three different regions of the plot(see Fig.\ref{fig:7e}) : origin  and two other points(quite distant from origin) on the y-axis. This indicates that upon measurement, three cases are most likely to happen : Firstly, both walkers can be found at origin. Secondly, one walker can be found at origin and the other one can be found positioned at a distant point on the positive side of the origin. Thirdly, one walker can be found at origin and the other one can be found positioned at a distant point on the negative side of the origin. The walkers become periodically correlated and anticorrelated with a period of one time step as shown in Fig. \ref{fig:7h1}. The oscillation amplitude increases with time. For \(\alpha_{1}=0,\alpha_{2}=0.5\), the probability of last two phenomenon decreases as can be seen from the flattening of the related two sharp peaks along \(X\) axis. The corresponding temporal variation of \(C_{12}\) is shown in Fig. \ref{fig:7f1}. \\

For \(|\psi^{+}\rangle\) initial state, the plots of the probability distributions \(P(x,y)\) are shown in Figs. \ref{fig:8a}-\ref{fig:8e}. For \(\alpha_{1}=0,\alpha_{2}=0\), the dynamics remains qualitatively similar to that obtained with \(|Sep\rangle\). As the values of \(\alpha_{1},\alpha_{2}\) are simultaneously increased from \(0\) to \(0.5\), the dynamics becomes very slow. The average separation reaches a value \(\sim 2.7\) after 100 steps \cite{supl}. Both \(P(x,y)\) and \(C_{12}\) evolves quite slowly with time (see Fig.\ref{fig:8b} and Fig.\ref{fig:8f1}). For \(\alpha_{1}=\alpha_{2}=1.25\), the probability distribution remains localized as expected (\(\Delta_{12}\)  \(\sim0.2\) after 100 steps, which is the smallest amongst all the studied cases with \(\hat{C}_{\alpha_{1},\alpha_{2}}(t)\) . The average separation performs oscillations of a quite small amplitude (\(\sim 0.1\)) \cite{supl}. The correlation function exhibits periodic oscillatory behavior (see Fig. \ref{fig:8g1}). The particles exhibit correlated evolution. In this way, this case is quite different from most of the ``localized" cases (i.e., cases with \(\alpha_{1}=\alpha_{2}=1.25\)) studied here where particles have been found to exhibit anti-correlated evolution. When \(\alpha_{1}=0, \alpha_{2}= 1.25\), the probability distribution \(P(x,y)\) again has three sharp peaks (see Fig.\ref{fig:8e}). The difference with the \(|Sep\rangle\) state is that here the height of the peak at origin is slightly smaller and the heights of the other two peaks are slightly higher than that obtained in case of \(|Sep\rangle\) state. The corresponding correlation function has an interesting temporal evolution (see Fig.\ref{fig:8h1}). Although, the particles mostly remain correlated during the evolution, \(C_{12}\) exhibits oscillations where the maximal value of correlation gradually increases with time. For \(\alpha_{1}=0, \alpha_{2}= 0.5\), the dynamics is quite slow (see Fig.\ref{fig:8f1}). It is interesting to note that for both \(\alpha_{1}=0.5, \alpha_{2}= 0.5\) and \(\alpha_{1}=0, \alpha_{2}= 0.5\), \(C_{12}\) decays in a qualitatively similar way although the parameter combinations are different from each other.\\

The results for \(|\psi^{-}\rangle\) initial state are quite similar to that obtained in the previously described case of two \(\mathbb{1}\)-interacting walkers starting from \(|\psi^{-}\rangle\) initial state. The two particles probability distributions are shown in the figures \ref{fig:9a}-\ref{fig:9e} and the temporal variations of the correlation functions are shown in the figures \ref{fig:9f1}-\ref{fig:9h1}.\\

\subsubsection{Overall diversity in the \(\hat{C}_{\alpha_{1},\alpha_{2}}(t)\) driven dynamics }
The time-dependent coin \(\hat{C}_{\alpha_{1},\alpha_{2}}(t)\) has generated rich variety of dynamical behavior. Here we briefly summarize the overall diversity in the dynamics for some sets of parameters.\\ 

For \(\alpha_{1}=\alpha_{2}=1.25\), although localization has been found for all different initial states and interactions, the character of localization for different initial states and interactions has been quite different. In some cases, we have observed correlated evolution whereas in some other cases anti-correlated evolution has been found. A ``dynamical'' kind of localization, where both the average separation and the correlation function exhibit periodic oscillations, has also been found in a particular case. 
On the other hand, we have also found a particular case of localization where the average separation is quite small(\(\sim0.2\)) in comparison to its typical values(\(\sim 1.5\)) \cite{supl}.\\

For \(\alpha_{1}=0,\alpha_{2}=1.25\), we have observed that depending on the initial states and interactions, the system can exhibit any one of the following three scenarios : (1) both the particles can be found simultaneously localized near origin, (2) only one of them can be found to be localized near origin, and (3) both of them can be found at positions quite distant from the origin.\\ 

For \(\alpha_{1}=\alpha_{2}=0.5\), both the decay rate of \(C_{12}\) and growth rate of \(\Delta_{12}\) have been found to change significantly for different initial states and interactions. For two \(\pi\)-phase interacting walkers starting from  \(|\psi^{+}\rangle\) state, both the rates attain their smallest values whereas for two \(\mathbb{1}\) interacting walkers starting from \(|\psi^{+}\rangle\) state, both the rates attain their highest values. On the contrary, such drastic change has not been found under similar conditions for the time independent coins (\(\alpha_{1}=\alpha_{2}=0\)).\\ 


\begin{figure}[th]
\centering

\includegraphics[scale=0.70]{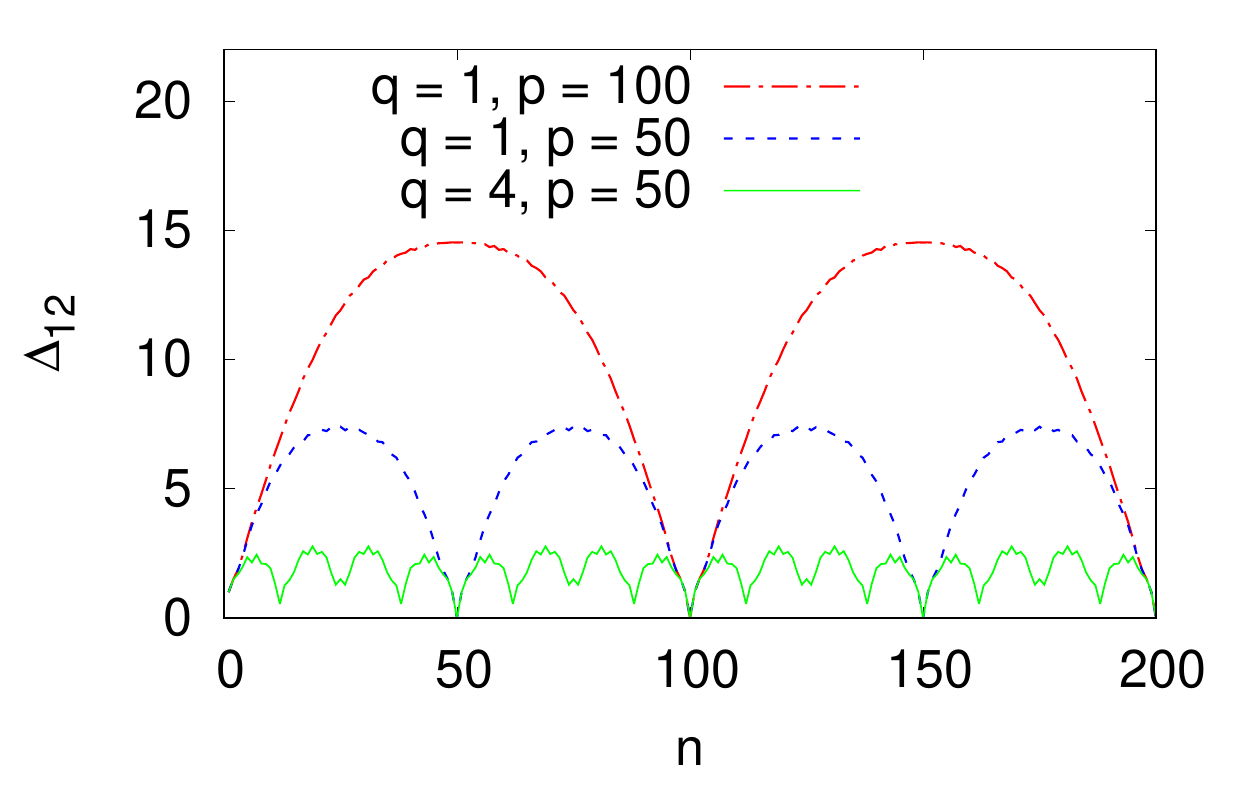}

\caption{\label{fig:10a}{The variations of the instantaneous average separation \(\Delta_{12}\) against dimensionless time \(n\) in case of two non interacting walkers starting from the \(|Sep\rangle\) state under influence of the coin  \(\hat{C}_{\phi}(t)\).  The considered values of the coin parameters \(q\) and \(p\) for different curves are described inside the figure
 }}
\end{figure}

\begin{figure*}[th]
\centering

\hspace*{-.15cm}\subfigure[\label{fig:11a} ]{\includegraphics[scale=0.37]{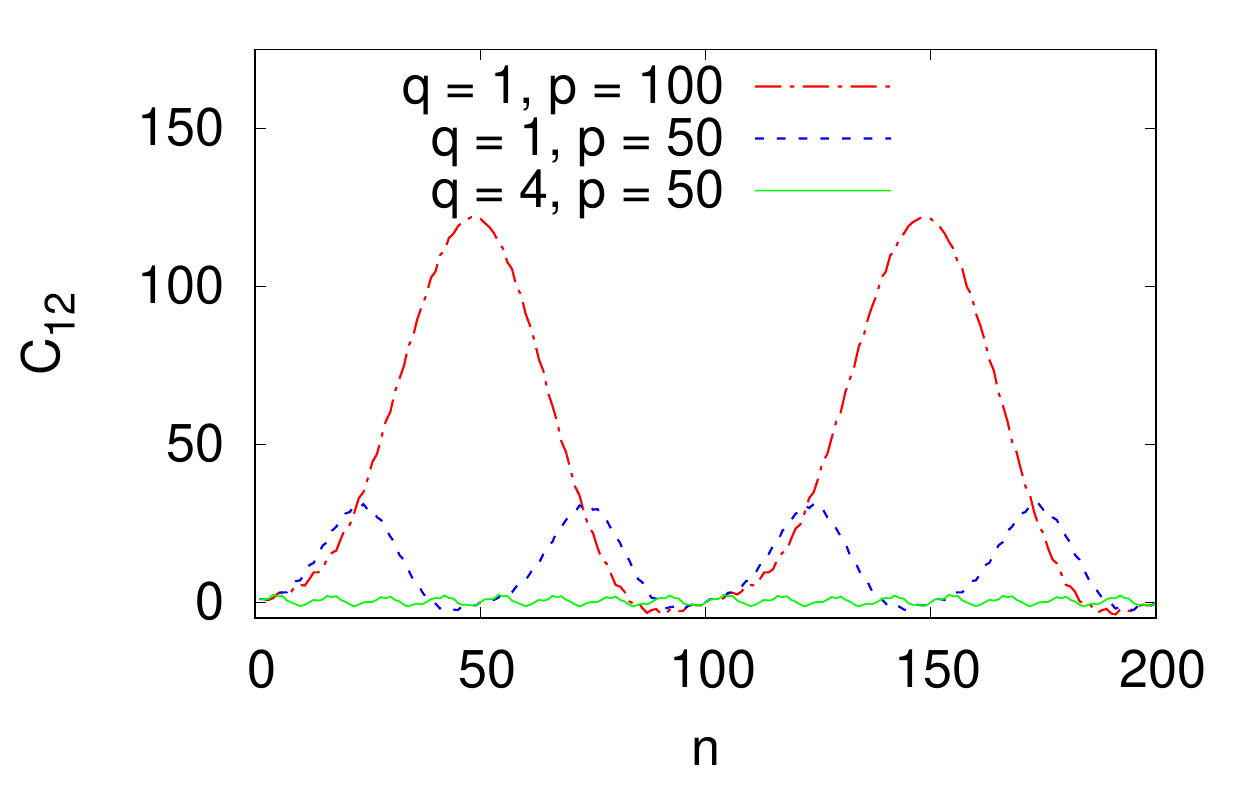}}\hspace*{-.15cm}
\subfigure[\label{fig:11b} ]{\includegraphics[scale=0.37]{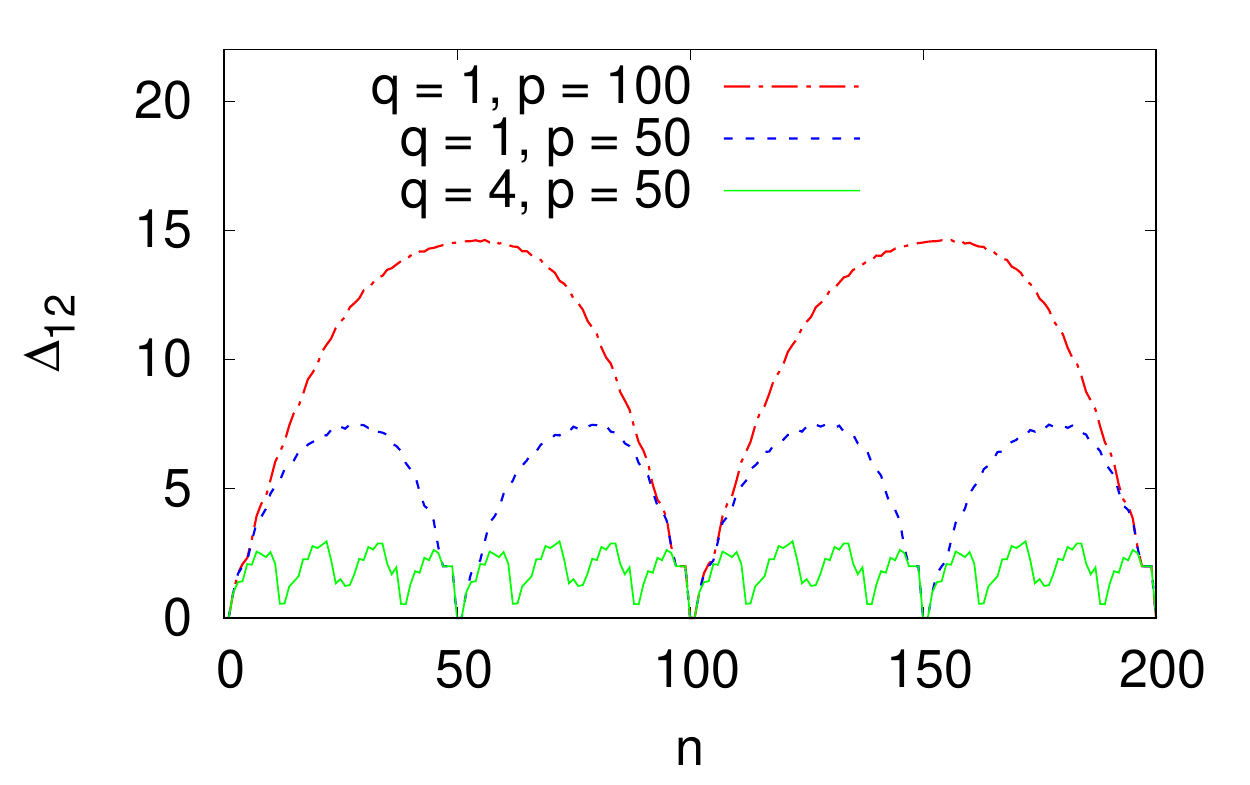}}\hspace*{-.15cm}
\subfigure[\label{fig:12a} ]{\includegraphics[scale=0.37]{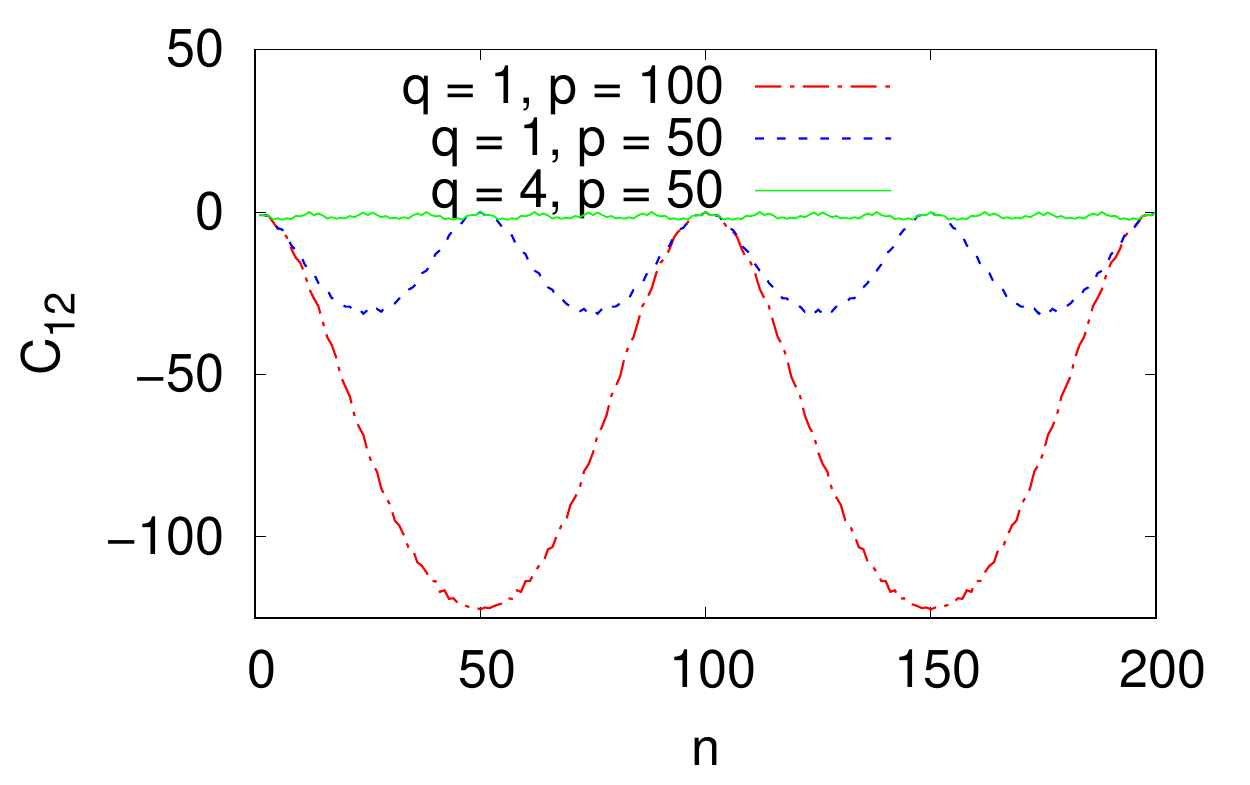}}\hspace*{-.15cm}
\subfigure[\label{fig:12b} ]{\includegraphics[scale=0.37]{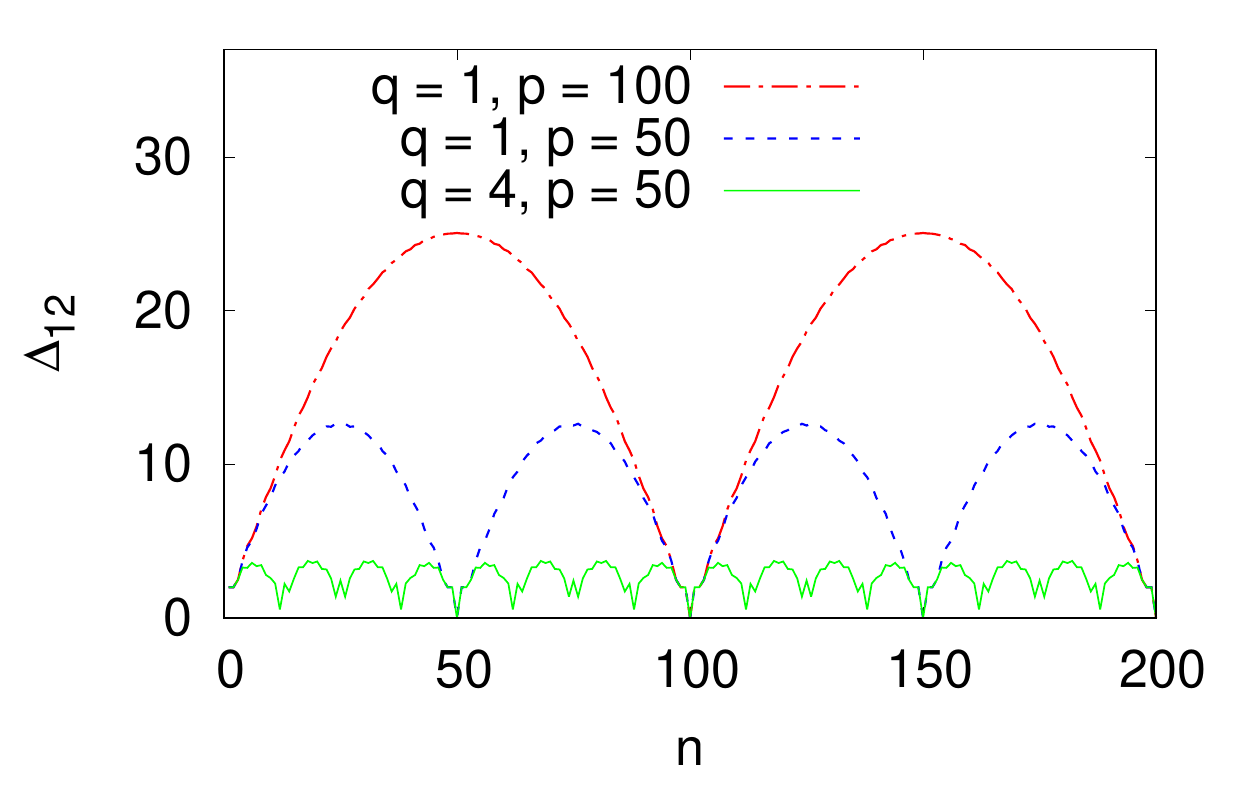}}\\
\hspace*{-.15cm}\subfigure[\label{fig:11c} ]{\includegraphics[scale=0.37]{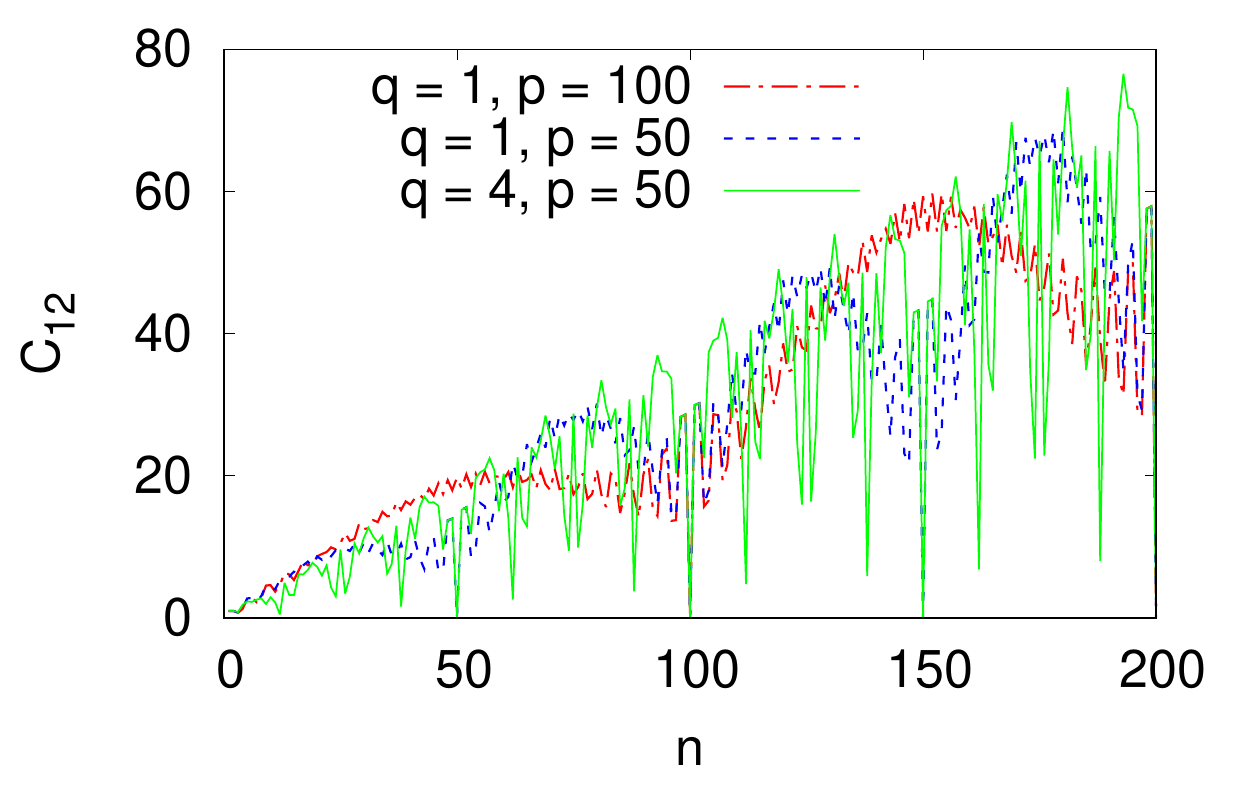}}\hspace*{-.15cm}
\subfigure[\label{fig:11d} ]{\includegraphics[scale=0.37]{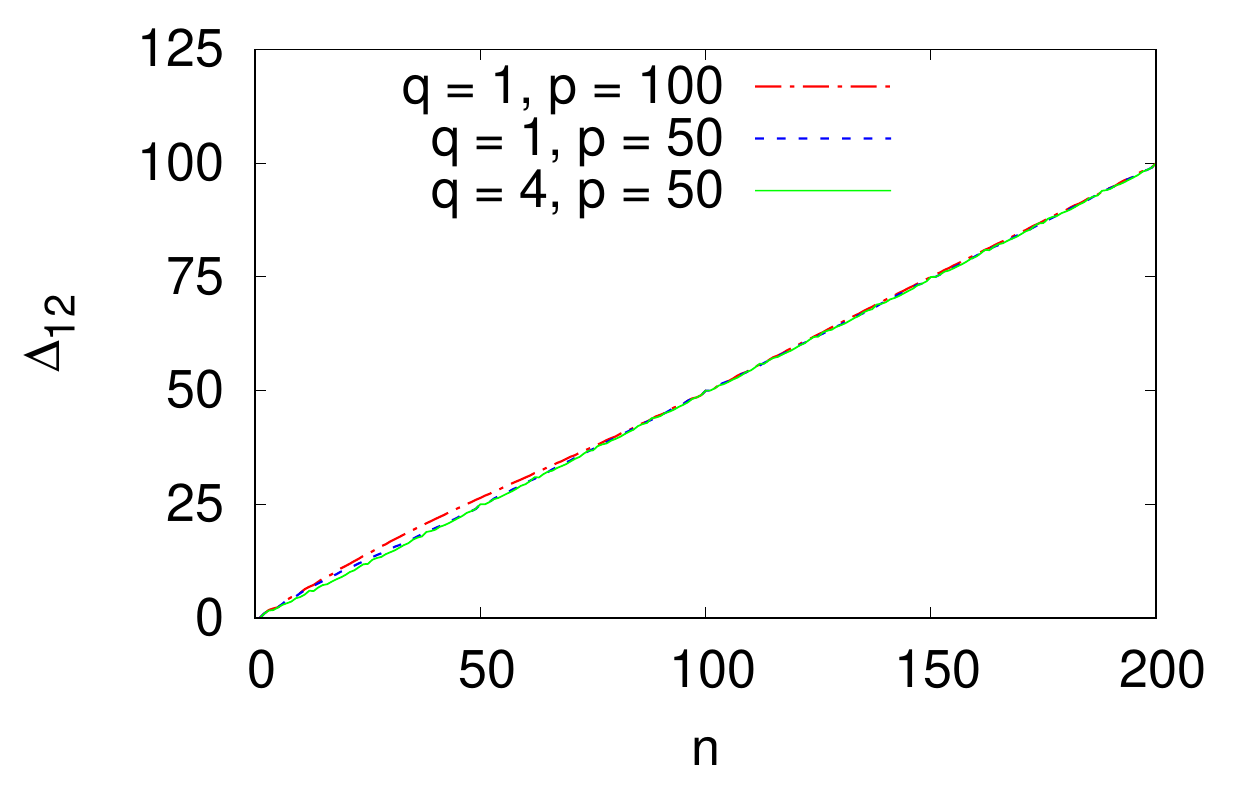}}\hspace*{-.15cm}
\subfigure[\label{fig:12c} ]{\includegraphics[scale=0.37]{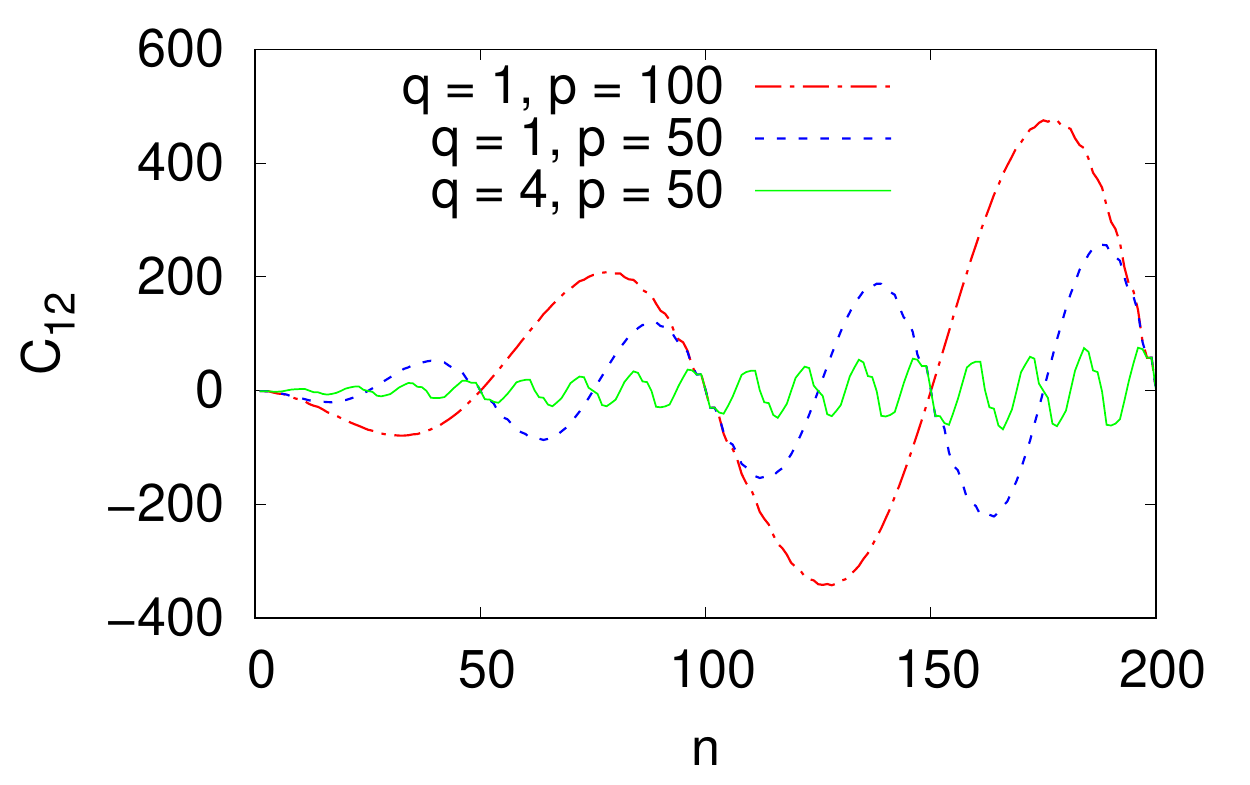}}\hspace*{-.15cm}
\subfigure[\label{fig:12d} ]{\includegraphics[scale=0.37]{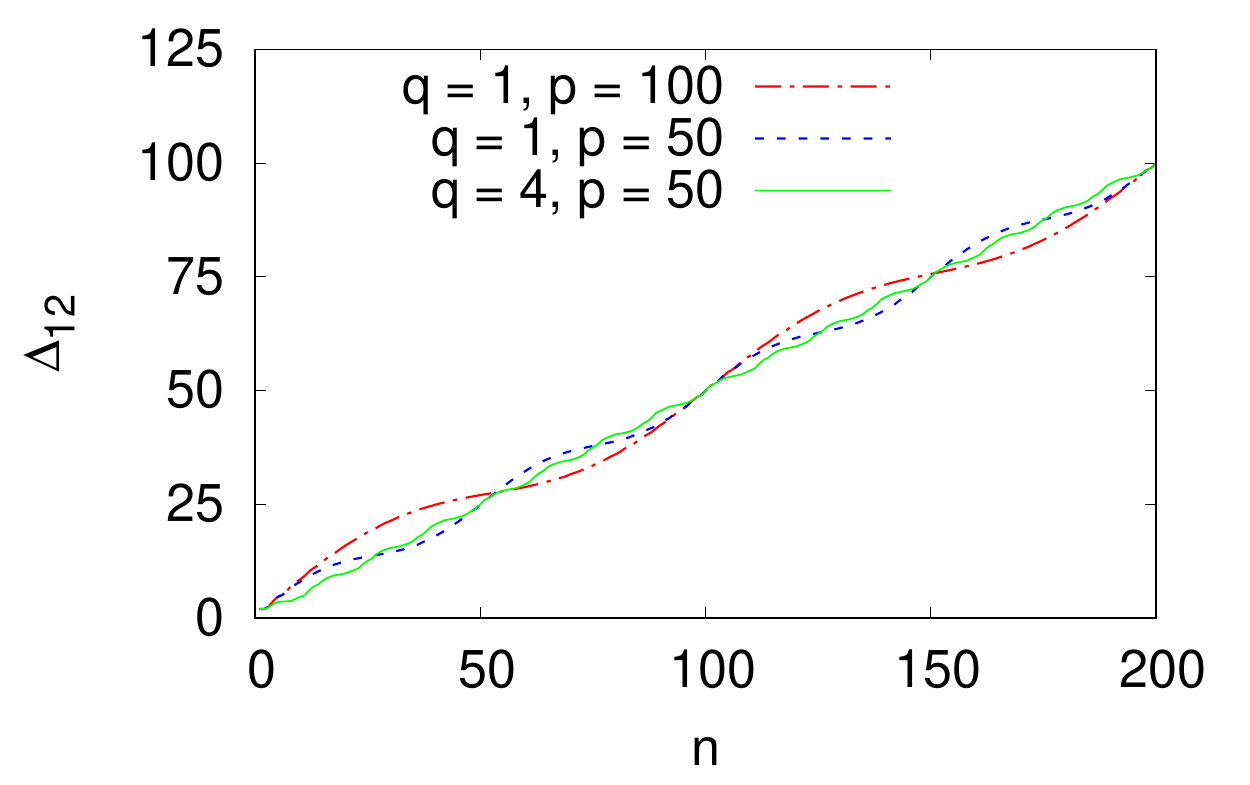}}

\caption{\label{fig:11}{The time-variations of the joint properties for two non-interacting quantum walkers with same coins : (a) Variations of  \(C_{12}\) in case of \(|\psi^{+}\rangle\); (b) Variations of \(\Delta_{12}\) in case of \(|\psi^{+}\rangle\), (c) Variations of \(C_{12}\) in case of \(|\psi^{-}\rangle\); (d) Variations of \(\Delta_{12}\) in case of \(|\psi^{-}\rangle\). The time-variations of the joint properties for two non-interacting quantum walkers with two different coins : (e) Variations of \(C_{12}\) in case of \(|\psi^{+}\rangle\); (f) Variations of \(\Delta_{12}\) in case of \(|\psi^{+}\rangle\), (g) Variations of \(C_{12}\) in case of \(|\psi^{-}\rangle\); (h) Variations of \(\Delta_{12}\) in case of \(|\psi^{-}\rangle\). The coin parameters have been mentioned inside the plots.}
 }
\end{figure*}

\begin{figure*}[th]
\centering

\subfigure[\label{fig:13a} ]{\includegraphics[scale=0.45]{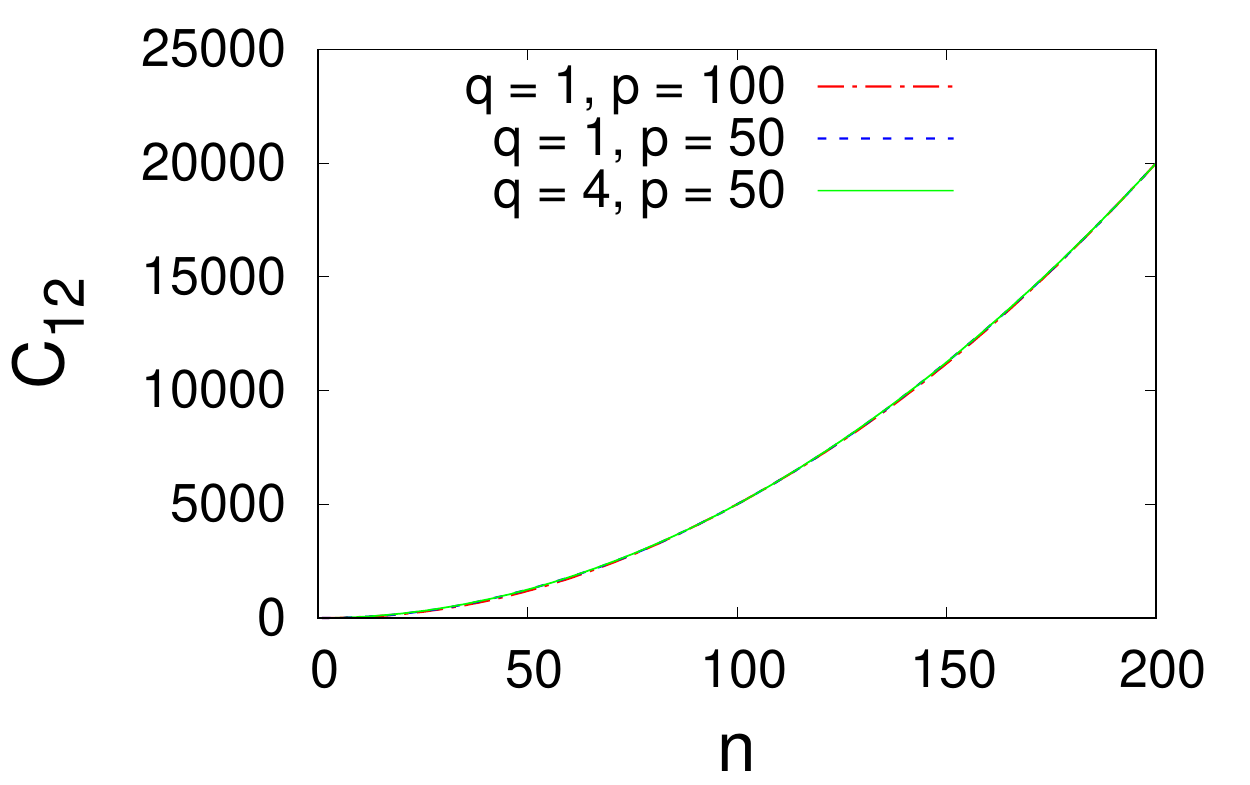}}
\subfigure[\label{fig:13b} ]{\includegraphics[scale=0.45]{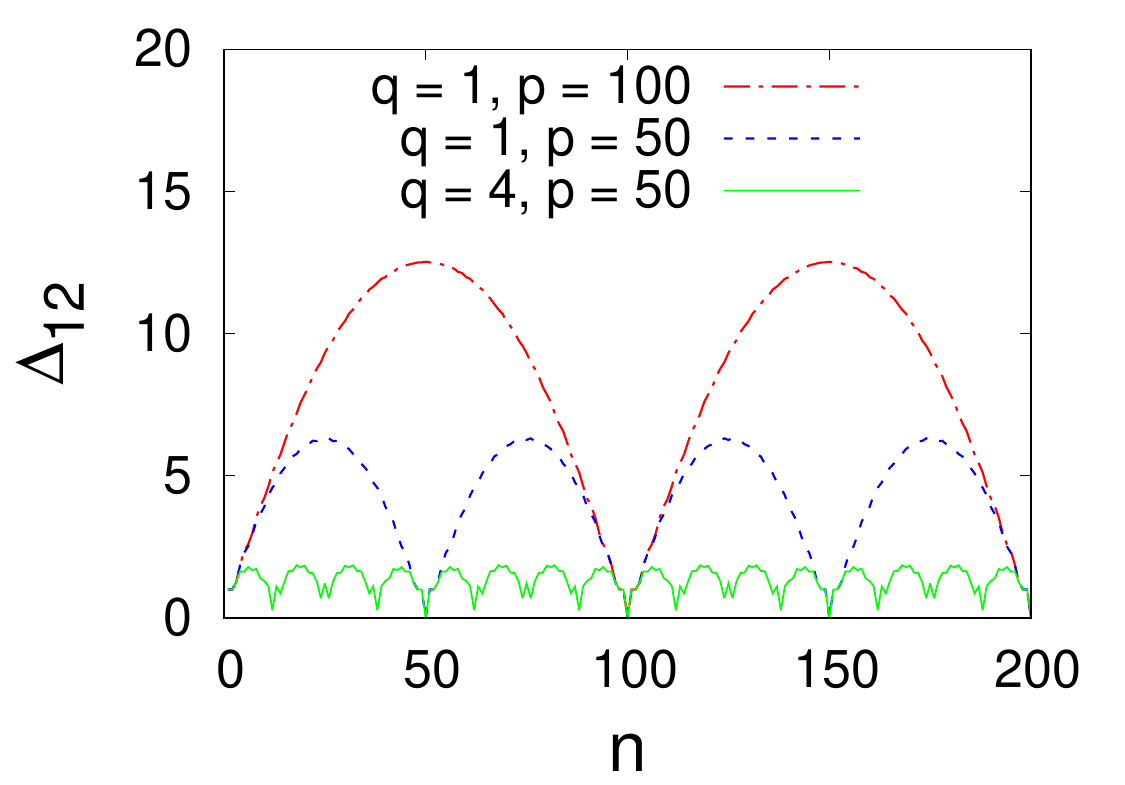}}
\subfigure[\label{fig:13c} ]{\includegraphics[scale=0.45]{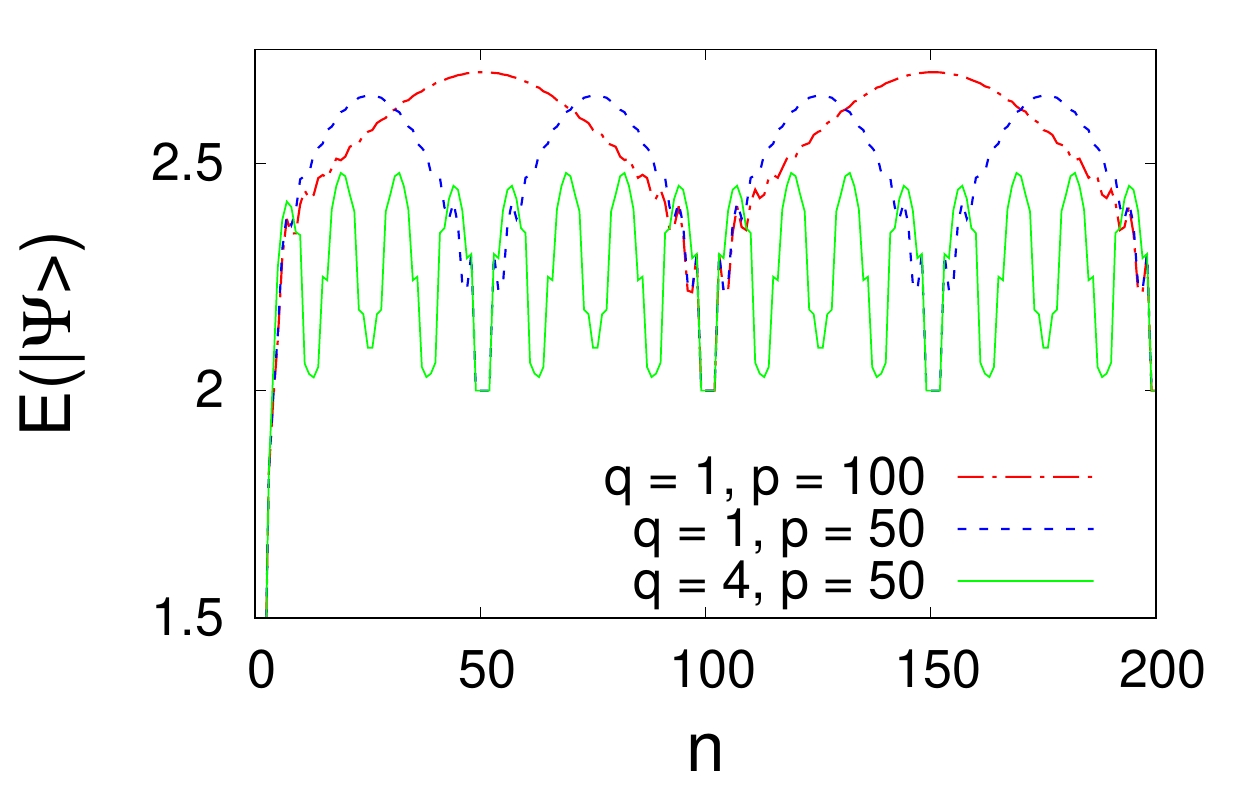}}
\subfigure[\label{fig:14a} ]{\includegraphics[scale=0.45]{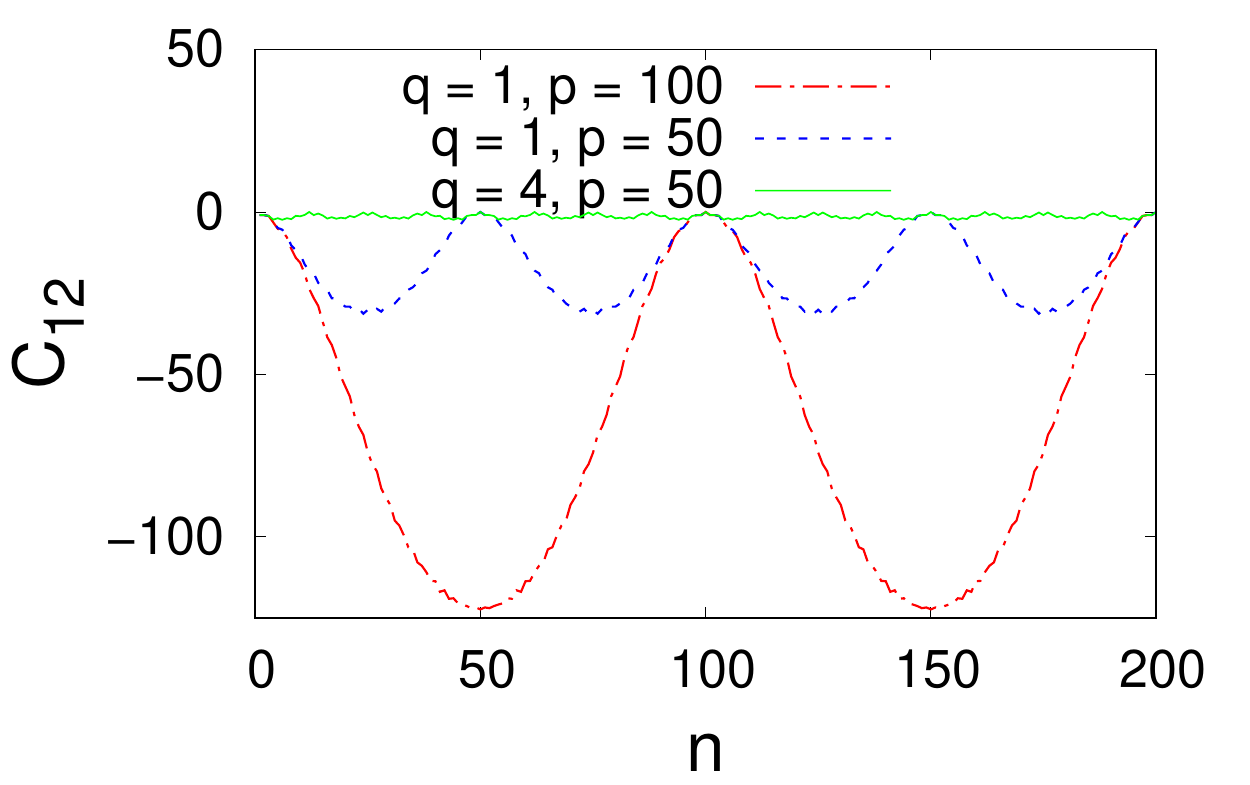}}
\subfigure[\label{fig:14b} ]{\includegraphics[scale=0.45]{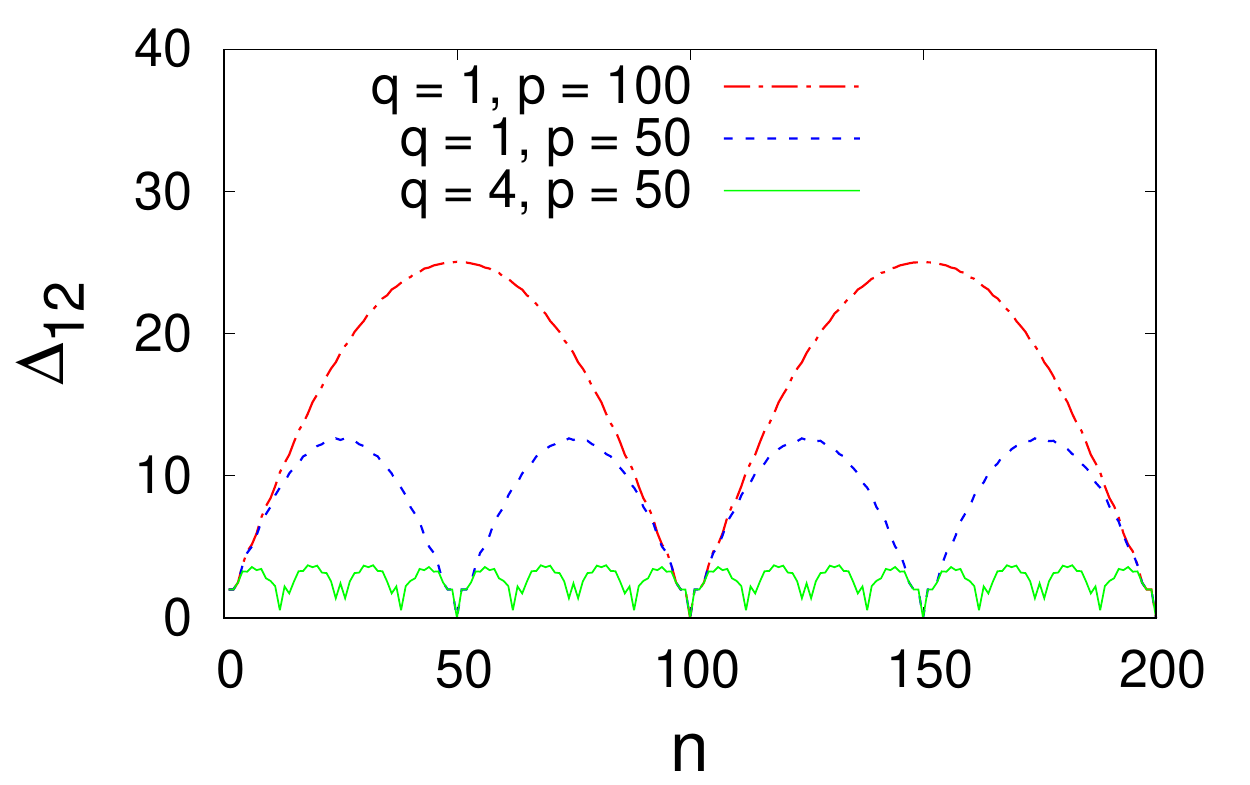}}
\subfigure[\label{fig:14c} ]{\includegraphics[scale=0.45]{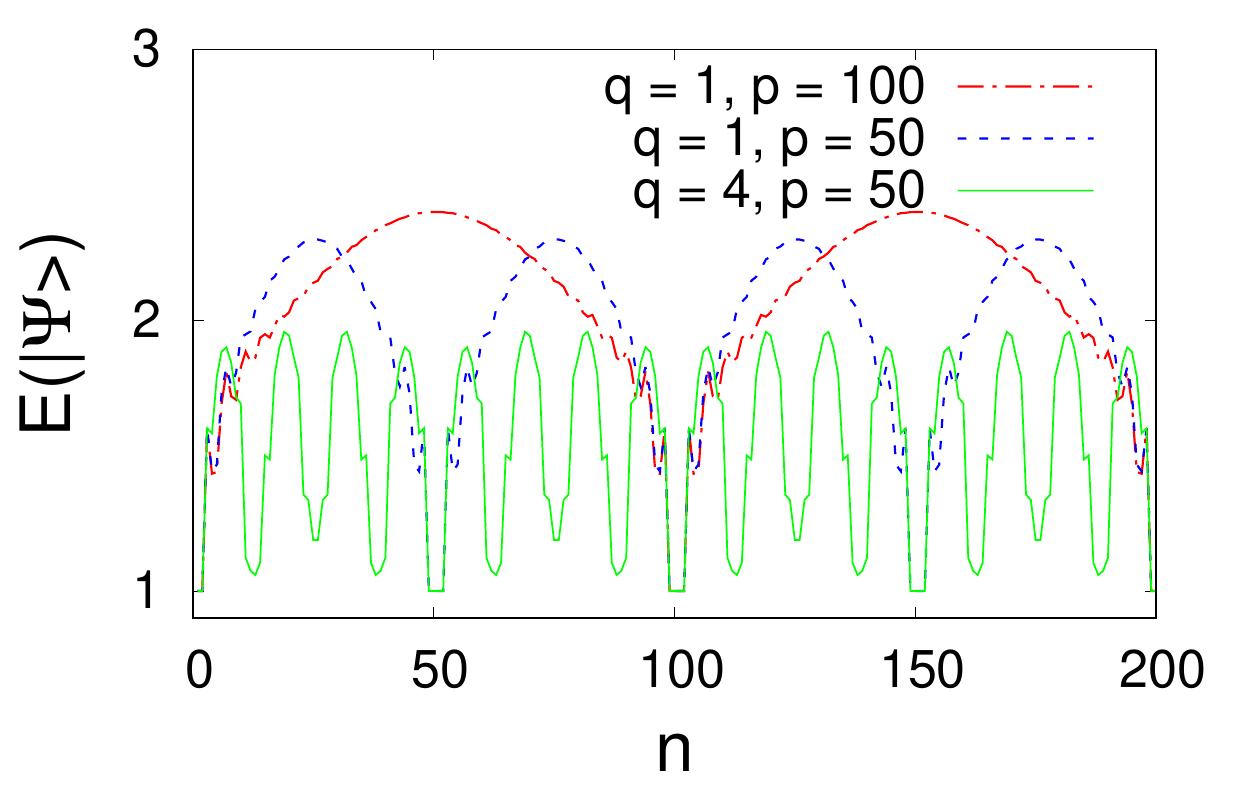}}
\subfigure[\label{fig:15a} ]{\includegraphics[scale=0.45]{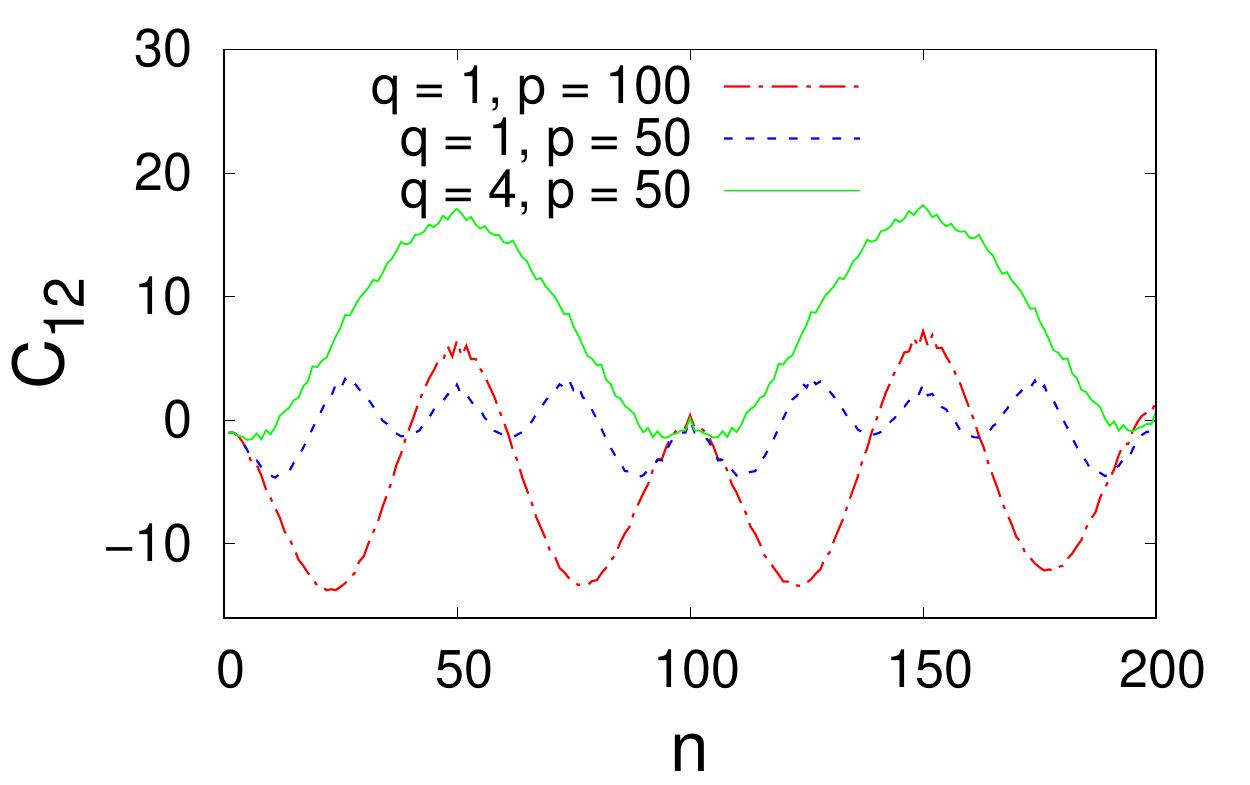}}
\subfigure[\label{fig:15b} ]{\includegraphics[scale=0.45]{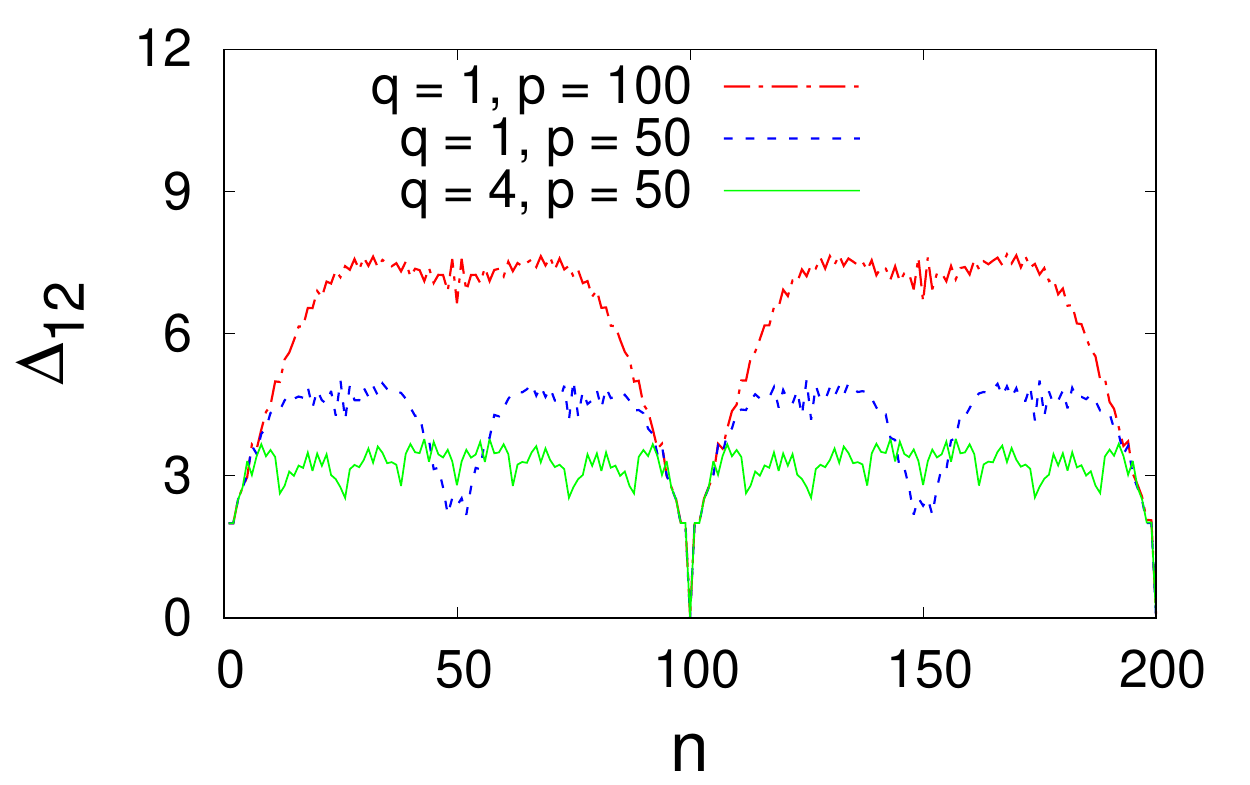}}
\subfigure[\label{fig:15c} ]{\includegraphics[scale=0.45]{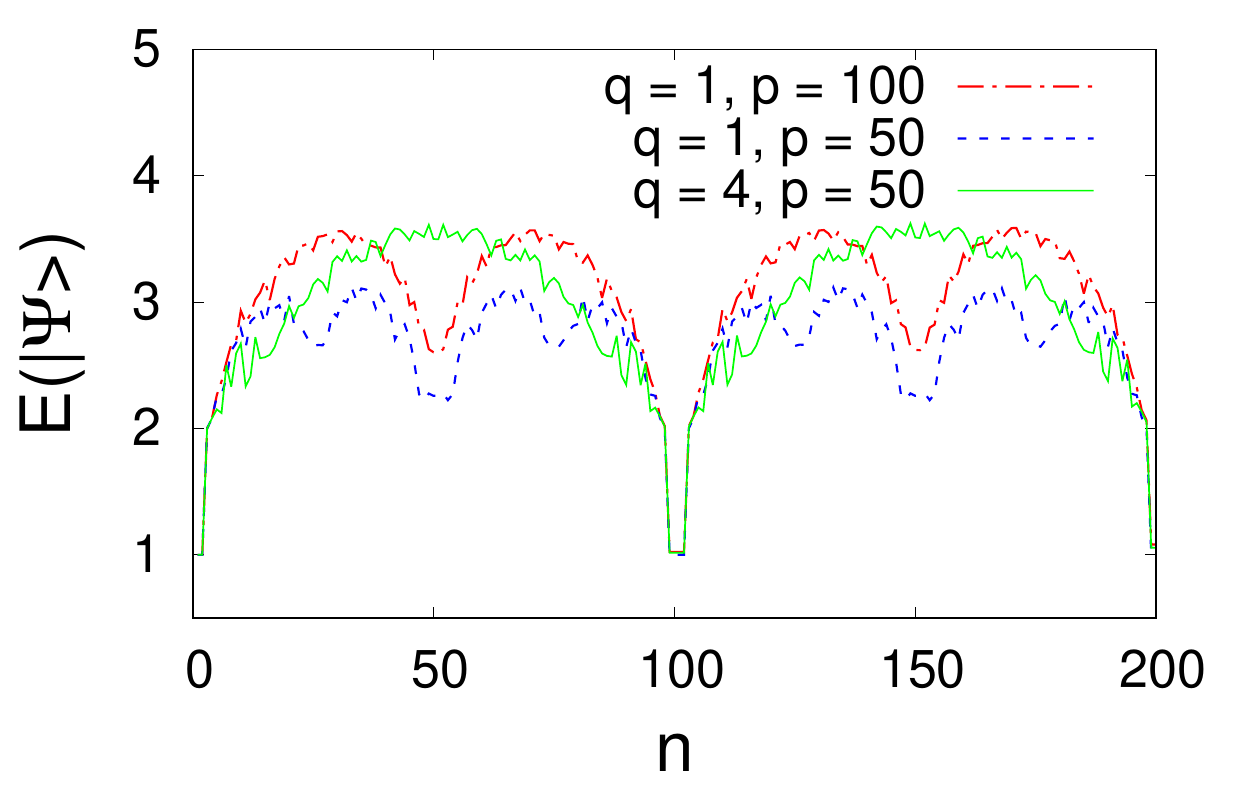}}

\caption{\label{fig:13}{The time-variation of the joint properties for two \(\mathbb{1}\)-interacting quantum walkers : (a) Variations of \(C_{12}\) in case of \(|Sep\rangle\); (b) Variations of \(\Delta_{12}\) in case of \(|Sep\rangle\), (c) Variations of \(E(|\psi\rangle)\) in case of \(|Sep\rangle\), (d) Variations of \(C_{12}\) in case of \(|\psi^{+}\rangle\); (e) Variations of \(\Delta_{12}\) in case of \(|\psi^{+}\rangle\), (f) Variations of \(E(|\psi\rangle)\) in case of \(|\psi^{+}\rangle\),(g) Variations of \(C_{12}\) in case of \(|\psi^{-}\rangle\); (h) Variations of \(\Delta_{12}\) in case of \(|\psi^{-}\rangle\), (i) Variations of \(E(|\psi\rangle)\) in case of \(|\psi^{-}\rangle\). The coin parameters are mentioned inside the plots.}
 }
\end{figure*}

\subsection{The case of time-dependent coin \(\hat{C}_{\Phi}(t)\)}

As described earlier in Sec. \ref{coins}, the time-dependent coin \(\hat{C}_{\Phi}(t)\) generates dynamical localization phenomenon in case of single particle QW. The fate of such behavior in case of two particle QW is a non-trivial question. How do the interactions and initial states influence the coin-parameter dependence of such QW evolution is also difficult to answer. Here we perform extensive numerical simulations in order to answer these questions. The related \(4\times4\) matrix \(\hat{C}_{\Phi,\Phi}\) has been described earlier in Sec. \ref{five}. We present the results mainly for three different sets of coin parameters \(q\) and \(p\) : (1) \(q=1,p=100\), (2) \(q=1,p=50\) and (3) \(q=4,p=50\) \cite{parameters2}.\\

The dynamical localization of a single quantum walker can be detected from the periodic variation of its probability distribution \(P(x)\) with time. Alternatively, one can also consider the periodic time-variation of the standard deviation as a signature of dynamic localization (see Fig.\ref{fig:110b} and related discussions in Sec. \ref{coins}). For the two particle case, we have shown below the temporal behavior of the collective dynamical properties of the walkers which bear clear signatures of both the presence and absence of the two body dynamical localization phenomena in different cases. In addition, some plots of \(P(x,y)\) are also provided in the supplementary material \cite{oscillatory_spreading}.

\subsubsection{Dynamics of two non interacting walkers under the influence of \(\hat{C}_{\Phi}(t)\) }

Let us first describe the results obtained in the simplest case, i.e., the case of two non-interacting walkers starting from \(|Sep\rangle\) initial state. In this case, the two-particle probability distribution \(P(x,y)\) is simply the product of two single walker probability distributions. Both the single walkers perform mutually independent dynamical localizations controlled by the coin parameters \(p\) and \(q\). So, as expected, the system exhibits coin-parameter dependent ``two-body dynamical localization".  \(P(x,y)\) has both bunching and anti-bunching peaks which start moving away from the origin with time and periodically return to the same point simultaneously after an interval of \(p\) steps. 
Therefore, \(\Delta_{12}\) exhibits periodic oscillations as the anti-bunching peaks contribute to \(\Delta_{12}\). We have shown the variation of \(\Delta_{12}\) against time for three different combinations of \(q\) and \(p\) in Fig. \ref{fig:10a}. The plots show that the period of oscillation is \(p\) and the number of secondary oscillation is \(q\) as was found in case of dynamical localization of single particle. The amplitude of the periodic oscillations depends on the values of both \(q\) and \(p\). It increases with \(p\). For a fixed value of \(p\), the amplitude decreases with increasing \(q\). Since there is no interaction, the correlation function \(C_{12}\) and the entropy \(E(|\psi\rangle)\) remain equal to zero.\\

The entangled bosonic and fermionic initial states generate positional correlations between the particles. We analyze the temporal variations of the correlation function \(C_{12}\) to distinguish and characterize the two-body dynamical localization phenomena generated by the two different initial states. Earlier, we demonstrated that \(|\psi^{-}\rangle\) state generates pure anti-bunching phenomena for the non-interacting walkers with time-independent Hadamard coins (the case of \(C_{\alpha_{1},\alpha_{2}}\) with \(\alpha_{1}=\alpha_{2}=0\) ). \(C_{12}\) remained negative throughout that evolution. Here also, we see that \(\hat{C}_{12}\) is negative during the evolution, but it exhibits distinct periodic oscillations of period \(p\) (see Fig.\ref{fig:12a}). Since the particles become anti-correlated, we call this phenomenon ``anti-correlated" two-body dynamical localization. The amplitude and period of oscillations of both \(C_{12}\) and \(\Delta_{12}\) are controlled by the parameters \(p\) and \(q\). 
The nature of time variation of \(\Delta_{12}\), shown in Fig. \ref{fig:12b}, is similar to that observed in case of \(|Sep\rangle\) state. The only difference is that the amplitude of oscillation of \(\Delta_{12}\), for fixed values of \(p\) and \(q\), is higher in case of \(|\psi^{-}\rangle\) as no terms of \(|\psi^{-}\rangle\) contributes to bunching behavior in difference to \(|Sep\rangle\).\\

\(|\psi^{+}\rangle\) state generates a quite different type of dynamical localization. The difference can be seen in the time-variation of the correlation function \(C_{12}\) shown in the figure \ref{fig:11a}.  In case of Hadamard coins, \(C_{12}\) remained positive for \(|\psi^{+}\rangle\) throughout the evolution. Here also, \(C_{12}\) mostly remains positive during the evolution. At a first glance, the plots of \(C_{12}\) (Fig. \ref{fig:11a}) may also look like inverted mirror image of the respective plots (Fig. \ref{fig:12a}) of \(C_{12}\) obtained in case of \(|\psi^{-}\rangle\). However, here \(C_{12}\) exhibits periodic oscillation between positive and negative values through zero value. So, the walkers periodically become mutually correlated-uncorrelated-anti-correlated with time. It is interesting to note that even for bosonic \(|\psi^{+}\rangle\) state, \(C_{\phi}(t)\) makes the particles mutually anti-correlated for certain short period of time. Since the particles mostly remain remain correlated, we call this phenomenon ``correlated" two-body dynamic localization. The related time variation of \(\Delta_{12}\), shown in Fig. \ref{fig:11b}, is slightly different in comparison to the above described case of \(|Sep\rangle\) state. 
The amplitude of \(\Delta_{12}\) becomes highest in case of \(|\psi^{-}\rangle\) among the three different initial states.\\


The system exhibits Interesting phenomena when the walkers are controlled by two different coins \(\hat{C}_{H}\) and \(\hat{C}_{\phi}(t)\). The first one is the Hadamard coin and the other one is the coin generating dynamical localization. The related \(4 \times 4\) matrix \(\hat{C}_{H,\Phi}\) has already been described in section \ref{five}. With \(\hat{C}_{H,\Phi}\), we have studied the QW evolution starting from \(|\psi^{+}\rangle\) and \(|\psi^{-}\rangle\) states. The average separation \(\Delta_{12}\), as shown in Figs. \ref{fig:11d} and \ref{fig:12d} respectively for \(|\psi^{+}\rangle\) and \(|\psi^{-}\rangle\) states , grows rapidly in both cases but there are clear signatures of oscillations in \(\Delta_{12}\), specially in Fig. \ref{fig:12d}. 
Such oscillatory behavior indicates dynamical localization of one of the particles. The corresponding variations of \(C_{12}\)  also show oscillations of increasing amplitude with a time period of \(p\). The plots of \(C_{12}\) are shown in Figs. \ref{fig:11c} and \ref{fig:12c} respectively for \(|\psi^{+}\rangle\) and \(|\psi^{-}\rangle\) states. For \(|\psi^{-}\rangle\), \(C_{12}\) periodically becomes positive and negative. It indicates that one of the particle is moving away from the origin with time whereas the other particle is performing dynamic localization around origin. For \(|\psi^{+}\rangle\), \(C_{12}\) remains mostly positive during evolution and its amplitude of oscillation increases with time. It indicates that one of the particles is going further away from the origin with time and the other particle is performing periodic oscillation with the center of oscillation a bit shifted from the origin in the direction of motion of the first particle.  It is interesting to note that even for \(|\psi^{-}\rangle\), the particles become correlated for certain amount of time.\\

Previously, we described the dynamics of the system under the influence of \(\hat{C}_{\alpha_{1},\alpha_{2}}(t)\) (\(\alpha_{1}=0,\alpha_{2}=1.25\)), (Figs. \ref{fig:2g2} and \ref{fig:3g2}), i.e. the case where the system was evolving under the influence of two coins of contrasting nature. 
In that case,  the particles became correlated, uncorrelated and anti-correlated periodically with time when they start evolving from a (\(|\psi^{+}\rangle\) initial state. On the contrary, for the combination of \(\hat{C}_{H}\) and \(\hat{C}_{\phi}(t)\), the particles remain correlated throughout the evolution (Fig. \ref{fig:11c}).\\



\subsubsection{Dynamics of two \(\mathbb{1}\) interacting walkers under the influence of \(\hat{C}_{\Phi}(t)\)}

Here we describe QW evolutions for \(\mathbb{1}\) interactions between two particles. \(\mathbb{1}\) interaction generates quite different evolutions for three different considered initial states. 
Let us first describe the QW evolution starting from \(|Sep\rangle\) initial state. In this interacting walk, the identity operator acts on the \(|0, \uparrow ; 0, \uparrow \rangle\) and \(|0, \downarrow ; 0, \downarrow \rangle\) terms of the \(|Sep\rangle\) state. The identity operator does not mix the coin states and as a result the particles are translated together in the same direction at each time step. This behavior is independent of the coin parameters as the coin operator does not get a chance to act on the above two terms. The related time variation of \(C_{12}\) is shown in Fig. \ref{fig:13a}. It is observed that \(C_{12}\) rapidly increases with time. Since this strong bunching behavior is independent of the coin parameters, all the curves for different parameter sets overlap each-other (see Fig.\ref{fig:13a}). The contribution of the other terms in \(|Sep\rangle\) state is much weaker than the contribution of the above terms. The contributions of the other terms can be seen in the variations of \(\Delta_{12}\) and \(E(|\psi\rangle)\). Such strong bunching behavior was also observed, due to the same reasons, in case of \(\hat{C}_{\alpha_{1},\alpha_{2}}(t)\) coins.\\ Here the particles become entangled as a result of the interaction and the entanglement entropy \(E(|\psi\rangle)\) oscillates periodically (see Fig.\ref{fig:13c}) about a certain value which appears to be a constant contribution to \(E(|\psi\rangle)\) coming from the \(|0, \uparrow ; 0, \uparrow \rangle\) and \(|0, \downarrow ; 0, \downarrow \rangle\) terms. The temporal variations of the average separation \(\Delta_{12}\) also show periodic oscillations. The figures \ref{fig:13b} and \ref{fig:13c} clearly indicate the influences of \(q\) and \(p\) on the amplitude and time-period of oscillations. For any set of coin parameters, the particles become more entangled as the average separation \(\Delta_{12}\) increases. The particles become minimally entangled when the average separation \(\Delta_{12}=0\).\\


For bosonic \(|\psi^{+}\rangle\) initial state, a unique scenario is obtained. All the three observables, including the entanglement entropy, exhibit coin parameter dependent tunable periodic oscillations for all the three sets of coin parameters (see Figs.\ref{fig:14a}-\ref{fig:14c} ). It is quite non-trivial that all the three observables exhibit such behavior. So, the signatures of two-walker dynamic localization becomes evident in all the joint properties of quantum origin. The particles start from the origin being uncorrelated (\(C_{12}=0\)) and minimally entangled (\(E(|\psi\rangle)=1\)). Then for the first half of the time-period, both the average separation \(\Delta_{12}\) and entanglement entropy \(E(|\psi\rangle)\) increases with time as the particles become more and more anti-correlated. Then during the next half of the time-period, both \(\Delta_{12}\) and \(E(|\psi\rangle)\) decreases simultaneously whereas \(C_{12}\) increases. At the end of each time-period, both the particles return to origin being uncorrelated and minimally entangled(\(E(|\psi\rangle)=1\)) (i.e., exactly the state of the system at time t=0). The time period and amplitude can be controlled by the coin parameters \(p\) and \(q\). The related figures are \ref{fig:14a}-\ref{fig:14c}\cite{oscillatory}. The tunable periodic oscillations of the entanglement entropy \(E(|\psi\rangle)\) is quite interesting. Here, the quasiperiod can be controlled by the parameter \(p\). There are also secondary oscillations controlled by \(q\). So, this is an example of a simple two particle system where we can generate controllable periodic oscillation of \(E(|\psi\rangle)\).\\

For \(|\psi^{+}\rangle\) initial state, the correlation function oscillates between zero and a certain negative value. So, the walkers mostly remain anti-correlated during the evolution. This is contrary to the case of time independent coins \(\hat{C}_{\alpha_{1},\alpha_{2}}\)( with \(\alpha_{1}=\alpha_{2}=0\)) where the particles remains correlated starting from \(|\psi^{+}\rangle\) initial state as the bunching behavior dominates over the anti-bunching phenomenon.\\ 

For \(|\psi^{-}\rangle\) initial state, the collective dynamics and its coin parameter dependence become more complex. Although, the system still exhibits  two-body dynamical localizations with tunable periodic oscillations of the three observables for some (\(q,p\)) combinations, the \(q,p\) dependences of the time-period and the number of secondary oscillations becomes more complex in comparison to that found in the previously described cases of \(|Sep\rangle\) and \(|\psi^{+}\rangle\) states. The temporal variations of the dynamical observables are shown in the figures \ref{fig:15a}-\ref{fig:15c}. The plots of \(\Delta_{12}\) show flatter peaks which is also a difference with the previously discussed cases. Fig. \ref{fig:15a} shows that the particles mostly remain correlated during their evolution for \(q=4,p=50\) whereas they mostly remain anti-correlated for \(q=1,p=100\). So, the nature of the correlations here depends strongly on the coin parameters. It can also be seen that the amplitudes of the periodic oscillations of \(C_{12}\) and \(\Delta_{12}\) are smaller than that found in the previously discussed cases (both for \(q=1,p=50\) and \(q=1,p=100\)). In the previously discussed cases of dynamical two body localization, the number of secondary oscillations of the dynamical observables was equal to the value of the parameter \(q\). The influence of the parameter \(q\) is more complex here. In order to understand the influence of the coin parameters, we have studied the dynamics for some more different values of  \(q\) and \(p\) \cite{parameters}. We find that in most cases the system exhibits dynamic localization where the time period of oscillation of the dynamic observables is either \(p\) or \(2p\). However, we couldn't see any such simple relationship between the number of secondary oscillation and \(q\). Similarly, the results do not indicate any simple dependence of the oscillation amplitudes on \(q,p\). It is interesting to note that a new phenomenon is found for some (\(q,p\)) combinations in which the probability distribution \(P(x,y)\) spreads with time in contrary to being localized. We call it ``oscillatory spreading" as \(P(x,y)\) exhibits time dependent oscillations. Such behavior is also observed for \(\pi\)-phase interactions. We describe the phenomenon in more detail in next paragraph with the help of figures \ref{fig:16a}-\ref{fig:17c}.\\

\subsubsection{Dynamics of \(\pi\)-phase interacting walkers under the influence of \(C_{\Phi}(t)\)}

\begin{figure*}[th]
\centering
\subfigure[\label{fig:16a} ]{\includegraphics[scale=0.45]{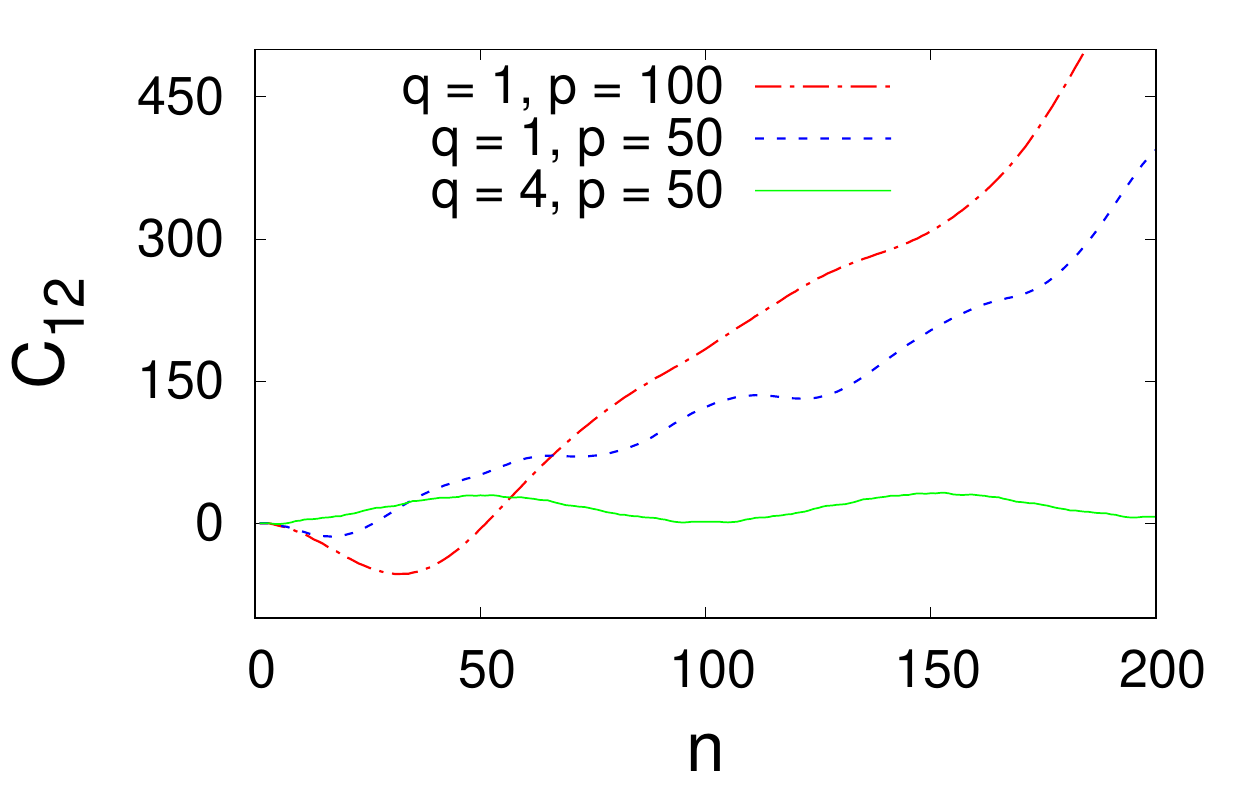}}
\subfigure[\label{fig:16b} ]{\includegraphics[scale=0.45]{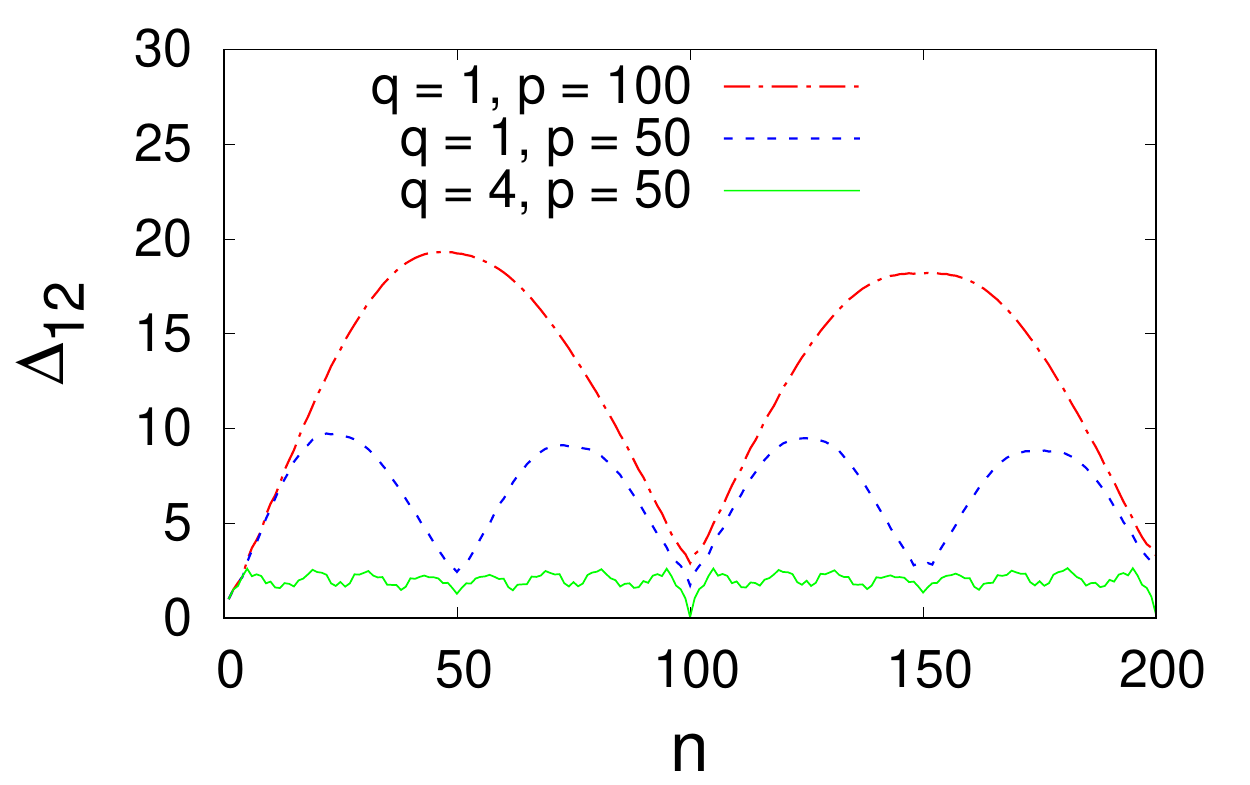}}
\subfigure[\label{fig:16c} ]{\includegraphics[scale=0.45]{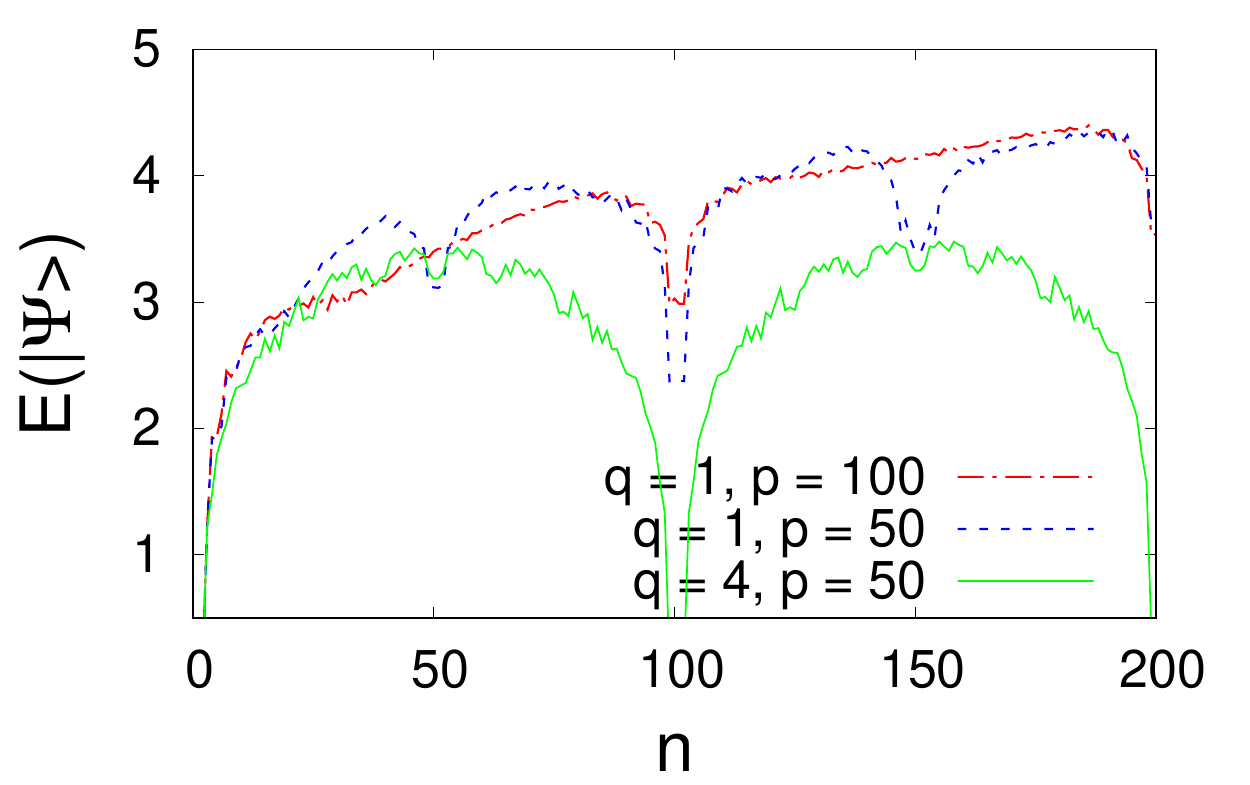}}
\subfigure[\label{fig:17a} ]{\includegraphics[scale=0.45]{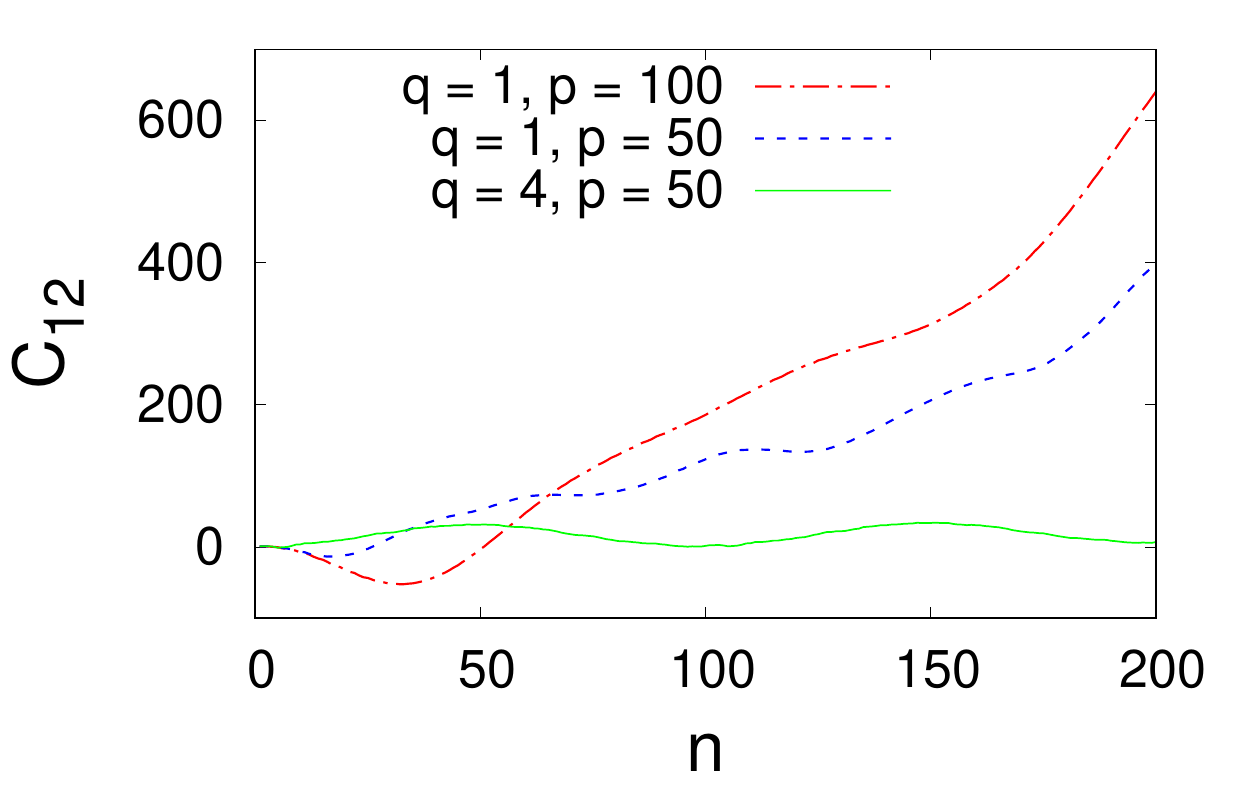}}
\subfigure[\label{fig:17b} ]{\includegraphics[scale=0.45]{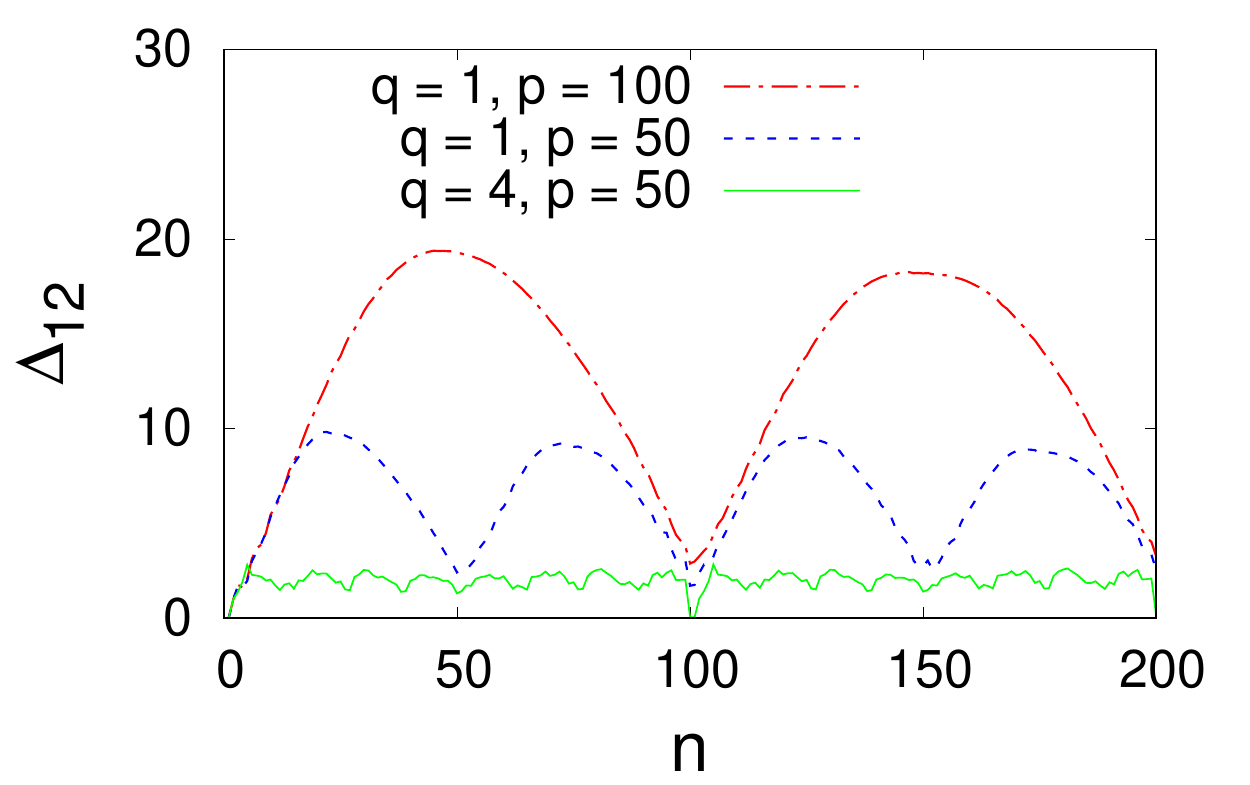}}
\subfigure[\label{fig:17c} ]{\includegraphics[scale=0.45]{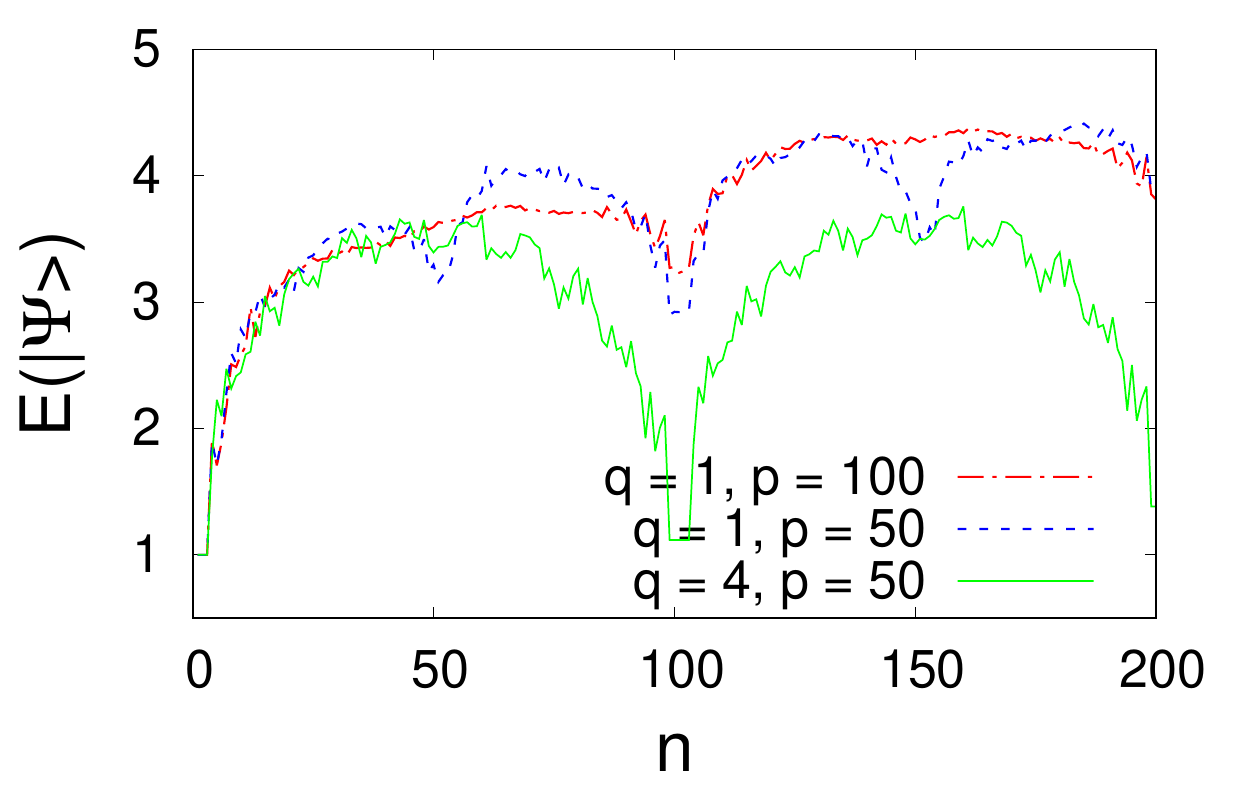}}
\subfigure[\label{fig:18a} ]{\includegraphics[scale=0.45]{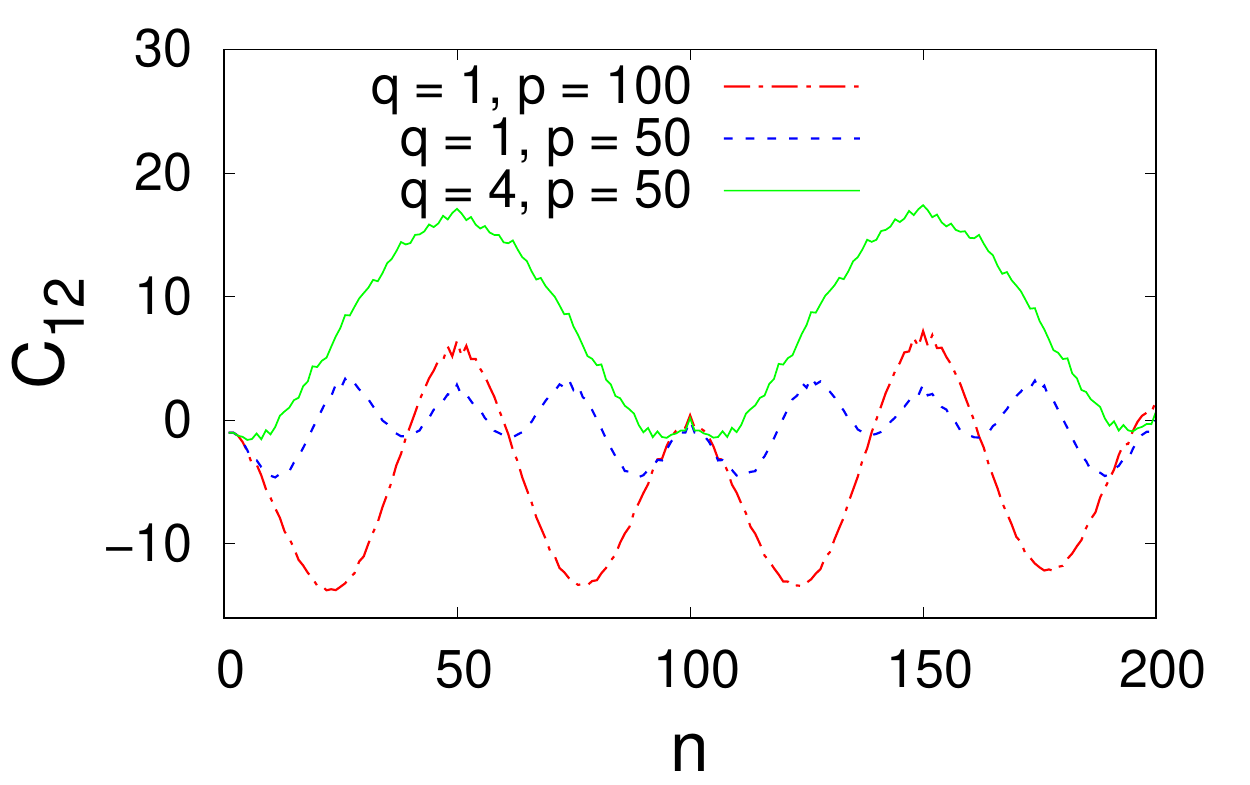}}
\subfigure[\label{fig:18b} ]{\includegraphics[scale=0.45]{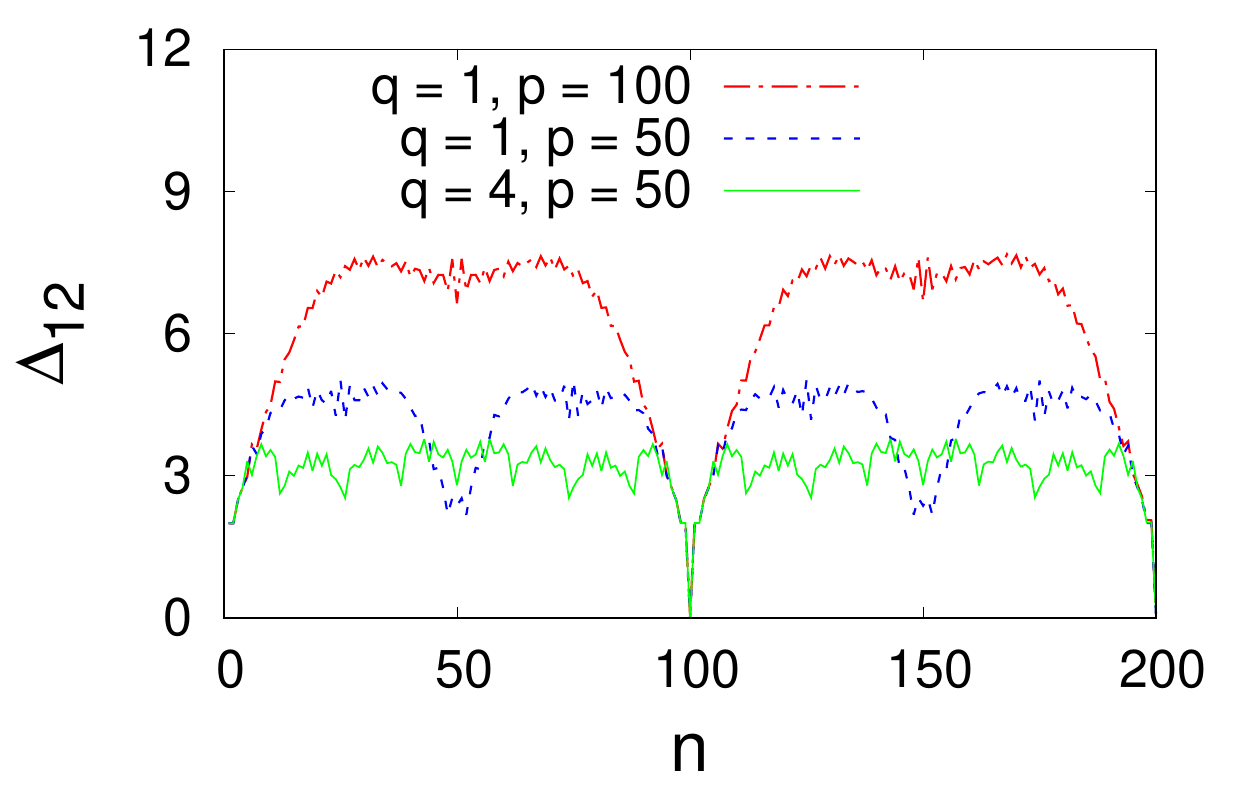}}
\subfigure[\label{fig:18c} ]{\includegraphics[scale=0.45]{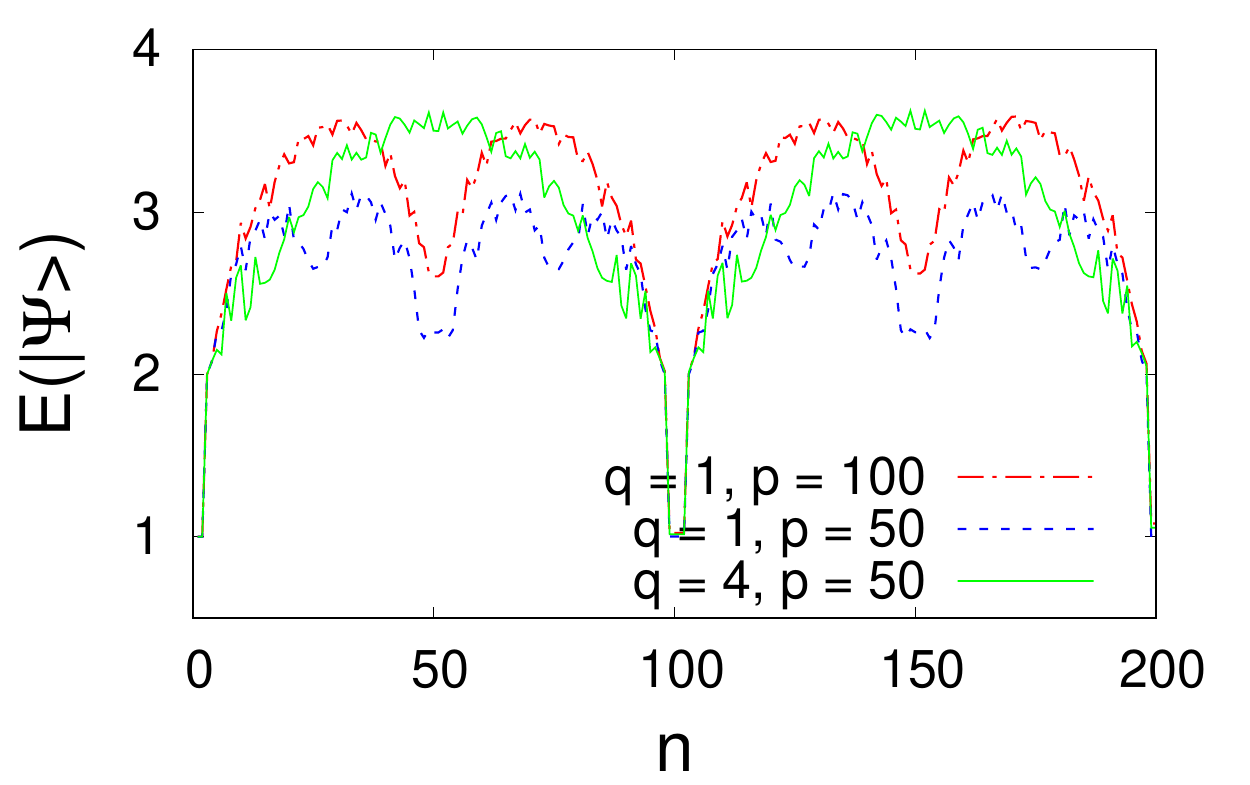}}
\caption{\label{fig:16}{The time-variation of the joint properties for two \(\pi\)-phase interacting quantum walkers : (a) Variations of  \(C_{12}\) in case of \(|Sep\rangle\); (b) Variations of \(\Delta_{12}\) in case of \(|Sep\rangle\), (c) Variations of \(E(|\psi\rangle)\) in case of \(|Sep\rangle\), (d) Variations of \(C_{12}\) in case of \(|\psi^{+}\rangle\); (e) Variations of \(\Delta_{12}\) in case of \(|\psi^{+}\rangle\), (f) Variations of \(E(|\psi\rangle)\) in case of \(|\psi^{+}\rangle\), (g) Variations of \(C_{12}\) in case of \(|\psi^{-}\rangle\); (h) Variations of \(\Delta_{12}\) in case of \(|\psi^{-}\rangle\), (i) Variations of \(E(|\psi\rangle)\) in case of \(|\psi^{-}\rangle\) initial state. The coin parameters are mentioned inside the plots.}
 }
\end{figure*}

When the particles interact via \(\pi\)-phase interaction and start from either \(|Sep\rangle\) or \(|\psi^{+}\rangle\) state, we find dynamic localization only for some specific combinations of the coin parameters \((q,p)\) among the different considered ones \cite{parameters}. For the other values of \((q,p)\), the particles remain correlated and the corresponding two particle probability distribution \(P(x,y)\) not only spreads with time but also exhibits oscillatory behavior. We call this phenomenon ``correlated oscillatory spreading" \cite{oscillatory_spreading}.  
The oscillatory spreading behavior is found for \(q=1,p=50\) and \(q=1,p=100\) (see Figs.\ref{fig:16a}-\ref{fig:17c}) \cite{oscillatory_spreading}. 
After some initial period, the two particles become more and more correlated with time as shown in the Figs. \ref{fig:16a} and \ref{fig:17a}. The entanglement entropy also increases slowly with time. Let us now discuss the dynamic localization behavior found in case of \(\pi\)-phase interactions. Here, \(q=4,p=50\) generates dynamic localization behavior (see Figs.\ref{fig:16a}-\ref{fig:17c}). However, the simple dependence (\(p\equiv\) period of oscillations, \(q\equiv\) number of secondary oscillations) on the coin parameter gets modified in this case. In order to understand the influence of the coin parameters, we have studied the dynamics for some more different values of  \(q\) and \(p\) \cite{parameters}. We find that only in few cases the system exhibits dynamic localization where the time period of oscillation of the dynamic observables is  \(2p\). However, we couldn't see any such simple relationship between the number of secondary oscillation and \(q\). Similarly the results do not indicate any simple dependence of the oscillation amplitudes on \(q,p\). Such modification was also found in the previously described case of \(\mathbb{1}\) interaction and \(|\psi^{-}\rangle\) initial state.  \\

When the particles start from the \(|\psi^{-}\rangle\) initial state, the dynamics is found to be quite similar to that observed in case of two \(\mathbb{1}\) interacting particles starting from \(|\psi^{-}\rangle\) state (see Figs.\ref{fig:18a}-\ref{fig:18c}). Even in case of time-independent \(C_{\alpha_{1},\alpha_{2}}\)(\(\alpha_{1}=\alpha_{2}=0\)) coins, \(|\psi^{-}\rangle\) initial state generated qualitatively similar dynamics in cases of \(\mathbb{1}\) and \(\pi\)-phase interactions. 
So, for \(\pi\)-phase interaction, the simple dependence  (\(p\equiv\) period of oscillations, \(q\equiv\) number of secondary oscillations) on the coin parameter gets modified for all the three different initial states.\\%

\subsubsection{Overall diversity in the \(\hat{C}_{\phi}(t)\) driven dynamics}
The time dependent coin \(\hat{C}_{\phi}(t)\) generates a wide-spectrum of two-body dynamical phenomena depending on the nature of interaction, initial state and the coin parameters.  In some cases, the system exhibits ``simple" dynamic localization where the \(q\),\(p\) dependence of the periodic oscillations of the dynamic observables is simple. Moreover, such phenomena can be divided into two category : (I) Correlated dynamic localization and (II) Anti-correlated dynamic localization depending on the nature of the two different entangled initial states. In some other cases, the system exhibits ``complex" dynamical localization where the \(q,p\) dependence of the periodic oscillations of the dynamic observables is more complicated. Moreover, the nature of positional correlation in these cases becomes dependent on the coin-parameters. Finally, the system also exhibits a third kind of dynamical phenomena : ``oscillatory spreading" where the system does not localize and the corresponding probability distribution shows oscillatory behavior alongwith spreading.\\

We have found particular cases of two body dynamic localization in which all three joint quantities (\(\Delta_{12}\),\(C_{12}\mbox{ and }E(|\psi\rangle)\)) perform periodic oscillations with coin-parameter dependent amplitude and time-period.\\ 

We have to remember that these are transient behavior, obtained for rational values of \(\frac{q}{p}\) \cite{coin2}. At long times we should expect ballistic dynamics at the level of individual particles. The dynamical localization phenomena can be further modified by using two different coins of kind \(\hat{C}_{\Phi}(t)\) with two different sets of \(q,p\) parameters. However, we have already presented a rich variety of dynamics by just focussing on the simpler cases which could be helpful for developing an understanding for more complex cases. Some of the cases studied here needs a better understanding which is left for future work.\\

\section{Conclusion \& Outlook\label{eight}}

Controlling QWs is a topic of current interest \cite{controlling}. Researchers have been prescribing new type of quantum coins (step-dependent coins \cite{step-dependent}, position dependent coins \cite{position-dependent}, a combination of two entangled coins \cite{two-entangled-coins} etc.) and different types of shift operators \cite{shift-operator} to manipulate QW dynamics of a single particle. We have presented here a first numerical simulation on controlling two-particle walks using time-dependent coins. Although various types of time-dependent coins can be constructed, we have considered here two specific time-dependent coins which enable a realization of wide spectrum of dynamical behavior. 
The results presented here can be considered as a generalization of two particle QW dynamics on a line. Some of the reported behaviors can be obtained in different situations even without the use of time-dependent coins. For example, two quantum walkers in presence of decoherences \cite{decoherence} are expected to generate dynamics qualitatively similar to that demonstrated using coin \(\hat{C}_{\alpha_{1},\alpha_{2}}(t)\) with \(\alpha_{1}=\alpha_{2}=0.5\). Position dependent phases \cite{15,16}, if employed in a system of two particles, are expected to generate two-body dynamical localization and oscillatory spreading phenomena qualitatively similar to that reported here.\\

One obvious extension of the present work would be to explore the dynamics on more complicated graphs. We have studied here the collective dynamics of two quantum particles. One can also study the dynamics of a multi-particle system using the time-dependent coins. We have considered here only \(\mathbb{1}\) and \(\pi\)-phase interactions. An interesting extension of the present work would be to consider other types of possible two-particle interactions such as long range interactions \cite{long-range-interactions}. Our work will form a base for understanding such results by comparing and contrasting those results with the results presented here.\\
We have also studied here the collective dynamics of two particles of different nature. Similar studies with multiple particles will help to answer the following question : How far can we manipulate the collective dynamical behavior of a group of quantum particles by tuning only the behavior of a single particle ? It is quite possible that a single particle with tunable dynamics in a group of particles can influence the collective dynamics in the presence of multi-particle quantum correlation and multipartite entanglement. However, an experimental realization of such a theoretically interesting idea may be a difficult problem.\\

\section{Acknowledgements}   
We thank Professor Parongama Sen for a careful reading of the manuscript and for her helpful comments and suggestions. T.K. Bose thanks Sudhindu Bikash Mandal for helpful discussions.\\

\bibliographystyle{revtex}
\bibliography{references}

\begin{thebibliography}{99}
\bibitem{parameters2}
The chosen set of parameters satisfy the conditions that the rational number \(q/p\) is an irreducible fraction and \(0\le \frac{q}{p}<\frac{1}{2}\). As described in Ref. \cite{coin2}, parameters satisfying such conditions are expected to generate dynamic localization behavior for a period of time \(p\) in case of single particle QW.
\end{thebibliography}
{}

\end{document}